\documentclass[%
onecolumn,
longbibliography,
amsmath,
amssymb,
aps,
]{revtex4-1}

\usepackage{graphicx}
\usepackage{dcolumn}
\usepackage{bm}
\usepackage{epsfig}
\usepackage{tabularx}
\usepackage{color}
\usepackage{verbatim}

\begin{document}
\preprint{APS/123-QED}

\title{Mesoscopic model for soft flowing systems with tunable viscosity ratio}

\author{Linlin Fei,$^{1,2}$  Andrea Scagliarini,$^{2}$ Andrea Montessori,$^{2,3}$ Marco Lauricella,$^{2}$ Sauro Succi,$^{2,4,5}$ and Kai H. Luo$^{1,6,}$\footnote{Corresponding author: K.Luo@ucl.ac.uk}}

\affiliation{$^1$ Center for Combustion Energy; Key laboratory for Thermal Science and Power Engineering of Ministry of Education; Department of Energy and Power Engineering, Tsinghua University, Beijing 100084, China\\
	$^2$ Istituto per le Applicazioni del Calcolo, Consiglio Nazionale delle Ricerche, Via dei Taurini 19, 00185 Rome, Italy\\
	$^3$ Department of Engineering, University of Rome “Roma Tre,” Via della Vasca Navale 79, 00141 Rome, Italy\\
	$^4$  Center for Life Nano Science at La Sapienza, Istituto Italiano di Tecnologia, 295 Viale Regina Elena, I-00161 Roma, Italy\\
	$^5$ Harvard Institute for Applied Computational Science, Cambridge, Massachusetts 02138, USA\\
	$^6$ Department of Mechanical Engineering, University College London, Torrington Place, London WC1E 7JE, UK\\
}

\date{\today}

\begin{abstract}
  We propose a mesoscopic model of binary fluid mixtures with tunable viscosity ratio based on a two-range pseudo-potential
  lattice Boltzmann method, for the simulation of soft flowing systems. 
In addition to the short range repulsive interaction between species in the classical single-range model, a competing mechanism 
between the short range attractive and mid-range repulsive interactions is imposed within each species.  
Besides extending the range of attainable surface tension as compared with the single-range model,  
the proposed scheme is also shown to achieve a positive disjoining pressure, independently of the viscosity ratio.
The latter property is crucial for many microfluidic applications involving a collection of disperse droplets with a
different viscosity from the continuum phase. As a preliminary application, the relative effective viscosity
of a pressure-driven emulsion in a planar channel is computed.
\end{abstract}                             

\maketitle

\section{Introduction}
Soft flowing systems, such as emulsions, foams, colloidal glasses, among others, are ubiquitous in nature and engineering,
and a better understanding of their rheology is crucial to the advancement of many fields of science and technology
\cite{larson1999structure,lyklema2005fundamentals,coussot2005rheometry,weaire2001physics,anna2006microscale,liu2011droplet,fu2017theoretical,bhateja2018rheology}.
Usually, soft flowing systems show very complex rheology, such as anomalous enhanced viscosity, structural and dynamical arrest, aging under moderate shear, etc, whose precise and quantitative description requires major extensions of non-equilibrium statistical mechanics
\cite{larson1999structure,sollich1997rheology}. These complex phenomena portray a complicated scenario, which is quite challenging even for the most advanced computational methods based on the solution of the Navier-Stokes (N-S) equations for nonideal fluids. Firstly, tracking the time evolution of complex interfaces between species and phases presents a serious hurdle for the macroscopic methods. Moreover, these methods are based on the continuum assumption, which makes it very challenging to capture the fundamental physics at micro and meso-scales \cite{li2016lattice}.
 
During the last three decades, a number of mesoscale methods based on kinetic theory have been developed \cite{mcnamara1988use,besold2000towards}. 
Among them, the lattice Boltzmann method (LBM) has met remarkable success for the simulation of complex flows and complex fluids \cite{mcnamara1988use,qian1992lattice,benzi1992lattice,shan1993lattice,bernaschi2010flexible,guo2013lattice,succi2015lattice,succi2018lattice,fei2018modeling,lauricella2018entropic}. In LBM simulations, the fluid is usually represented by populations of fictitious particles colliding locally and streaming to adjacent nodes along the links of a regular lattice. The scale-bridging nature of LBM allows its natural incorporation of microscopic and/or mesoscopic physics, while the efficient “collision-streaming” algorithm makes it computationally appealing \cite{qian1992lattice,succi2018lattice,li2016lattice}. Among the existing LBM for multiphase and multicomponent systems, a very popular and widely used scheme is the 
pseudopotential model, originally proposed by Shan and Chen (S-C) \cite{shan1993lattice}. In the original pseudopotential model (also named S-C model), the interactions between populations of molecules are modeled by a pseudo-interaction between the fictitious particles based on a density-dependent pseudopotential, and the phase separation is achieved through a short range attraction between the two fluid phases (liquid/gas in the case of multiphase single component systems, and liquid/liquid for mixture of immiscible fluids, as in the present case). A two-range pseudopotential
  A LBM has been proposed \cite{falcucci2007lattice} that was proved able to simulate flowing soft-glassy materials \cite{benzi2009mesoscopic, benzi2009mesoscopic2}, through the competition between the standard S-C short range interaction and an added middle range interaction. This method has obtained success in reproducing many features of the physics of these
  systems, such as structural frustration, aging, elastoplastic rheology, in confined and unbounded flows of microemulsions \cite{sbragaglia2012emergence,benzi2015internal,dollet2015two,scagliarini2015non}.

However, the two-range mesoscopic LBM suffers 
the problem of  a spurious viscosity-dependence in the pressure tensor, due to the discrete lattice effects introduced 
by the velocity shift forcing scheme \cite{benzi2009mesoscopic,benzi2009mesoscopic2}. 
As analyzed by Benzi \textit{et al.} \cite{benzi2009mesoscopic2}, the kinetic part of the pressure tensor includes $\tau$-dependent  terms (see Eq. (12) therein), which means that the surface tension and equilibrium densities in the model depend on the viscosities of the fluid components. 
In addition, the model usually suffers numerical instabilities for  multi-component flows with different viscosities. Guo \textit{et al.} argued that the discrete lattice effects must be considered in the introduction of the force field into LBM, and they proposed an alternative representation of the forcing term \cite{guo2002discrete}.
In this work, we propose, for the first time, a merge between these two techniques, which proves capable of achieving  
a new scheme with i) tunable surface tension over a sizeable range of values, ii) a positive disjoining pressure and  iii) no spurious dependence on the fluid viscosity. 
The new scheme is then used to compute the  relative effective viscosity of pressure-driven emulsions in a planar flow at non-unit viscosity ratios.

The remainder of this paper is structured as follows. In Sec. \ref{sec.2}, we present the proposed two-range pseudopotential model in detail. Section \ref{sec.3} gives extensive numerical experiments that validate and highlight the most salient features of the model, including an application to pressure-driven emulsions in a planar flow with different
dynamic viscosities. Finally, concluding remarks are given in Sec. \ref{sec.4}.

\section{Two-range pseudopotential lattice Boltzmann model with tunable viscosity ratio}\label{sec.2}
\subsection{Forcing scheme}\label{sec.2a}
\begin{figure}
	\includegraphics[width=0.4\textwidth]{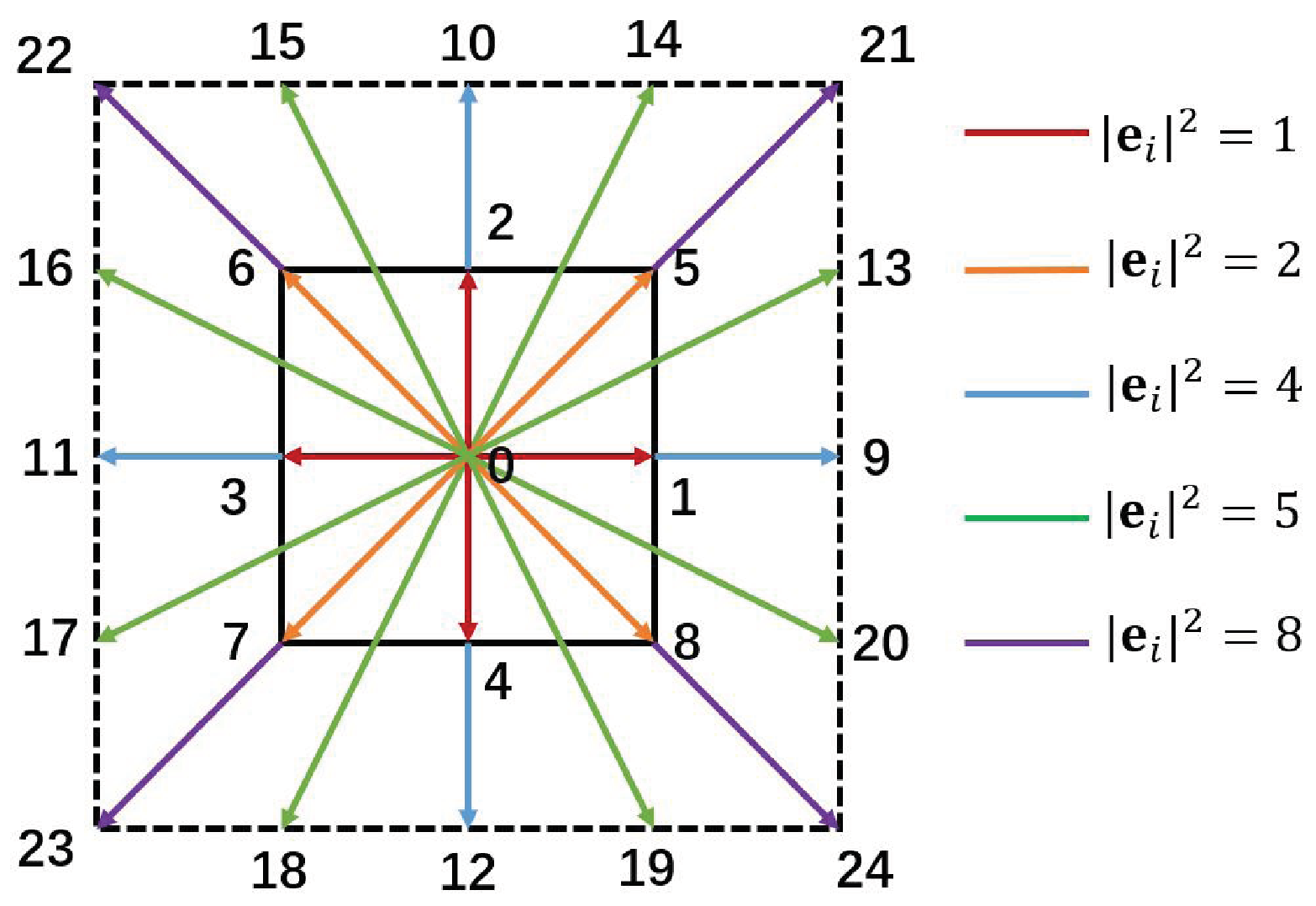}
	\caption{\label{DVM} The discrete lattice used in this work. The fluid lives in the D2Q9 lattice while the interactions extend to the full D2Q25.}
\end{figure}

In the two-range pseudopotential model for complex flows, the motion of the fluid is represented by a set of populations of fictitious particles (distribution functions) $ {f_{k ,i}}({\mathbf{x}},t) $ at position $ {\mathbf{x}}
 $ and time $ t $, where the subscripts $k$ and $i$ denote fluid component and discrete velocity direction, respectively. In this paper,  two-dimensional flow problems are considered and the D2Q9 lattice $ {{\bf{e}}_i} = \left[ {\left| {{e_{ix}}} \right\rangle  ,\left| {{e_{iy}}} \right\rangle } \right] $ ($ i = 0,1,...,8 $, see Fig. \ref{DVM}) is used. The lattice speed  $ c=\Delta{x}/\Delta {t}=1 $ and the lattice sound speed $ c_{s}=1/\sqrt{3} $ are adopted, in which $ \Delta{x} $ and $ \Delta {t} $ are the lattice spacing and time step. The evolution equation for the distribution functions (DFs) is given by 
\begin{equation}\label{e2}
{f_{k,i}}({\bf{x}} + {{\bf{e}}_i},t + \Delta t) - {f_{k,i}}({\bf{x}},t) =  - \frac{{\Delta t}}{{{\tau _k}}}[{f_{k,i}}({\bf{x}},t) - f_{k,i}^{eq}({\rho _{k,}}{{\bf{u}}^{eq}})] + \Delta t{C_{k,i}}
\end{equation}
where ${{\tau _k}} $ is the relaxation time for each 
component which is related to the kinematic viscosity by ${\nu _k} = ({\tau _k} - 0.5\Delta t)c_s^2$, and $ {C_{k,i}} $ is the forcing term by which a force field $ {{\bf{F}}_k} $
 is incorporated into the LBM. The local equilibrium distribution function (EDF) is usually given by a low-Mach-number truncation form as
\begin{equation}\label{e3}
f_{k,i}^{eq}({\rho _{k,}}{{\bf{u}}^{eq}}) = w({\left| {{{\bf{e}}_i}} \right|^2}){\rho _k}\left[ {1 + \frac{{{{\bf{u}}^{eq}}\cdot{{\bf{e}}_i}}}{{c_s^2}} + \frac{{{{\bf{u}}^{eq}}{{\bf{u}}^{eq}}:({{\bf{e}}_i}{{\bf{e}}_i} - c_s^2{\bf{I}})}}{{c_s^4}}} \right]
\end{equation}
where $ {\rho _k} $ is the density for each component, the weights are $ w(0) = 4/9 $, $ w(1) = 1/9 $, and $ w(2) = 1/36 $, and ${{{\bf{u}}^{eq}}}$ is an effective velocity. To conserve the total momentum of particles of all components in the absence of interparticle forces, the effective velocity must be given as \cite{shan1995multicomponent},
\begin{equation}
{{\bf{u}}^{eq}} = {{\sum\limits_k {\frac{{{\rho _k}{{\bf{u}}_k}}}{{{\tau _k}}}} } \mathord{\left/
		{\vphantom {{\sum\limits_k {\frac{{{\rho _k}{{\bf{u}}_k}}}{{{\tau _k}}}} } {\sum\limits_k {\frac{{{\rho _k}}}{{{\tau _k}}}} }}} \right.
		\kern-\nulldelimiterspace} {\sum\limits_k {\frac{{{\rho _k}}}{{{\tau _k}}}} }},
\end{equation}
where ${\rho _k}{{\bf{u}}_k}$ is the $k$ th component momentum. 
In the original formulation proposed by Shan \textit{et al.} \cite{shan1993lattice, shan1994simulation} and the recent two-range model \cite{benzi2009mesoscopic,benzi2009mesoscopic2,sbragaglia2012emergence,benzi2015internal}, the force field is implemented via a shift of the velocity in the EDF of each component, while the forcing term  $ {C_{k,i}} $ is not written explicitly. Using the general form in Eq.  (\ref{e2}), it can be shown that the velocity-shift method corresponds to an explicit forcing term \cite{huang2011forcing},
\begin{equation}
C_{k,i}^{SC} = \frac{1}{{{\tau _k}}}[f_{k,i}^{eq}({\rho _{k,}}{{\bf{u}}^{eq}} + {{\bf{F}}_k}{\tau _k}/{\rho _k}) - f_{k,i}^{eq}({\rho _{k,}}{{\bf{u}}^{eq}})].
\end{equation}
In addition, the component momentum in the original S-C model is calculated as
${\rho _k}{{\bf{u}}_k} = \sum\nolimits_i {{f_{k,i}}{{\bf{e}}_i}} $, which is independent of the forcing field. The velocity-shift method can be regarded as a first-order approach (in time), because the change in the momentum due to the force field is calculated using the first-order numerical integration. As shown by different researchers \cite{benzi2009mesoscopic2,huang2011forcing,li2012forcing,lycett2015improved}, some additional viscosity-dependent terms (related to the viscous stress tensor) are recovered in the macroscopic equations by the velocity shift method, while the method by Guo \textit{et al.} \cite{guo2002discrete} eliminates this effect by introducing extra compensating terms.
It may be noted that the method by Guo \textit{et al.} is not the only way to 
accurately incorporate the force field into the lattice Boltzmann equation. It is also possible to construct a forcing scheme using the second-order trapezoidal integration of the change of the distribution due to the force field \cite{he1998novel, he1998discrete}. On the other hand, it has been shown recently that a forcing scheme can be constructed based on the second-order time-splitting scheme \cite{dellar2013interpretation,hajabdollahi2018symmetrized}, where the changes in the momentum due to the force field, obtained via integration using a second-order Crank-Nicholson scheme, are introduced by means of two half forcing time steps around the collision step. More information on the recent discussions about the forcing schemes can be found in \cite{li2016revised,fei2017consistent,huang2018eliminating,fei2018three} and references therein.

In the present work we adopt the methodology of Guo \textit{et al.}, and the forcing term in Eq. (\ref{e2}) is given by,
\begin{equation}\label{e4}
C_{k,i}^{Guo} = \left( {1 - \frac{{\Delta t}}{{2{\tau _k}}}} \right)w({\left| {{{\bf{e}}_i}} \right|^2})\left[ {\frac{{{{\bf{e}}_i} - {{\bf{u}}^{eq}}}}{{c_s^2}} + \frac{{({{\bf{e}}_i}\cdot{{\bf{u}}^{eq}})}}{{c_s^4}}{{\bf{e}}_i}} \right]\cdot{{\bf{F}}_k}.
\end{equation}
In the forcing term, the velocity is the same as the effective velocity in the EDF,
as suggested by Guo \textit{et al.}, while the density and momentum for the $k$th component are calculated as
\begin{equation}\label{e5}
{\rho _k} = \sum\limits_{i = 0}^8 {{f_{k,i}}} , ~~~{\rho _k}{{\bf{u}}_k} = \sum\limits_{i = 0}^8 {{f_{k,i}}} {{\bf{e}}_i} + \Delta t{{\bf{F}}_k}/2.
\end{equation}
The barycentric velocity of the fluid mixture is
\begin{equation}
{\bf{u}} = \sum\limits_k {{\rho _k}{{\bf{u}}_k}} /\rho 
\end{equation}
where the total density is $\rho  = \sum\nolimits_k {{\rho _k}} $. It is seen that the barycentric velocity is not equal to ${{{\bf{u}}^{eq}}}$ unless all the components have the same viscosity. In the original method proposed by Guo \textit{et al.} \cite{guo2002discrete} where the single-component flow is considered,  the actual fluid velocity, the velocity in the forcing term and the velocity in the EDF coincide. We wish to point out that, by replacing  ${{{\bf{u}}^{eq}}}$ in Eqs. (\ref{e3}) and (\ref{e4}) by ${\bf{u}} $, unphysical deformation of a dispersed droplet is observed unless a unity viscosity ratio is used.

\subsection{Competing mechanism and pressure tensor}\label{sec.2b}

A crucial characteristic of the two range pseudopotential model is the competing mechanism between the short
range attractive and middle range repulsive interactions. The short range interactions act between the nearest-neighbor lattice nodes (connecting through the D2Q9 lattice ), while the middle range interactions act between the next-to-nearest-neighbor lattice nodes extending up to a D2Q25 
lattice $ {{\bf{e}}_j} = \left[ {\left| {{e_{jx}}} \right\rangle ,\left| {{e_{jy}}} \right\rangle } \right] $
(${\rm{j  =  0,}}...{\rm{,24}}$, see Fig. \ref{DVM}). The competing interaction force is explicitly written as  \cite{benzi2009mesoscopic,benzi2009mesoscopic2}
\begin{equation}\label{e7}
{\bf{F}}_k^c =  - {G_{k,1}}{\psi _k}({\bf{x}})\sum\limits_{i = 0}^8 {w({{\left| {{{\bf{e}}_i}} \right|}^2}){\psi _k}({\bf{x}} + {{\bf{e}}_i})} {{\bf{e}}_i} - {G_{k,2}}{\psi _k}({\bf{x}})\sum\limits_{j = 0}^{24} {p({{\left| {{{\bf{e}}_j}} \right|}^2}){\psi _k}({\bf{x}} + {{\bf{e}}_j})} {{\bf{e}}_j},
\end{equation}
where $ {G_{k,1}} $ and $ {G_{k,2}} $ are the strength coefficients for the short-range and middle-range interactions respectively, and the weights for D2Q25 lattice are  $ p(0) = 247/420 $,
$ p(1) = 4/63 $,  $ p(2) = 4/135 $,  $ p(4) = 1/180 $,  $ p(5) = 2/945 $ and  $ p(8) = 1/15120 $. The pseudopotential originally suggested by SC \cite{shan1993lattice,shan1994simulation}, $ {\psi _k}({\rho _k}) = {\rho _0}(1 - {e^{ - {\rho _k}/{\rho _0}}}) $ (with a uniform reference density $ {\rho _0}=1.0 $ for each component) is adopted.

\begin{figure}
	\includegraphics[width=0.4\textwidth]{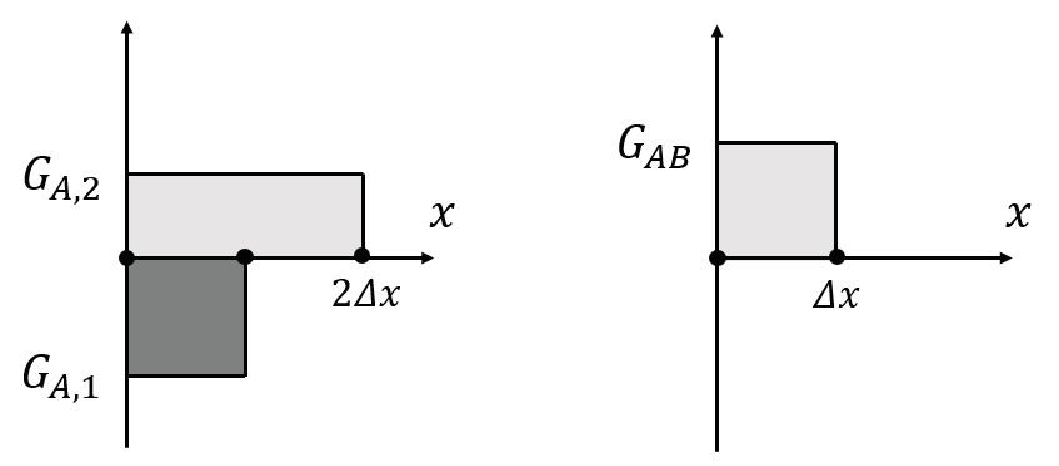}
	\caption{\label{interaction} A sketch of different interactions on component $ A $. Left panel: short-range attractive ($  {G_{A,1}} < 0  $) and middle-range repulsive ($  {G_{A,2}} > 0  $) interactions within component $ A $. Right Panel: short-range repulsive ($  {G_{AB}} > 0  $) interaction between components $ A $ and $ B $.}
\end{figure}

For a multicomponent fluid system, a repulsive force acts among all component as usual \cite{shan1993lattice},
\begin{equation}\label{e8}
{\bf{F}}_k^r =  - {\rho _k}({\bf{x}}){\sum _{\bar k}}{G_{k\bar k}}\sum\limits_{i = 0}^8 {w({{\left| {{{\bf{e}}_i}} \right|}^2}){\rho _{\bar k}}({\bf{x}} + {{\bf{e}}_i})} {{\bf{e}}_i},
\end{equation}
where $ {G_{k\bar k}}{\rm{ = }}{G_{\bar kk}} $ is the strength coefficients for the inter-component interaction. Supplemented with the body force $ {\bf{F}}_k^b $, the total force in Eq. (\ref{e4}) is 
$ {{\bf{F}}_k} = {\bf{F}}_k^c + {\bf{F}}_k^r + {\bf{F}}_k^b $. 
Although the model is in principle able to simulate a system with a generic number of components, we will restrict ourselves to the most
common case of a two fluids mixture (i.e., $ k = A,B $) with the same strength coefficients
	($ {G_{A,1}} = {G_{B,1}} $, $ {G_{A,2}} = {G_{B,2}} $, $ {G_{AB}} = {G_{BA}} $) in practice \cite{benzi2009mesoscopic,benzi2009mesoscopic2}. More specifically, a positive (negative) strength coefficient represents a repulsive (attractive) interaction. For the competing interactions in Eq. (\ref{e7}) imposed within each species, phase-separating fluids (vapor and liquid phases for each species) are confined to the condition \cite{falcucci2010lattice} $ {G_{A,1}} + {G_{A,2}} < 0 $ , while in the present paper we set $ {G_{A,1}} + {G_{A,2}} =  - 1 $. The repulsive interactions between species are used as usual, $ {G_{AB}} > 0 $. A sketch of different interactions on component $ A $ is shown in Fig. \ref{interaction}.

To capture the complex rheological features of the soft flowing systems, it is crucial to analyze the pressure tensor associated with the two-range pseudopotential model. According to the analysis by Benzi\textit{ et al.} \cite{benzi2009mesoscopic2}, the pressure tensor is defined as the sum of an interaction part plus a kinetic part,
\begin{equation}\label{e9}
{P_{ab}} = P_{ab}^{{\mathop{int}} }+P_{ab}^{kin}, 
\end{equation}
where $ a $ and $ b $ run over the spatial coordinates. The bulk equation of state reads as follows:
\begin{equation}
{p_b}({\rho _A},{\rho _B}) = \sum\limits_{k = A,B} {[{\rho _k} + \frac{1}{2}} ({G_{k,1}} + {G_{k,2}}){\psi _k^2}]c_s^2 + {G_{AB}}{\rho _A}{\rho _B}c_s^2.
\end{equation}

The interaction tensor is defined by the condition
\begin{equation}\label{e10}
{\partial _b}P_{ab}^{int} =  - \sum\limits_k {{{\bf{F}}_{ka}}}
\end{equation}
which is independent of the specific forcing scheme and can be obtained based on the expression proposed by Shan \cite{shan2008pressure}.
Calculation details of $ P_{ab}^{{\mathop{\rm int}} } $ can be found in Dollet {\it et al} \cite{dollet2015two} and
Sbragaglia \& Belardinelli \cite{sbragaglia2013interaction}.

For the kinetic part, however, this is not the case. For the static interface problem considered in the next subsection, the kinetic part in Eq. (\ref{e9}) is given by
\begin{equation}\label{e15}
P_{ab}^{kin} = \sum\limits_{k,i} {{f_{k,i}}{e_{ia}}{e_{ib}}}.
\end{equation}
In the original two-range model \cite{benzi2009mesoscopic,benzi2009mesoscopic2}, the explicit expression of $ P_{ab}^{kin} $ is,
$ P_{ab}^{kin} = \sum\limits_k {{\rho _k}} c_s^2{\delta _{ab}} + K_{ab}^\tau $, where $ K_{ab}^\tau $ denotes the following extra $ \tau$-dependent terms,
\begin{equation}\label{ekab}
K_{ab}^\tau {\rm{ = }}{\left( {\tau  - \frac{1}{2}} \right)^2}\frac{{{F_a}{F_b}}}{\rho } + c_s^4\frac{{{\rho _A}{\rho _B}}}{\rho }{\left( {\tau  - \frac{1}{2}} \right)^2}\left( {\frac{{{\partial _a}{\rho _A}}}{{{\rho _A}}} - \frac{{{\partial _a}{\rho _B}}}{{{\rho _B}}}} \right)\left( {\frac{{{\partial _b}{\rho _A}}}{{{\rho _A}}} - \frac{{{\partial _b}{\rho _B}}}{{{\rho _B}}}} \right),
\end{equation}
where ${F_a} = ({F_{A,a}} + {F_{B,a}})$ and $\tau  = ({\rho _A}{\tau _A} + {\rho _B}{\tau _B})/\rho $. Due to the $ \tau$-dependent terms, the original two-range model suffers numerical instability for two-component flow with a relative large viscosity ratio.
Moreover, the surface tension and disjoining pressure are dependent on the viscosity ratio. Remarkably, the $ \tau$-dependent terms can be removed in the present model, due to proper consideration of discrete lattice effects in the forcing scheme in Eqs. (\ref{e4} and \ref{e5}). As a result, the present two-range model shows significant improvements compared with the original two-range model in terms of tunable viscosity ratio and also in terms of eliminating the dependence of surface tension and disjoining pressure on the viscosity ratio (see Sec. \ref{sec.3}). In addition, the total mass of both species in the present model is conserved to machine accuracy.

\subsection{Macroscopic effect of the mesoscopic interaction: the Disjoining Pressure}\label{sec.2c}
Paradigmatic soft materials, such as foams and emulsions, consist of dispersion of one fluid 
(a gas in the case of foams,
a liquid for emulsions) in a liquid; the dispersion is stabilized against full phase separation by the presence of surfactants, 
that lower the interfacial energy thus inhibiting droplet (bubble) coalescence.
A mechanical {\it translation} of this (microscopic) stabilization effect of surfactants at mesoscopic level can be
done borrowing the concept of disjoining (or Derjaguin) pressure $\varPi$ from the theory of thin liquid films \cite{derjaguin1978question}. 
In such framework, the disjoining pressure emerges as a repulsive force per unit area between opposing interfaces, due 
to interface-interface interactions \cite{derjaguin1978question}, that stabilizes the thin film.
Analogously, for a thin film formed between two droplets/bubbles to be stable, a (positive) disjoining pressure has to overcome the 
capillary pressure at the curved interface \cite{bergeron1999forces}. 
As analyzed by Sbragaglia \textit{et al.} \cite{sbragaglia2012emergence}, the pure short-range interaction in the original S-C model 
(corresponding to $ {G_{k,1}}={G_{k,2}} = 0 $ in Eq. (\ref{e7})) always yields a negative disjoining pressure. 
Notably, a positive disjoining pressure can be achieved in the two-range model by tuning the strength coefficients at a fixed  
$ {G_{k,1}} + {G_{k,2}} $.

\begin{figure}
	\includegraphics[width=0.4\textwidth]{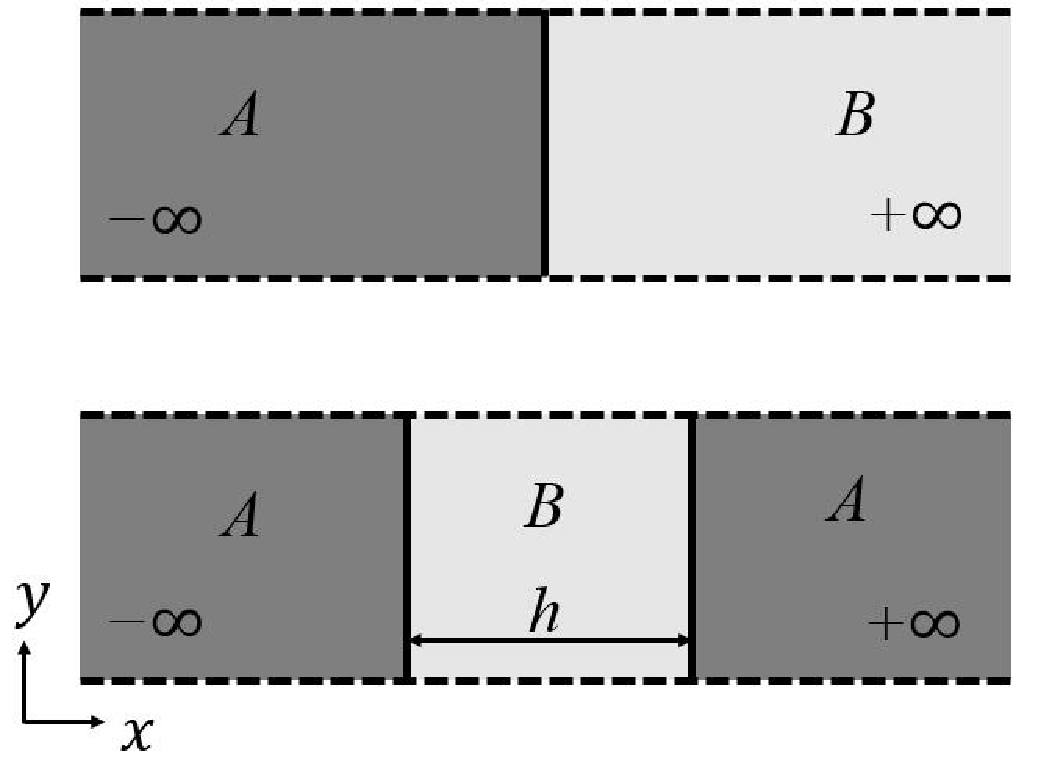}
	\caption{\label{interface}
	A sketch for one and two interfaces 
	between components $ A $ and $ B $. Top panel, one interface case for the calculation of the surface tension ${\gamma}$ in Eq.  (\ref{e16}). 
	Bottom panel, two interfaces separated by a distance $ h $ for the calculation of the overall line tension ${\gamma _f}(h)$ in Eq. (\ref{e17}). When the distance $ h $ is large enough, ${\gamma _f}(h)$ is convergent to $2\gamma $. By setting a series of  $ h $, the disjoining pressure can be calculated through Eq. (\ref{e19}).}
\end{figure}

To quantitatively determine $\varPi$ in our model, let us first consider a one-dimensional problem: 
 a flat interface orthogonal to the $ x $ coordinate (see Fig. \ref{interface}). The surface tension can, then, be defined 
as the integral of the mismatch between the normal ($ {P_{xx}} $) and tangential ($ {P_{yy}} $) components for the pressure tensor,
\begin{equation}\label{e16}
\gamma  = \int_{ - \infty }^{+\infty}  {({P_{xx}} - {P_{yy}})}dx,
\end{equation}
where the $ ({P_{xx}} - {P_{yy}}) $ can be explicitly obtained via  Eq. (\ref{e9}).
We then consider two flat interfaces, separated by a distance $ h $ (see Fig. \ref{interface}), now the integral in Eq. (\ref{e16}) is defined as the overall line tension,
\begin{equation}\label{e17}
{\gamma _f}(h) = \int_{ - \infty }^{+\infty}  {({P_{xx}} - {P_{yy}})}dx.
\end{equation}
As we can see, $ {\gamma _f} $ is a function of $ h $, and the limit condition is $ {\gamma _f}(h \to \infty ) = 2\gamma $. According to Refs. \cite{bergeron1999forces,sbragaglia2012emergence}, the disjoining pressure $ \varPi $ is defined as
\begin{equation}\label{e18}
\int_{\varPi (h = \infty )}^{\varPi (h)} h d\varPi  = s(h),
\end{equation}
where $ s(h) = {\gamma _f}(h) - 2\gamma $.
A simple differentiation of Eq. (\ref{e18}) yields $ ds(h)/dh = hd \varPi /dh $. Supplementing with the boundary condition $ \varPi (h \to \infty )=0 $, the disjoining pressure finally reads as follows,
\begin{equation}\label{e19}
\varPi (h) = \frac{{s(h)}}{h} - \int_h^\infty  {\frac{{s(\tilde h)}}{{{{\tilde h}^2}}}d\tilde h}.
\end{equation}
By setting a series of interface distances, Eq. (\ref{e19}) can be calculated using standard numerical integration method.

\section{NUMERICAL SIMULATIONS}\label{sec.3}
In this section, numerical simulations are conducted to verify the aforementioned arguments and highlight the main features of the present two-range pseudopotential model. Unless otherwise specified, the strength coefficients in Eq. (\ref{e7}) are set the same for both components ($ {G_{A,1}} = {G_{B,1}} $ and $ {G_{A,2}} = {G_{B,2}} $) with ${G_{A,1}} + {G_{A,2}} =  - 1$ \cite{sbragaglia2012emergence,dollet2015two}, and
the coefficients $ {G_{AB}} $ is chosen such that the interface width is about 4 lattice spacings \cite{porter2012multicomponent}.
\begin{figure}
	\includegraphics[width=0.45\textwidth]{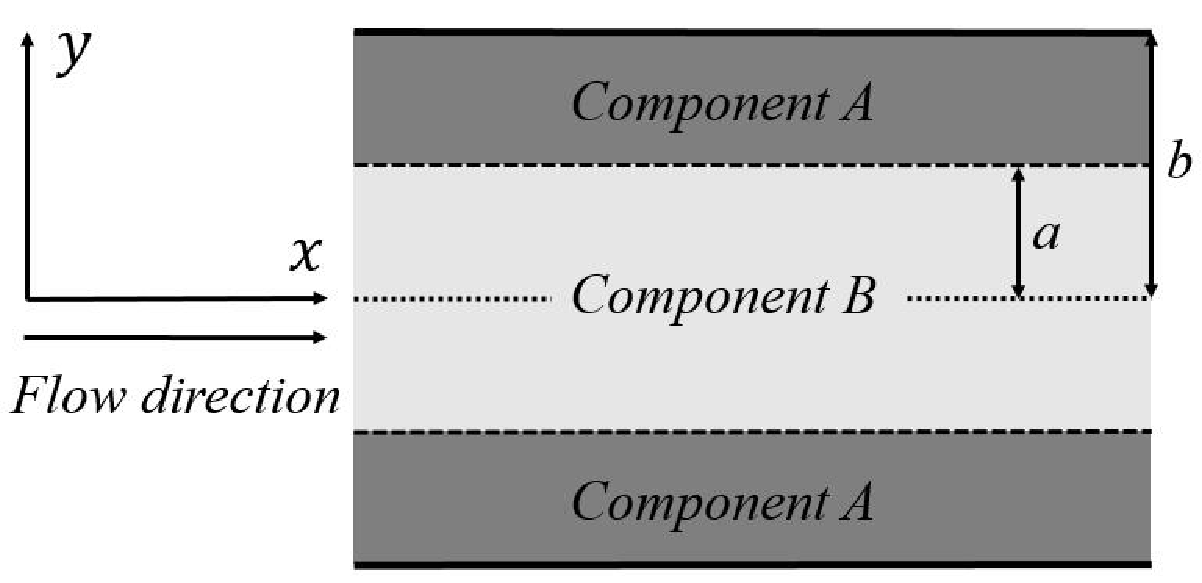}
	\caption{\label{cocurrent} Sketch of the two-component Poiseuille flow.}
\end{figure}

\subsection{Two-component Poiseuille flow}\label{sec.3a}
Firstly, a two-component Poiseuille flow  (along the $ x $ direction) driven by a body force is studied (see also Fig.~\ref{cocurrent}). In this flowing system, the nonwetting phase (component $ B $) flows in the central region of the channel, $ 0 < \left| y \right| < a $, the wetting phase  (component $ A $) flows between the nonwetting phase and the walls, $ a < \left| y \right| < b $, and the dynamic viscosity ratio is $M = \frac{{{\mu _B}}}{{{\mu _A}}} = \frac{{{\nu _B}}}{{{\nu _A}}}$ ($ M $ simplifies to the kinematic viscosity ratio due to the unit density ratio considered).
 In our simulations, periodic boundary conditions are applied along the $ x $ direction, the half-way bounce-back boundary scheme is applied to the up and bottom walls, and the computational domain is covered by $ 10 \times 160 $ lattice nodes. The analytical solution for the problem is given by \cite{porter2012multicomponent},
\begin{equation}
u(y){\rm{ = }}\left\{ \begin{array}{l}
\frac{{{F_b}}}{{2{\mu _B}}}({a^2} - {y^2}) + \frac{{{F_b}}}{{2{\mu _A}}}({b^2} - {a^2}),~0 < \left| y \right| < a \\ 
\frac{{{F_b}}}{{2{\mu _A}}}({b^2} - {y^2}),~a < \left| y \right| < b. \\ 
\end{array} \right.
\end{equation}
where the body force $ {{F_b}} $ acts on both components and $ b $ is set to $80\Delta x$, with $b=2a$.

\begin{figure*}[!ht]
	\center {
		{\epsfig{file=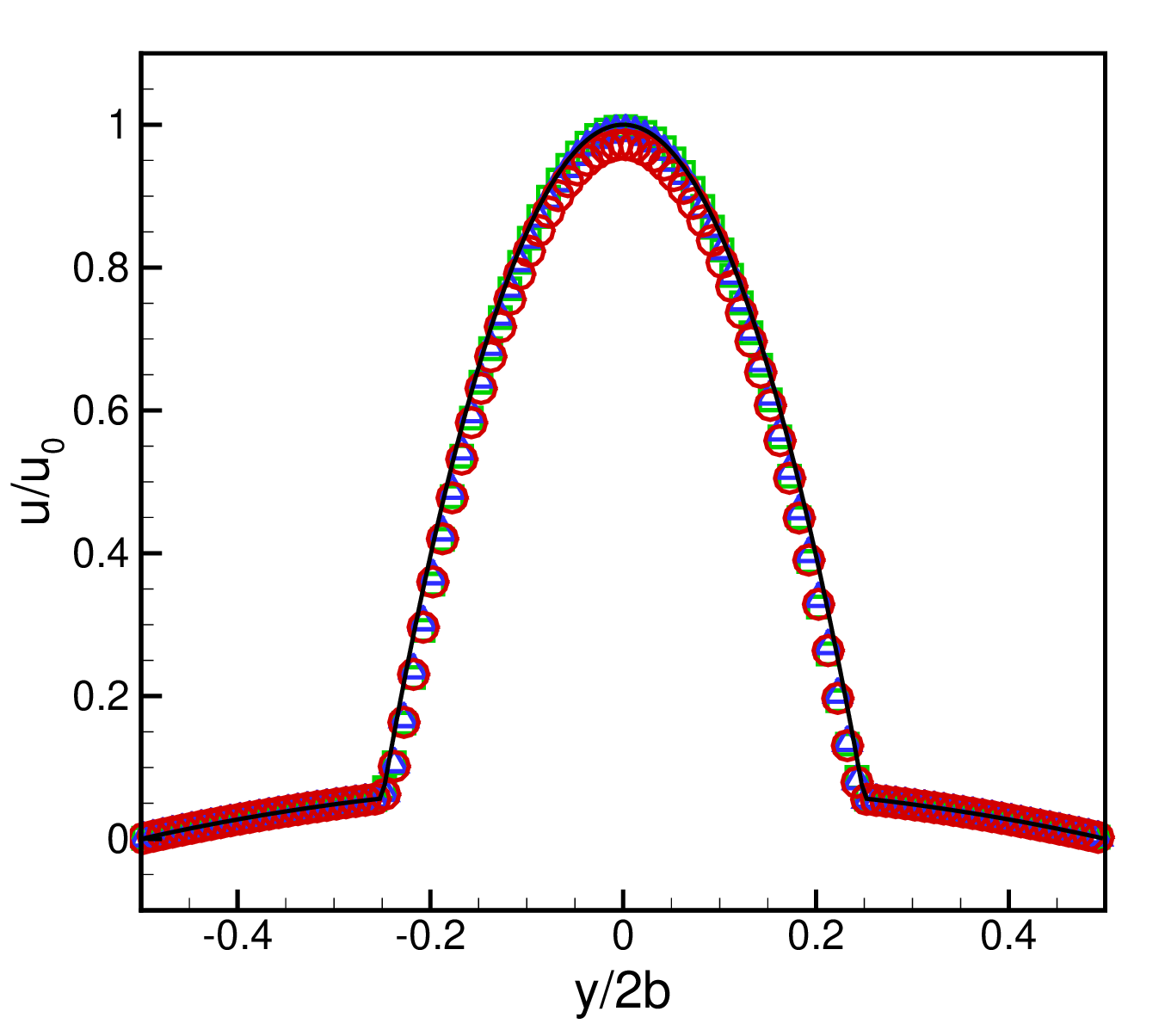,width=0.32\textwidth,clip=}}\hspace{0.0cm}  
		{\epsfig{file=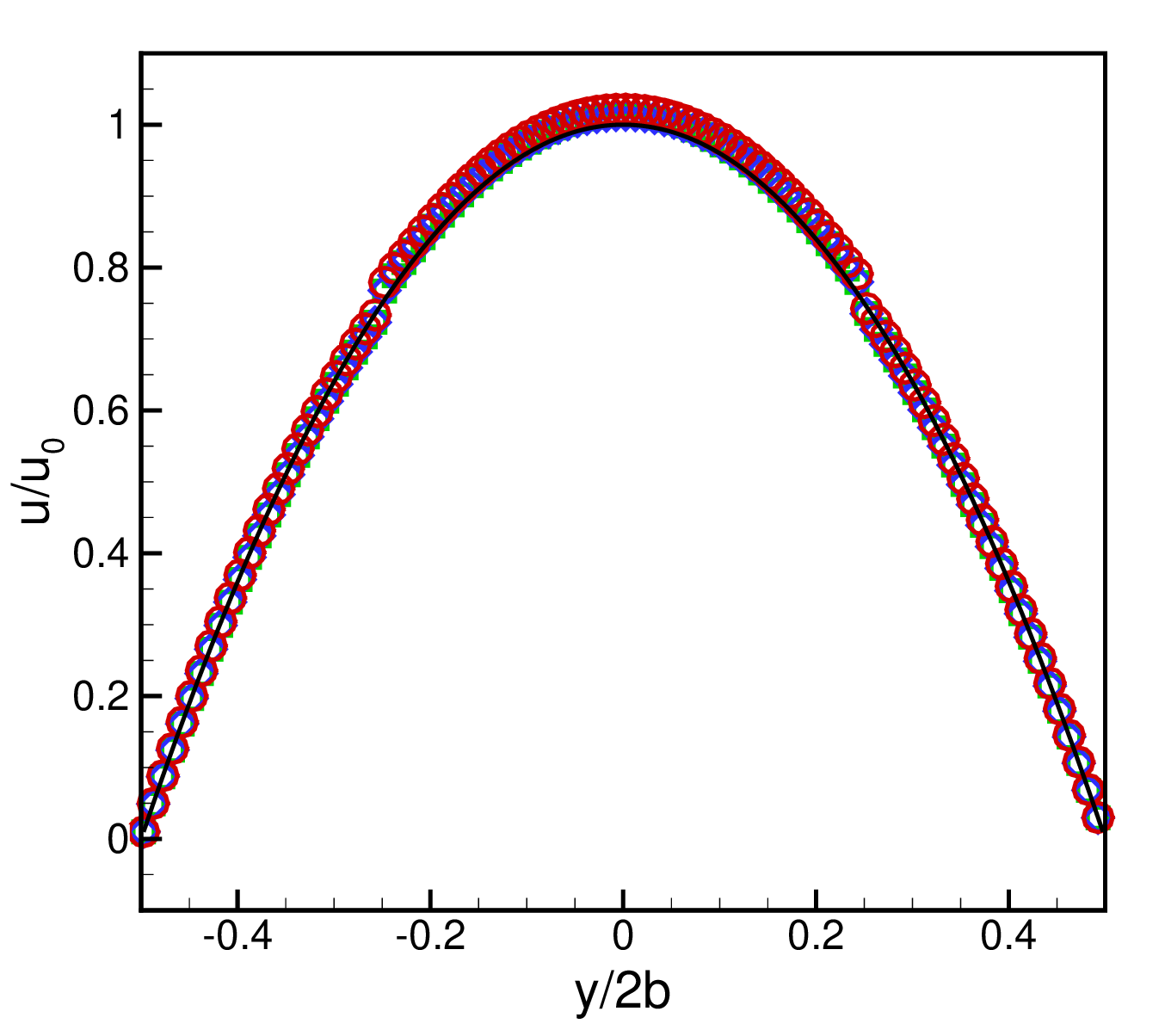,width=0.32\textwidth,clip=}}\hspace{0.0cm}  
		{\epsfig{file=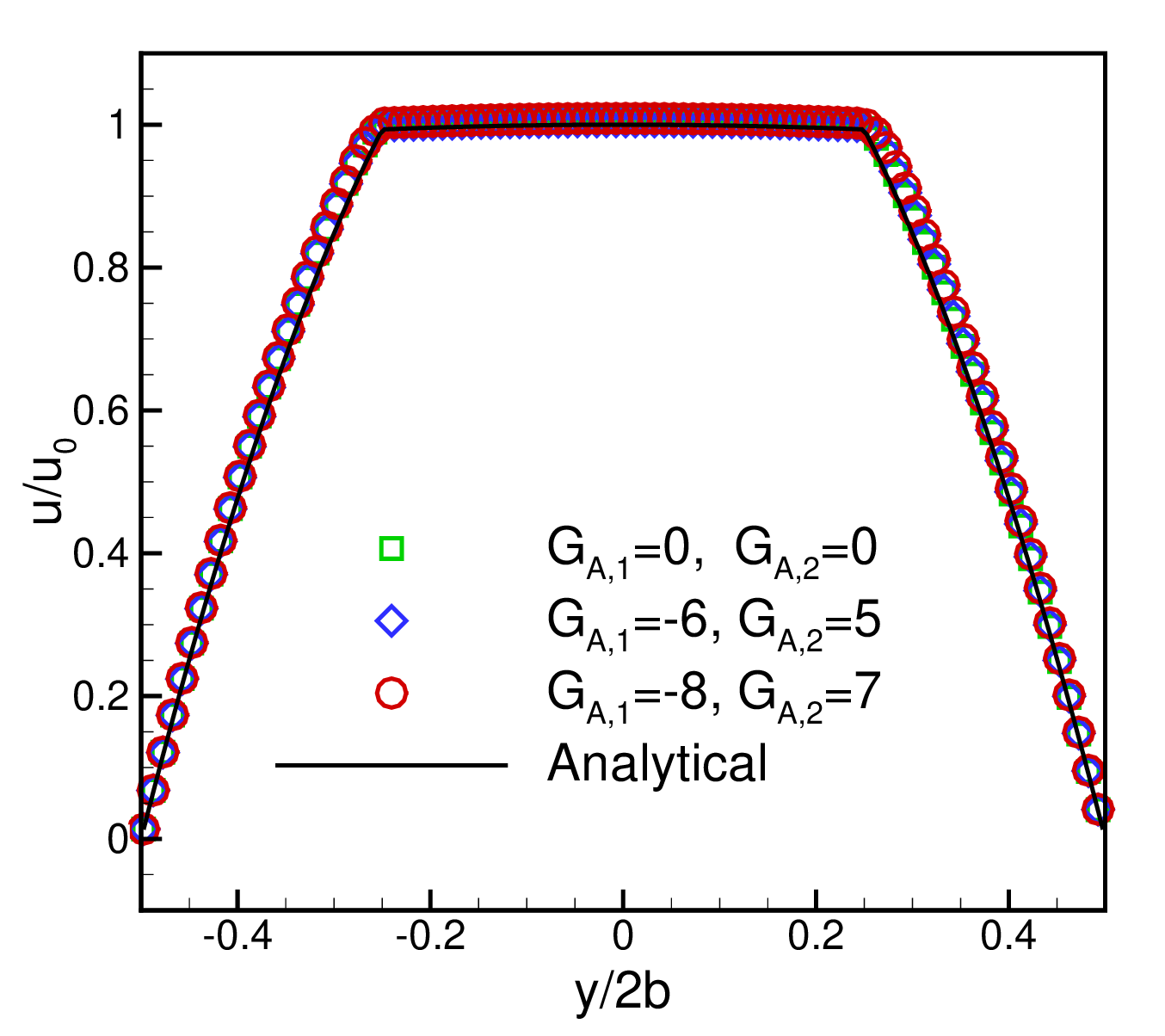,width=0.32\textwidth,clip=}}\vspace{0.0cm}\\
	}
	\caption{Steady velocity profiles for two-component Poiseuille flow with different viscosity ratios, $M = \left\{ {1/50,1,50} \right\}$ (from left to right); the profiles are normalized by the centreline velocity $u_0 = 0.05$ (lbu).
          The symbols are numerical results by the present two-range
	pseudopotential models, and the lines are the analytical solutions.
	}
	\label{FIG1}
\end{figure*}
It is known that the velocity shift forcing scheme suffers numerical instability for two-component flow with a relatively high
kinematic viscosity ratio $M$ \cite{huang2015multiphase}. All implementations of the
 original two-range model so far were at $M=1$ \cite{benzi2009mesoscopic,benzi2009mesoscopic2,sbragaglia2012emergence}. 
 Figure \ref{FIG1} shows the comparison between numerical results obtained with the present two-range model and the
 analytical solutions at different viscosity ratios, $M = \left\{ {1/50,1,50} \right\}$
  (from left to right). For the original two-range model, the simulation is unstable for $M > 5$, thus it is not shown in the figure. For the present model, we consider three cases: two of them corresponding to positive disjoining pressures ($ {G_{A,1}} =  - 6 $ and -8) and a reference case ($ {G_{A,1}} ={G_{A,2}}= 0 $ ). It can be seen that the numerical results are in good agreement with the analytical solutions, except for some small discrepancies near the interface. As analyzed in Refs. \cite{porter2012multicomponent,li2014deformation}, the small discrepancies may be related the diffused interface feature in LBM and have no significant effects 
 on the flow away from the interface. These results prove that, with our model, the viscosity ratio can be tunable over a relatively large range,
 still reproducing physically correct results. 

\subsection{Tunable surface tension}\label{sec.3b}
\begin{figure}
	\includegraphics[width=0.4\textwidth]{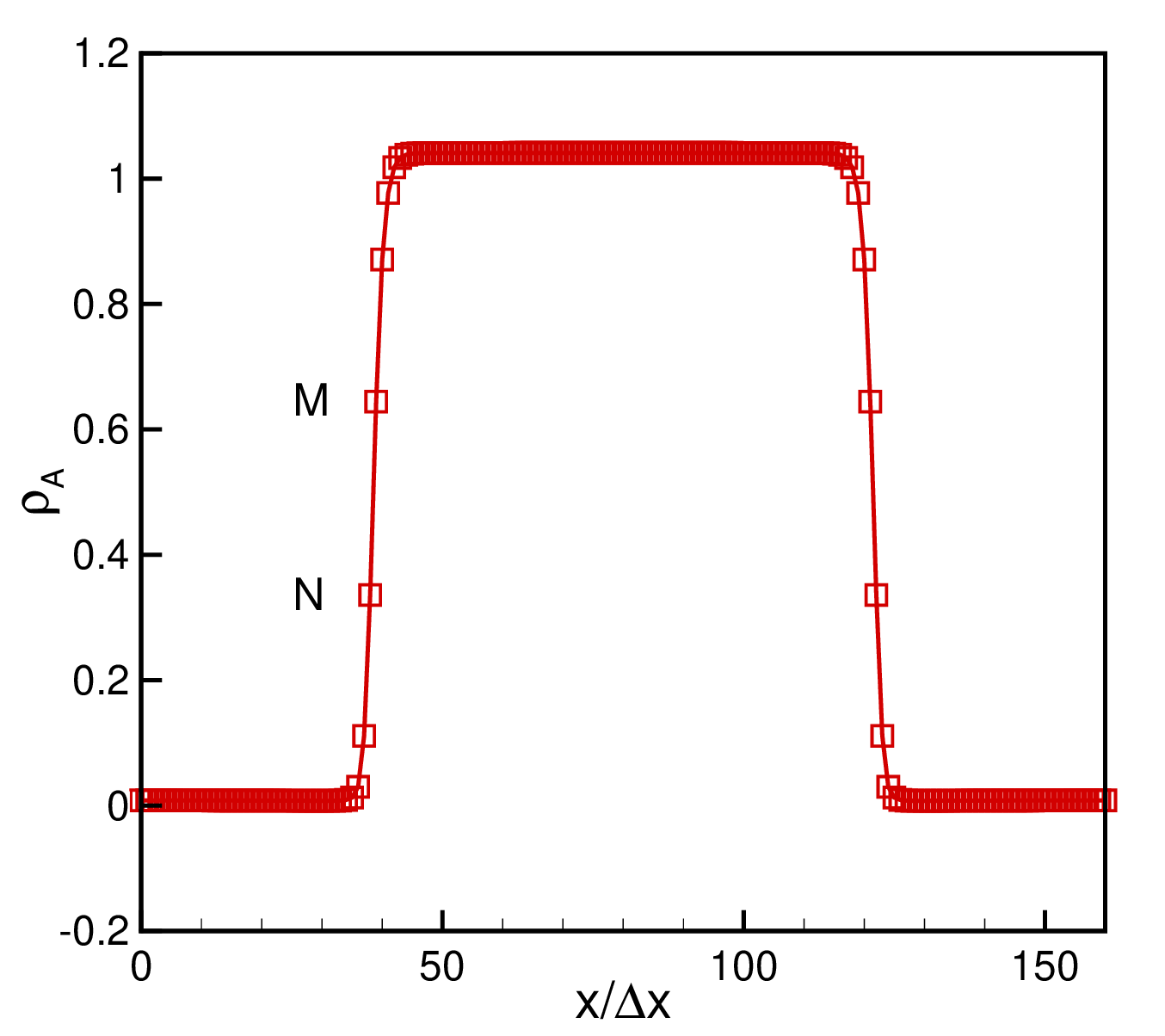}
	\caption{\label{FIG2}
		The density profile along the horizontal centerline for
		a steady droplet (component $ A $) with radius $R = 42\Delta x$ in the center of a periodic box: we measure the interface thickness as, $W = ({\rho _{A,\max }} - {\rho _{A,\min }})\Delta x/[{\rho _A}(M) - {\rho _A}(N)]$
		, where $ M $ and $ N $ are the two points near the location  where ${{\bar \rho }_A} = ({\rho _{A,\max }} + {\rho _{A,\min }})/2$.
    In this figure, the interface thickness is $ W = (1.041 - 0.008)\Delta x/(0.644 - 0.336) = 3.35\Delta x $.}
\end{figure}

In the single-range pseudopotential model, the surface tension can only be varied through the parameter $ {G_{AB}} $.
Thanks to the introduction of the 
competing mechanism, the surface tension in the present two-range model can be also adjusted by tuning the parameters $ {G_{A,1}} $ and $ {G_{A,2}} $. To confirm this statement, a series of Laplace's tests are carried out to measure the surface tension under different conditions. Four different cases
($ {G_{A,1}} = 0,~{G_{A,2}} = 0 $; $ {G_{A,1}} = -10,~{G_{A,2}} =9 $; $ {G_{A,1}} = 0,{G_{A,2}} = -1 $ and
$ {G_{A,1}} = 10,{G_{A,2}} = -11 $) are considered, in which the first case corresponds to the classical single-range model. The tunable range of $ {G_{AB}} $ and the achievable range of $ \gamma $ are shown in Table~\ref{TAB1}. The parameter $ {G_{AB}} $ is tuned progressively with a 0.1 interval to find a range over which the measured interface thickness $ W $ is within
  $ {\rm{2}}\Delta x \le W \le {\rm{5}}\Delta x $; such range is costrained since too sharp interfaces ($W < 2\Delta x$) suffer from numerical instability and too wide ones ($W > {\rm{5}}\Delta x$) deteriorate the numerical accuracy near the interface. The interface width  $ W $ is defined by fitting a hyperbolic tangent curve to the density profile \cite{lycett2015improved}, which can be rewritten as
$W = ({\rho _{A,\max }} - {\rho _{A,\min }})/(\partial {\rho _A}/\partial x){|_{{{\bar \rho }_A}}}$ with ${{\bar \rho }_A} = ({\rho _{A,\max }} + {\rho _{A,\min }})/2$ and solved using the finite difference method (see Fig. \ref{FIG2}).

It can be seen from Table~\ref{TAB1} that the achievable range for the single-range model is $0.031 \le \gamma  \le 0.14$. For the two-range model, we have more freedom to tune the surface tension due to the additional competing interactions, and for the four cases considered in this work, the achievable range is $6 \times {10^{{\rm{ - }}4}} \le \gamma  \le 0.27$. Specifically, larger (smaller) $ {G_{A,1}}$ gives larger (smaller) surface tension, because, at fixed  $ {G_{k,1}} + {G_{k,2}}$, the diagonal elements of the kinetic pressure tensor $ P_{ab}^{{\mathop{int}} }  $ are proportional to $ {G_{A,1}}$ (see Eqs. (14-16) in \cite{benzi2009mesoscopic2}).

In addition, unlike the original two-range model \cite{benzi2009mesoscopic2}, where the surface tension is  viscosity-dependent, in the present model the surface tension and viscosity ratio are decoupled, as it will be discussed in the next subsection.

\begin{table}
	\renewcommand\arraystretch{1.2}
	\caption{\label{TAB1}%
		Tunable range of $ {G_{AB}} $ and the corresponding range of the surface tension $\gamma$ for the present model under different cases. The case without two-range interactions ($ {G_{A,1}} = 0,~{G_{A,2}} = 0 $) is shown for comparison.
	}
	\begin{ruledtabular}
		\begin{tabular}{ccccc}
			\textrm{ case }
			&
			\textrm{$ {G_{A,1}} = 0,~{G_{A,2}} = 0 $}&
			\textrm{$ {G_{A,1}} = -10,~{G_{A,2}} = 9 $ }&
			\textrm{$ {G_{A,1}} = 0,~{G_{A,2}} = -1 $ }&
			\textrm{$ {G_{A,1}} = 10,~{G_{A,2}} = -11 $ }
			\\
			\colrule
			range of ${G_{AB}}$ &$[2.8,5.0]$
			& $[2.0,3.4]$
			& $[2.3,5.1]$
			& $[2.9,7.2]$
			\\
			
			range of $\gamma$  & $[0.031,0.14]$
			& $[0.0006,0.041]$
			& $[0.028, 0.16]$
			& $[0.081,0.27]$
			\\
		\end{tabular}
	\end{ruledtabular}
\end{table}

\subsection{Independence of surface tension and disjoining pressure on the viscosity ratio}\label{sec.3c}
\begin{figure}
	\includegraphics[width=0.4\textwidth]{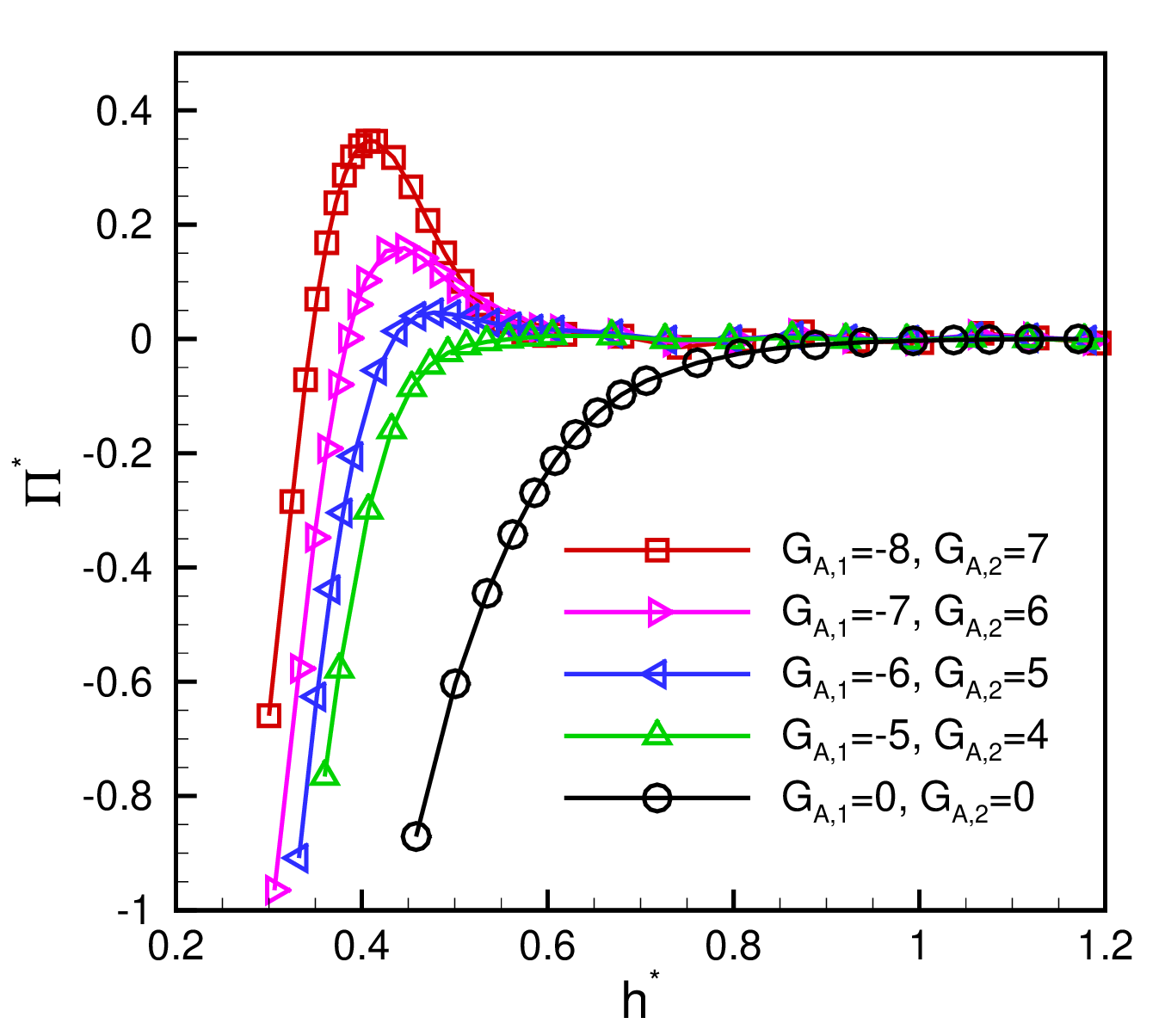}
	\caption{\label{FIG3a} Dimensionless disjoining pressure $ {\varPi ^{\rm{*}}}$ as a function of dimensionless distance $ {h^*} $ for different cases. For the case
		$ {G_{A,1}} = 0,{G_{A,2}} = 0 $	(without two-range competing interactions), the disjoining pressure decreases with ${h^*}$, thus it cannot support a stable thin film between two interfaces. For other cases (with two-range competing interactions), 
		with the decrease of ${h^*}$, the disjoining pressure increases gradually to a peak and then goes down, which stabilizes the thin film. The peak value of disjoining pressure increases with decreasing  $ {G_{A,1}} $.}
\end{figure}

The emergence of positive disjoining pressure is a unique feature of the two-range model, which supports the stable thin film between two interfaces and distinguishes the two-range model from the classical single-rang. To measure the disjoining pressure $ \varPi $, we consider two planar interfaces, separated by a distance $h$ (see also Fig. \ref{interface}).
After $ {\gamma _f}(h) $ is obtained through Eq. (\ref{e17}) for various $h$'s,
the disjoining pressure can be calculated according to Eq. (\ref{e19}).
 To be general, we introduce a dimensionless disjoining pressure
$ {\varPi ^{\rm{*}}}{\rm{ = }}\varPi {h_0}/\gamma $, where
$ {h_0}\sim O(10\Delta x) $ is a length scale defined as $ {\gamma _f}(h > {h_0}) = 2\gamma $.
Firstly, we choose $ {G_{AB}} = 3.0 $, and calculate the dimensionless disjoining pressure as a function of the dimensionless distance $ {h^*} = h/{h_0} $ for the present model with different pairs of $ ({G_{A,1}},{G_{A,2}}) $, as shown in Fig.~\ref{FIG3a}.
From the figure, we can see that  for the case  ($ {G_{A,1}}{\rm{ = }}0,{G_{A,2}}{\rm{ = }}0 $), corresponding to the classical single-range model,
the disjoining pressure is always negative and decreases with  ${h^*}$. For other cases with the two-range competing interactions, the disjoining pressure increases with the decrease of ${h^*}$ firstly and then goes down, attaining positive values inbetween.
\begin{figure*}[!ht]	        
	\center {			
		{\epsfig{file=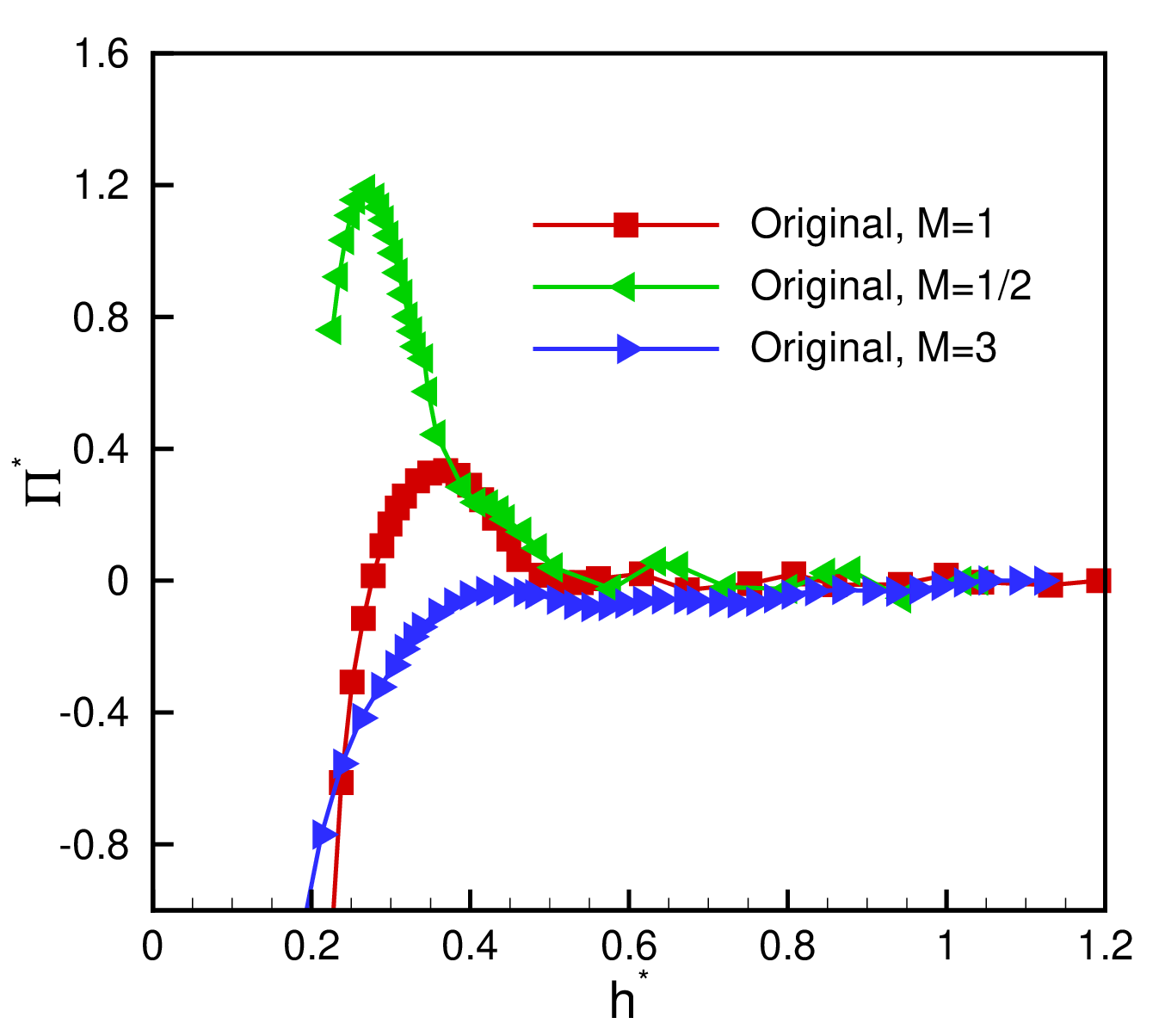,width=0.4\textwidth,clip=}}\hspace{0.0cm}  
		{\epsfig{file=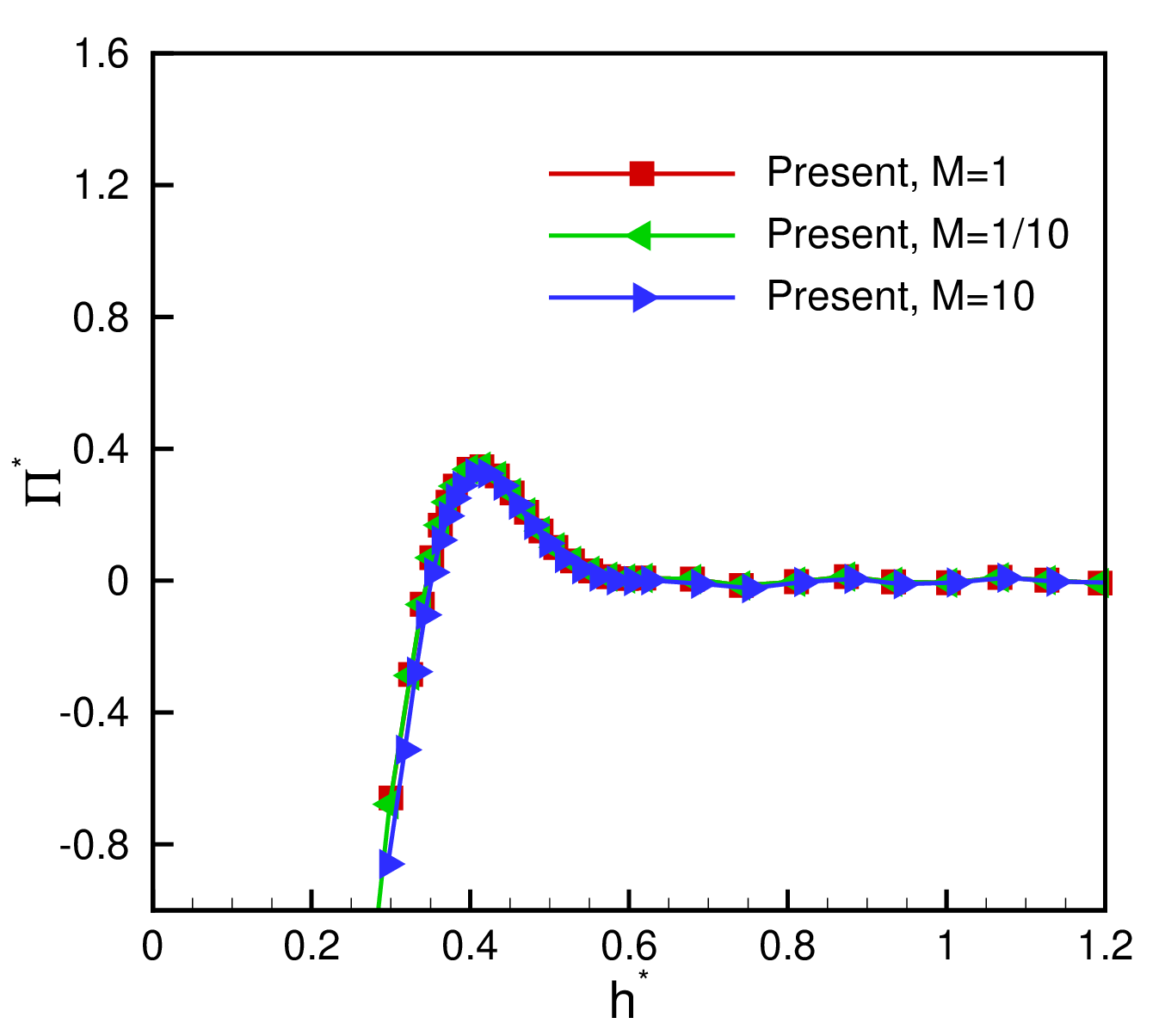,width=0.4\textwidth,clip=}}\hspace{0.0cm}  \\
	}
	\caption{\label{FIG3b} 
	Comparison of the obtained disjoining pressure $ {\varPi ^{\rm{*}}}$ at different viscosity ratio $ M $ by the original (left) and present (right) two-range models. For the original model \cite{benzi2009mesoscopic,benzi2009mesoscopic2},
	$ {\varPi ^{\rm{*}}}$ changes substantially and even changes the sign when $ M $ varies. For the present model, $ {\varPi ^{\rm{*}}}$ is independent of $ M $.
	It may be noted that the profiles for the two models at $ M=1 $ are 
	approximately the same.
	}
\end{figure*}

\begin{figure*}[!ht]
	\center {
		\includegraphics[width=0.14\textwidth,angle=270]{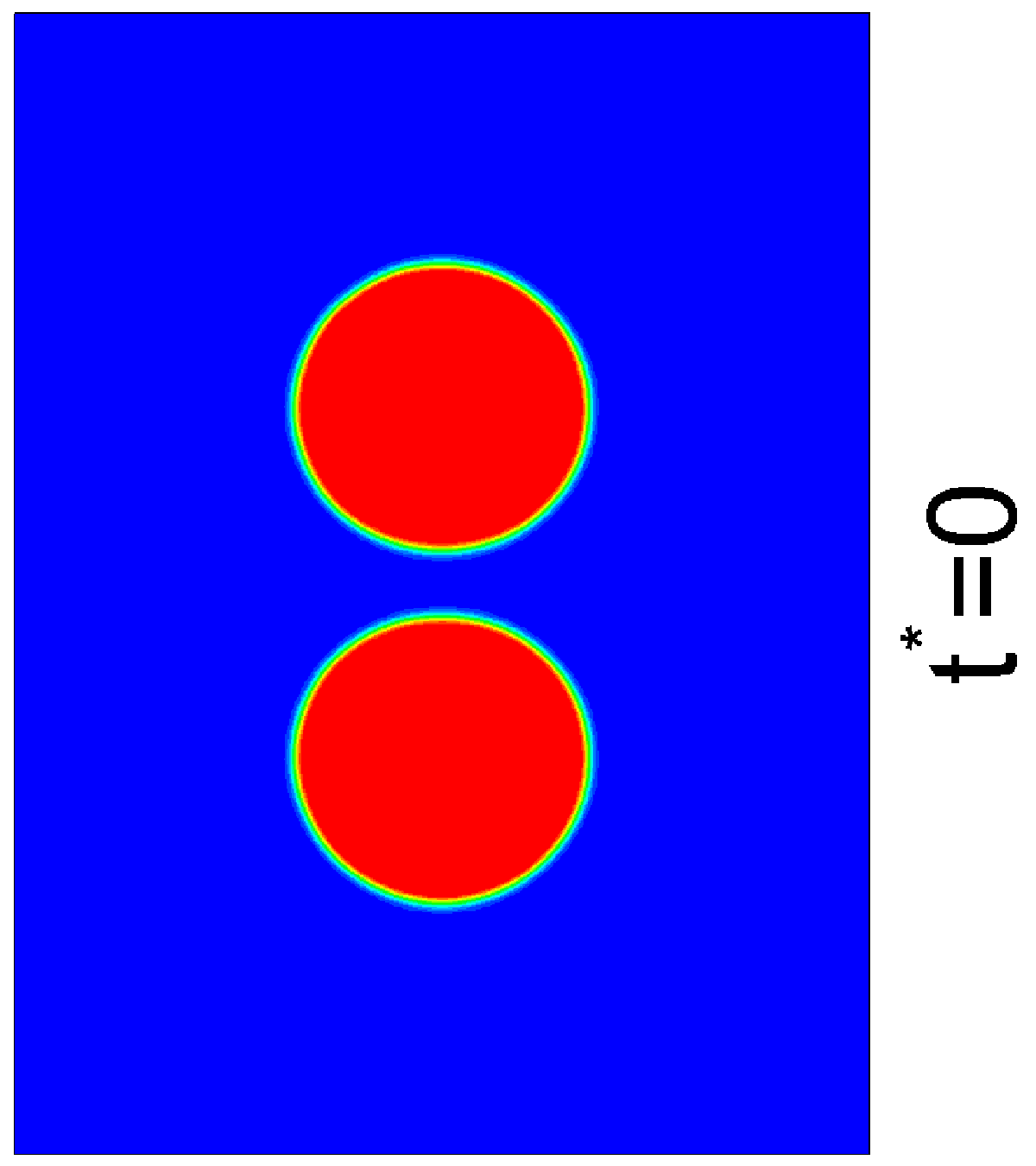}
		\includegraphics[width=0.14\textwidth,angle=270]{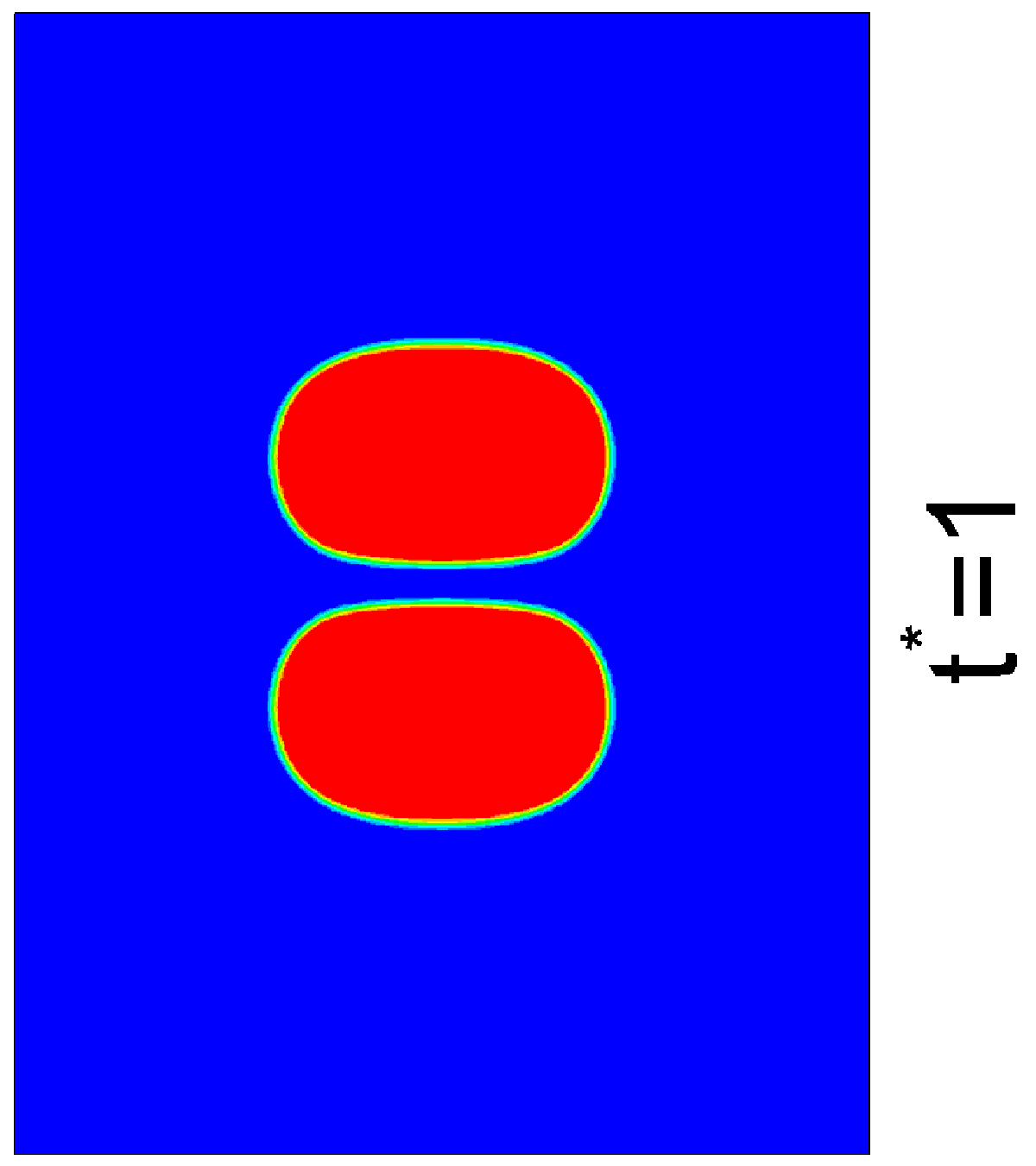}
		\includegraphics[width=0.14\textwidth,angle=270]{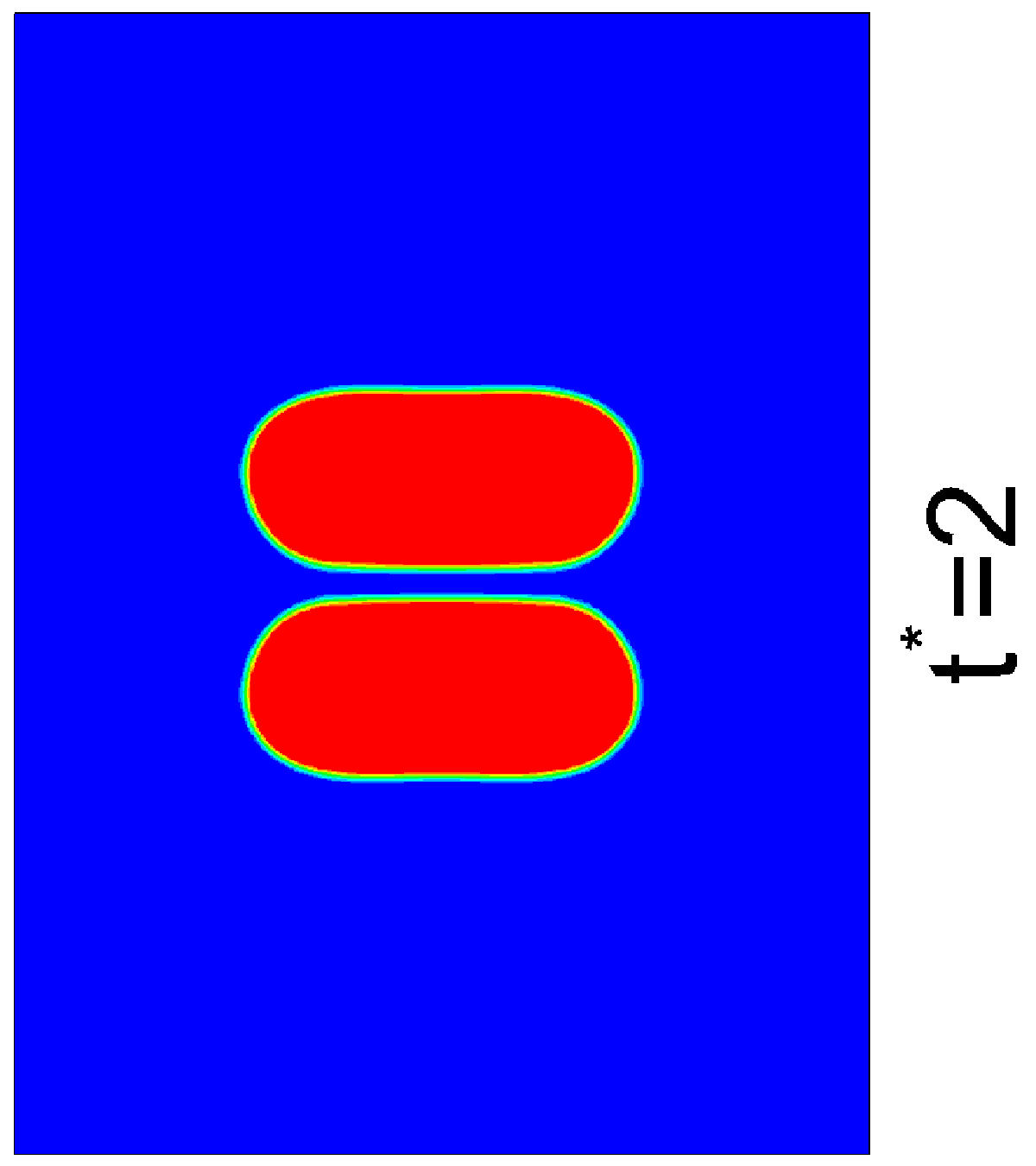}
		\includegraphics[width=0.14\textwidth,angle=270]{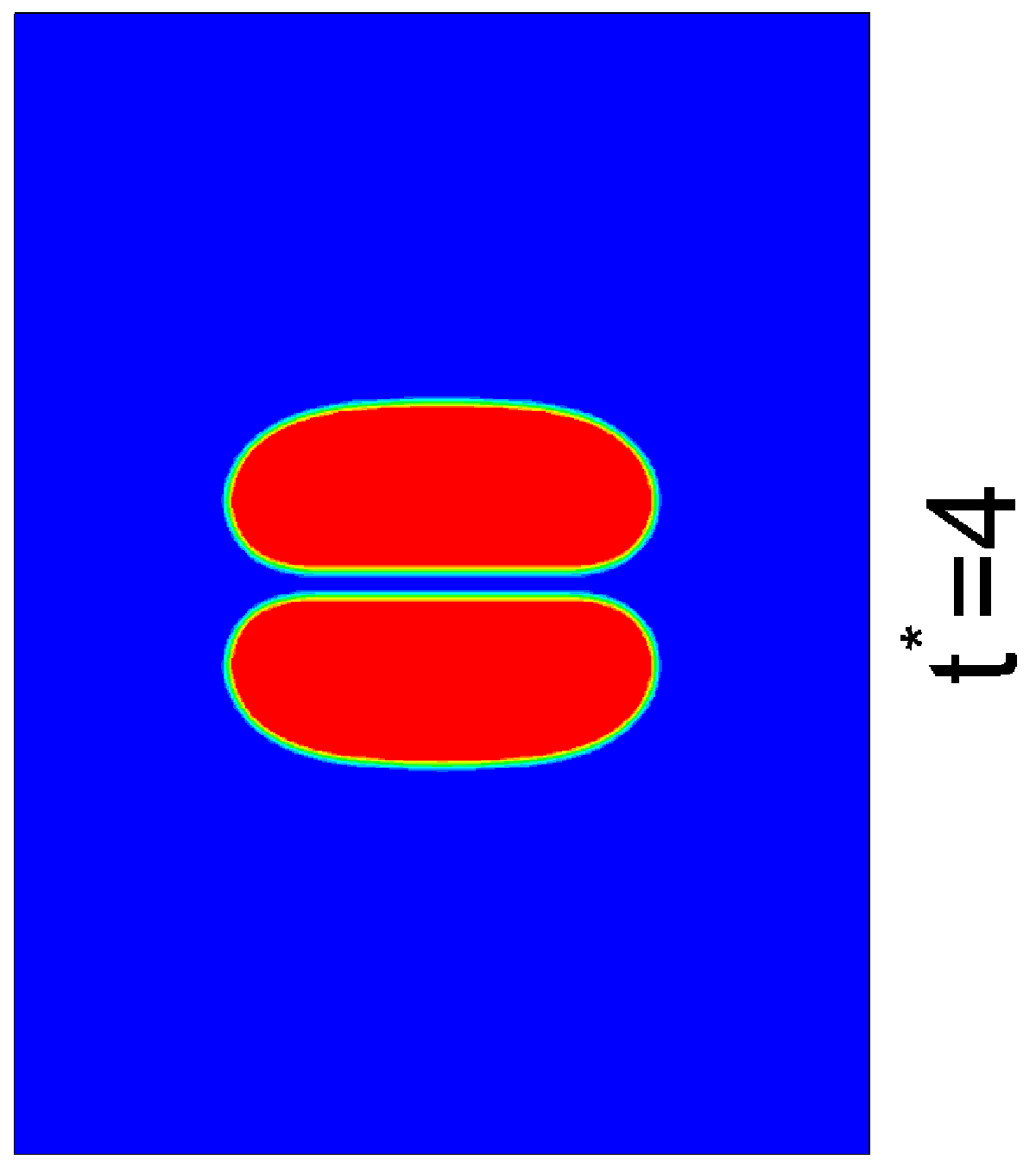}
		\includegraphics[width=0.14\textwidth,angle=270]{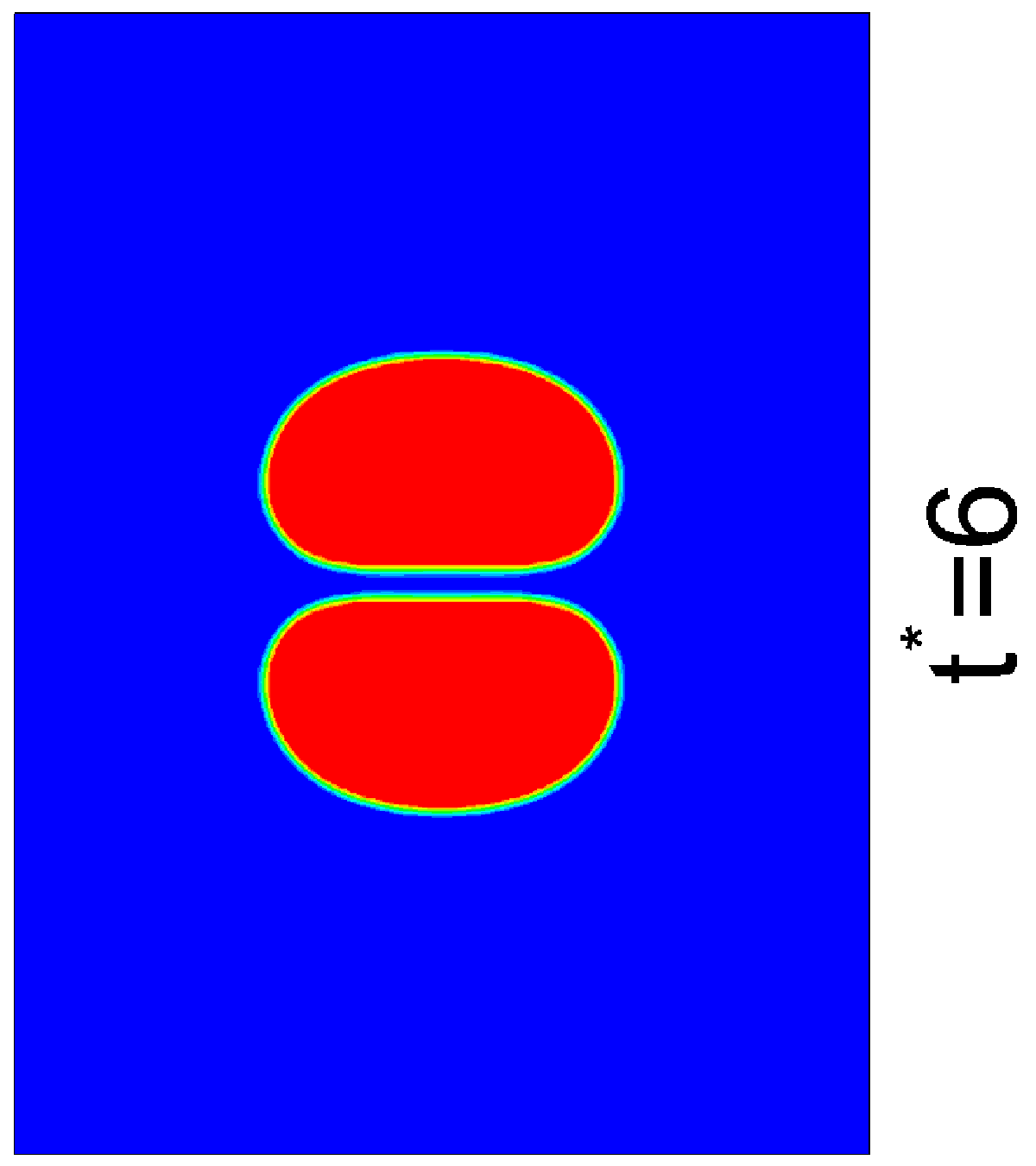}
		\includegraphics[width=0.14\textwidth,angle=270]{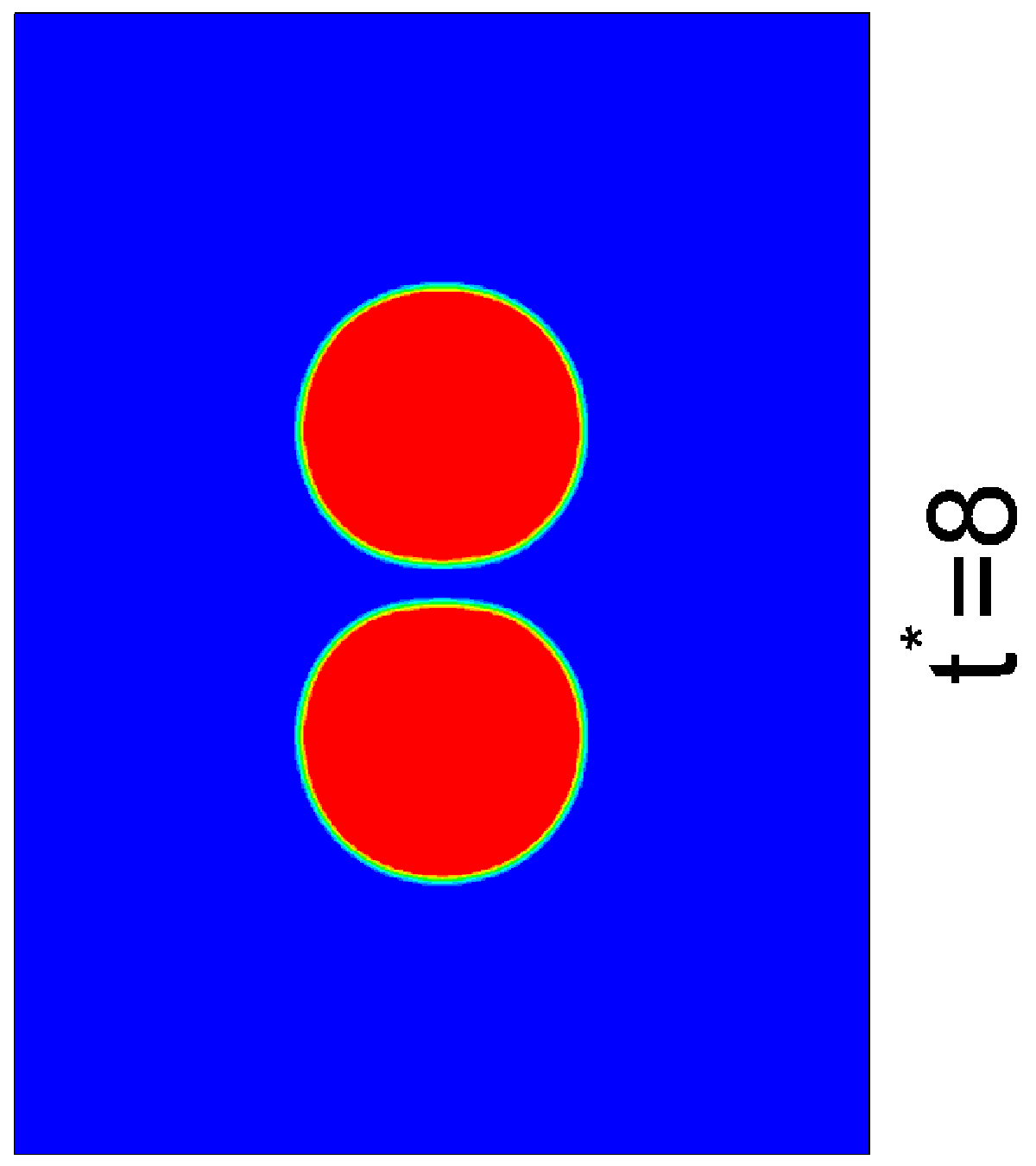}
		\\
	    {(a)}
	    \vspace{0.3cm}\\
	  		\includegraphics[width=0.14\textwidth,angle=270]{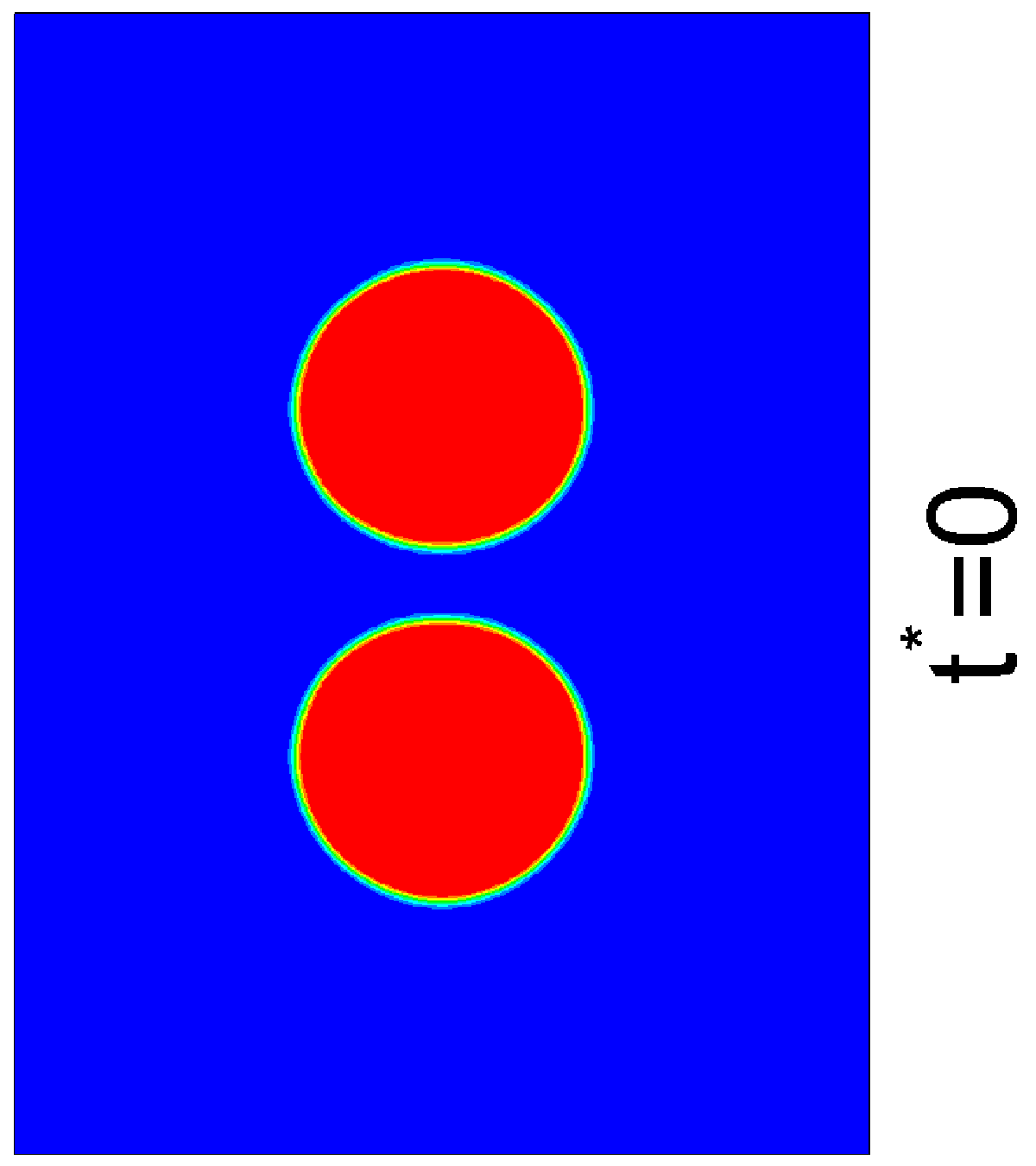}
	  		\includegraphics[width=0.14\textwidth,angle=270]{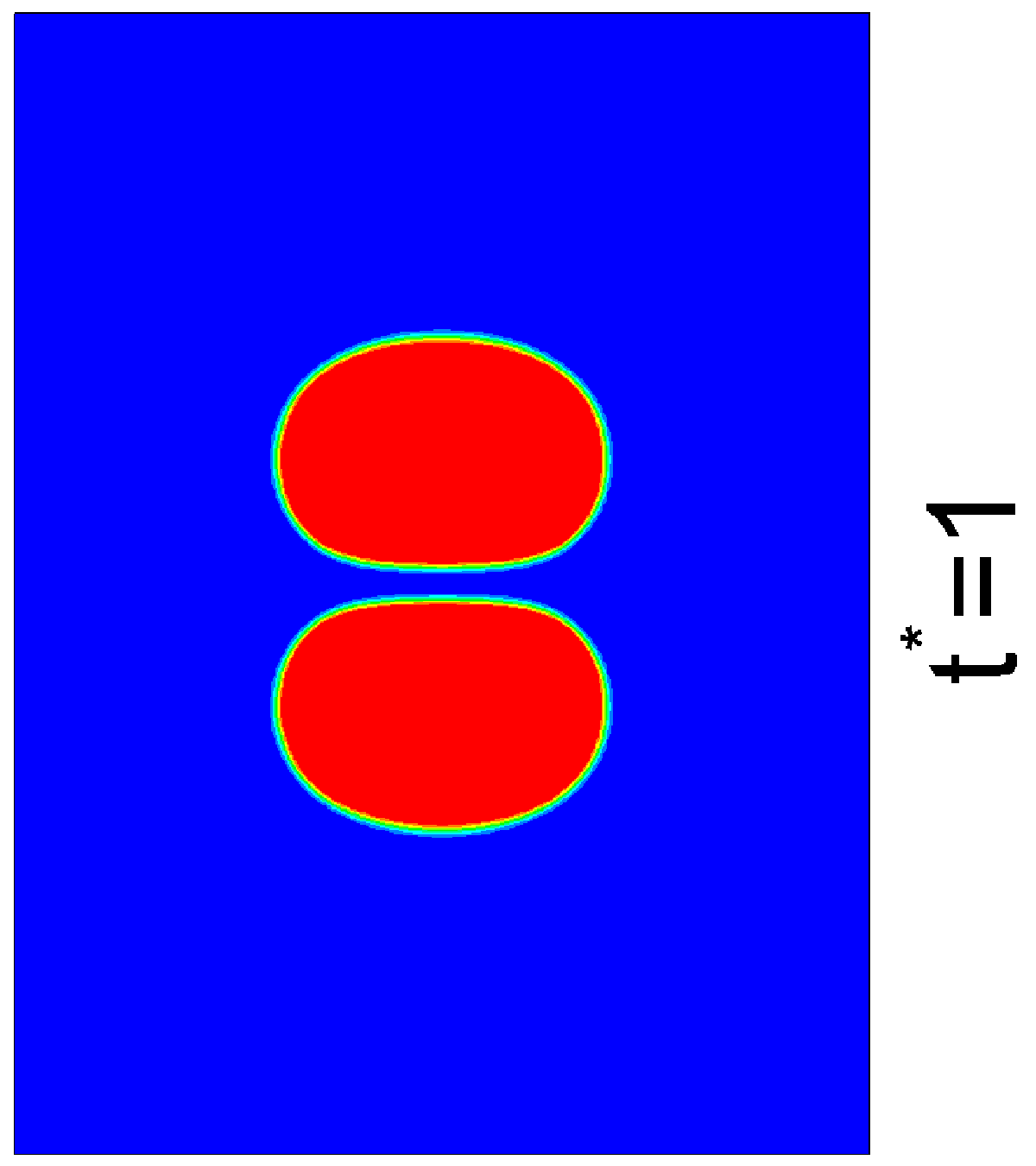}
	  		\includegraphics[width=0.14\textwidth,angle=270]{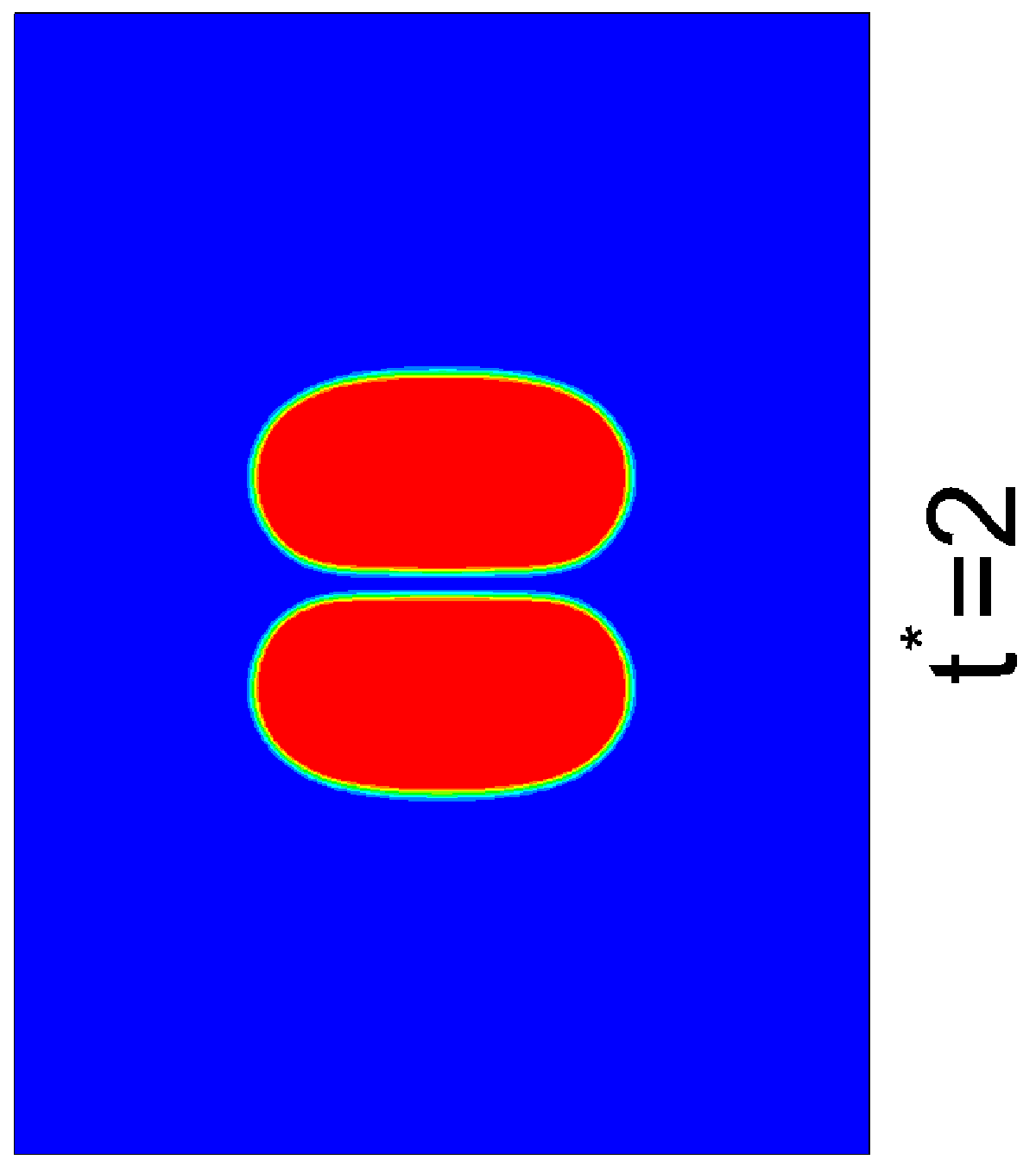}
	  		\includegraphics[width=0.14\textwidth,angle=270]{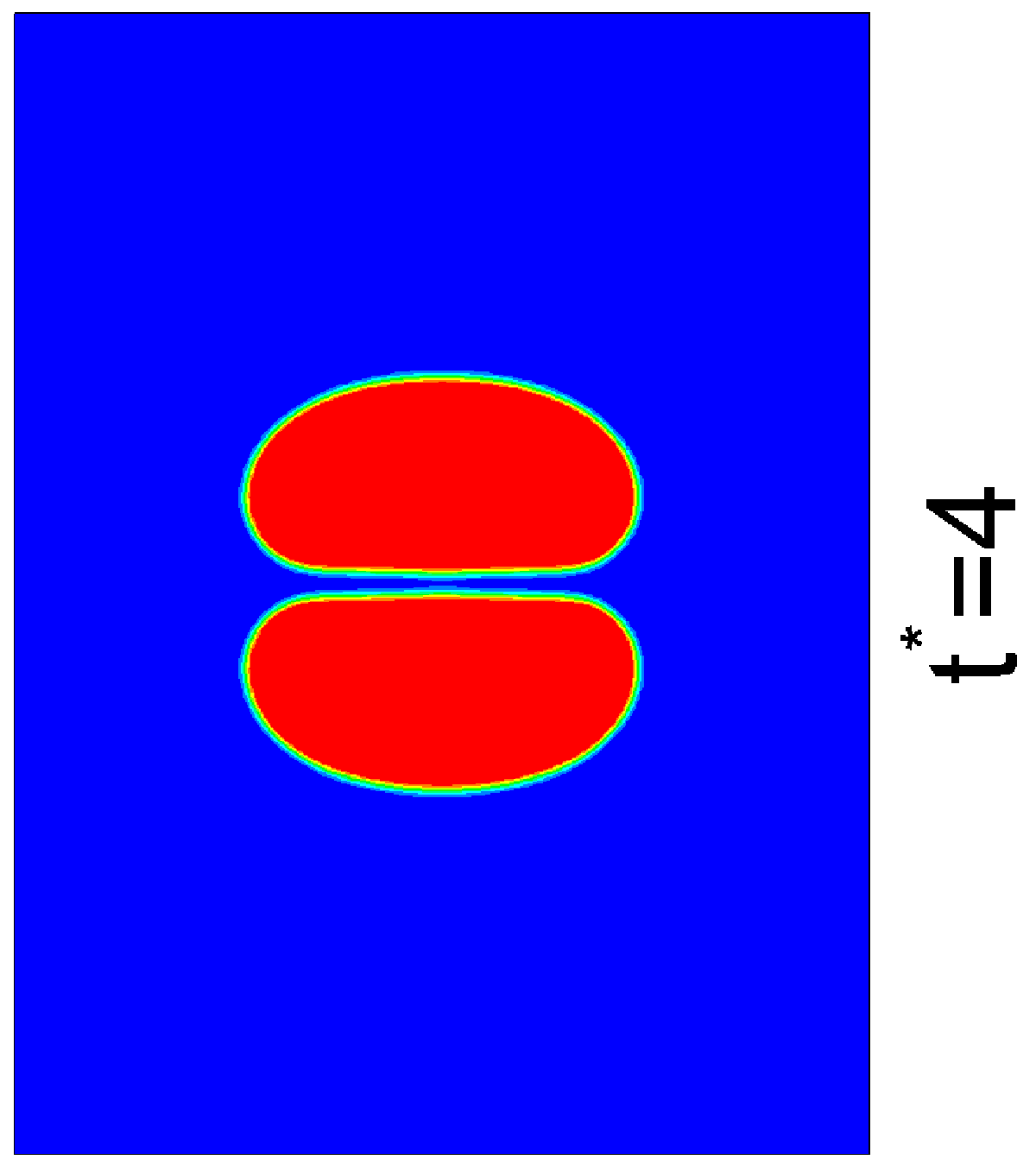}
	  		\includegraphics[width=0.14\textwidth,angle=270]{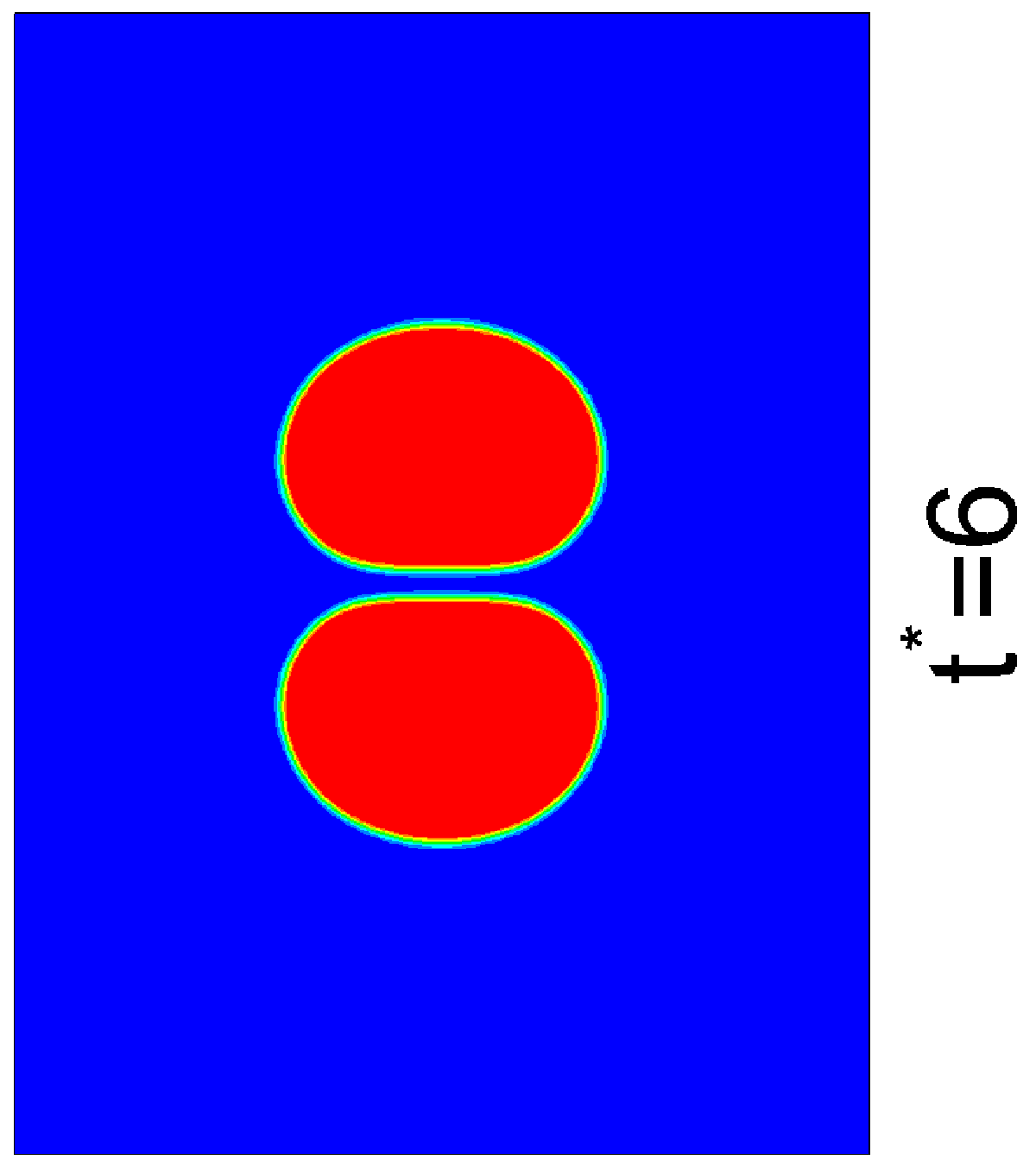}
	  		\includegraphics[width=0.14\textwidth,angle=270]{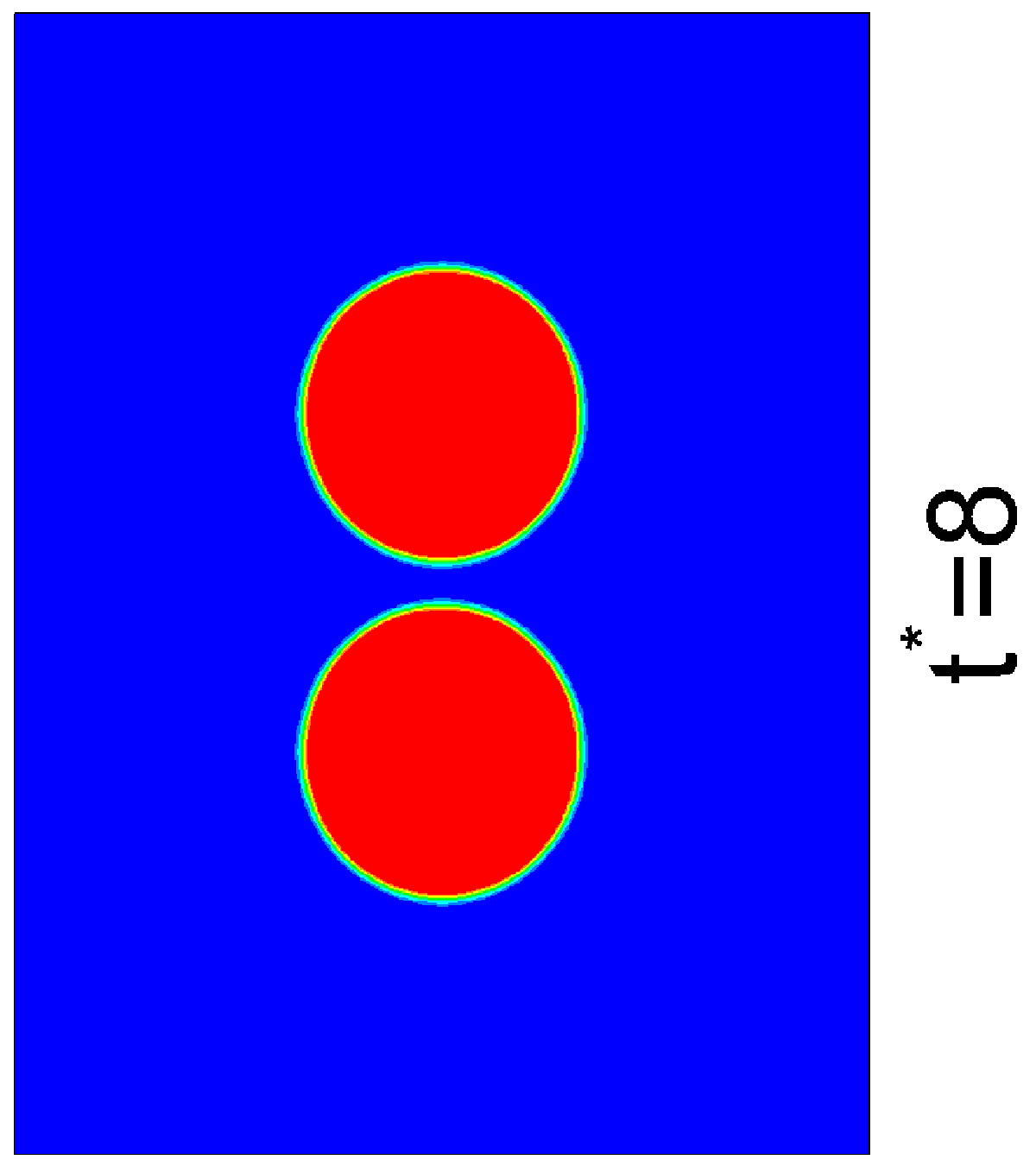}
		\\
		{(b)}
		\\
	}
	\caption{
			Snapshots $ ({t^*} = tU/D) $ of two equal-sized droplets collision at viscosity ratio $ M=1 $: (a) the original two-range model and (b) present two-range model.
	}
	\label{FIG5}
\end{figure*}

\begin{figure*}[!ht]
	\center {
	   \includegraphics[width=0.14\textwidth,angle=270]{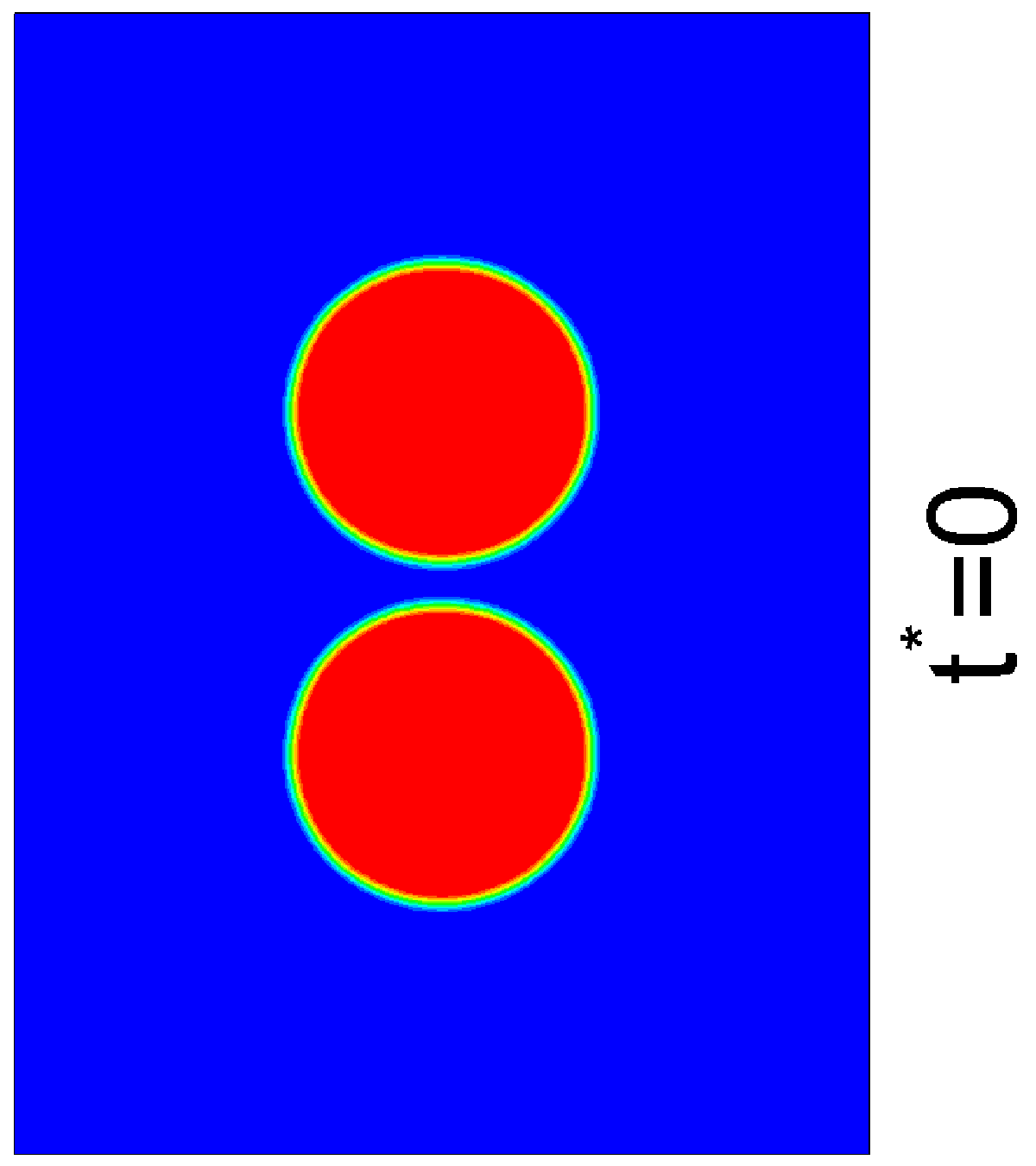}
	   \includegraphics[width=0.14\textwidth,angle=270]{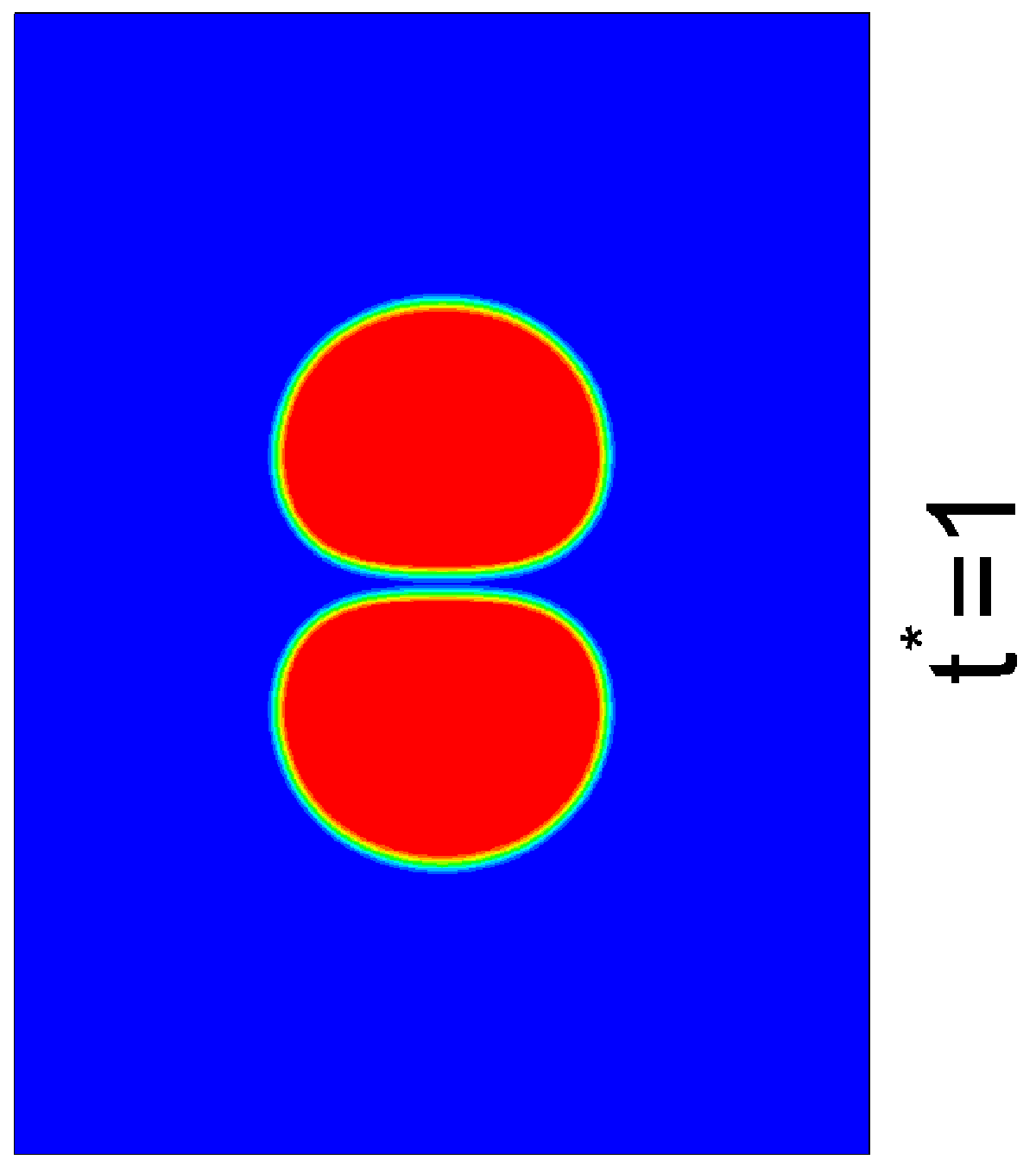}
	   \includegraphics[width=0.14\textwidth,angle=270]{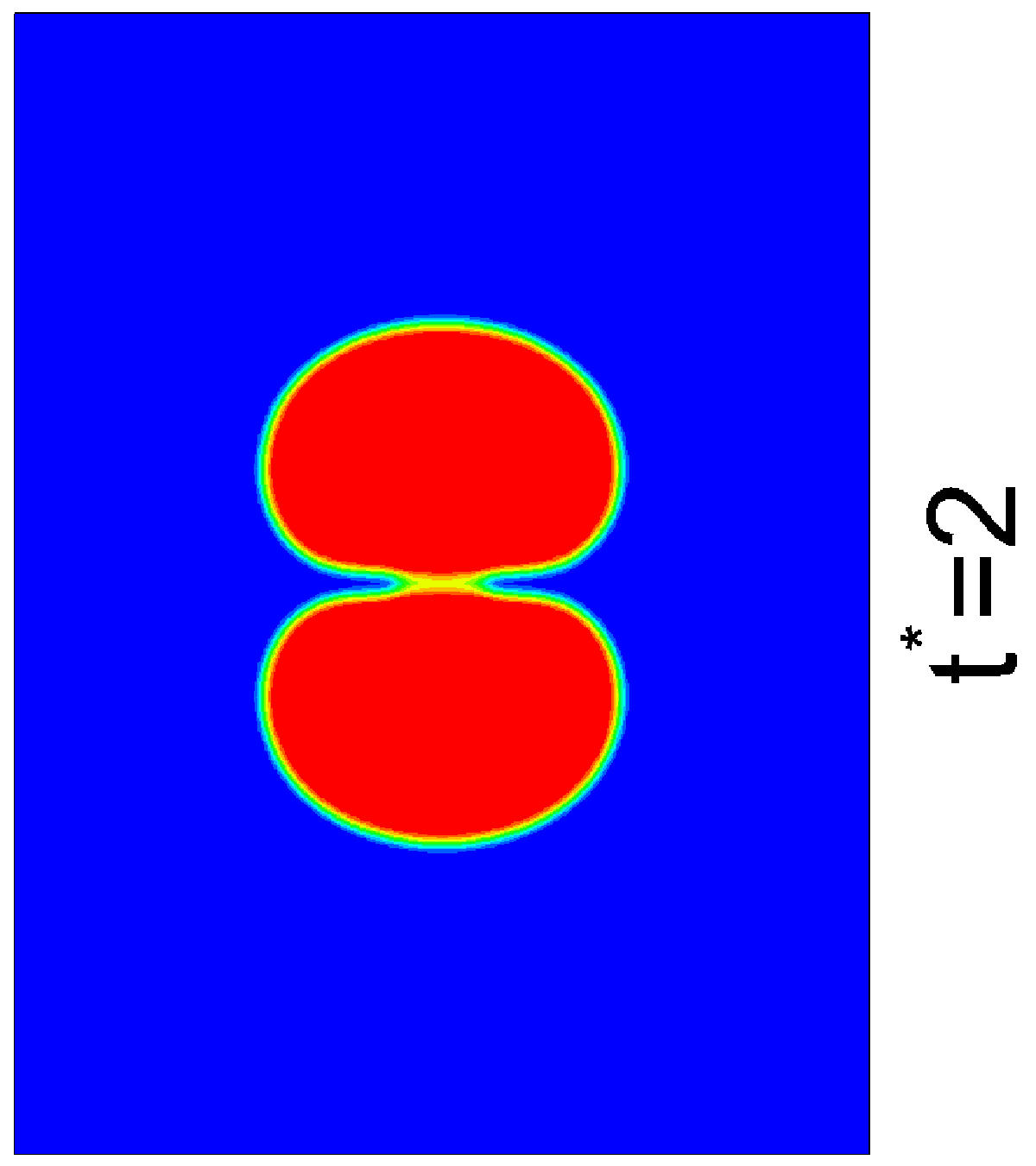}
	   \includegraphics[width=0.14\textwidth,angle=270]{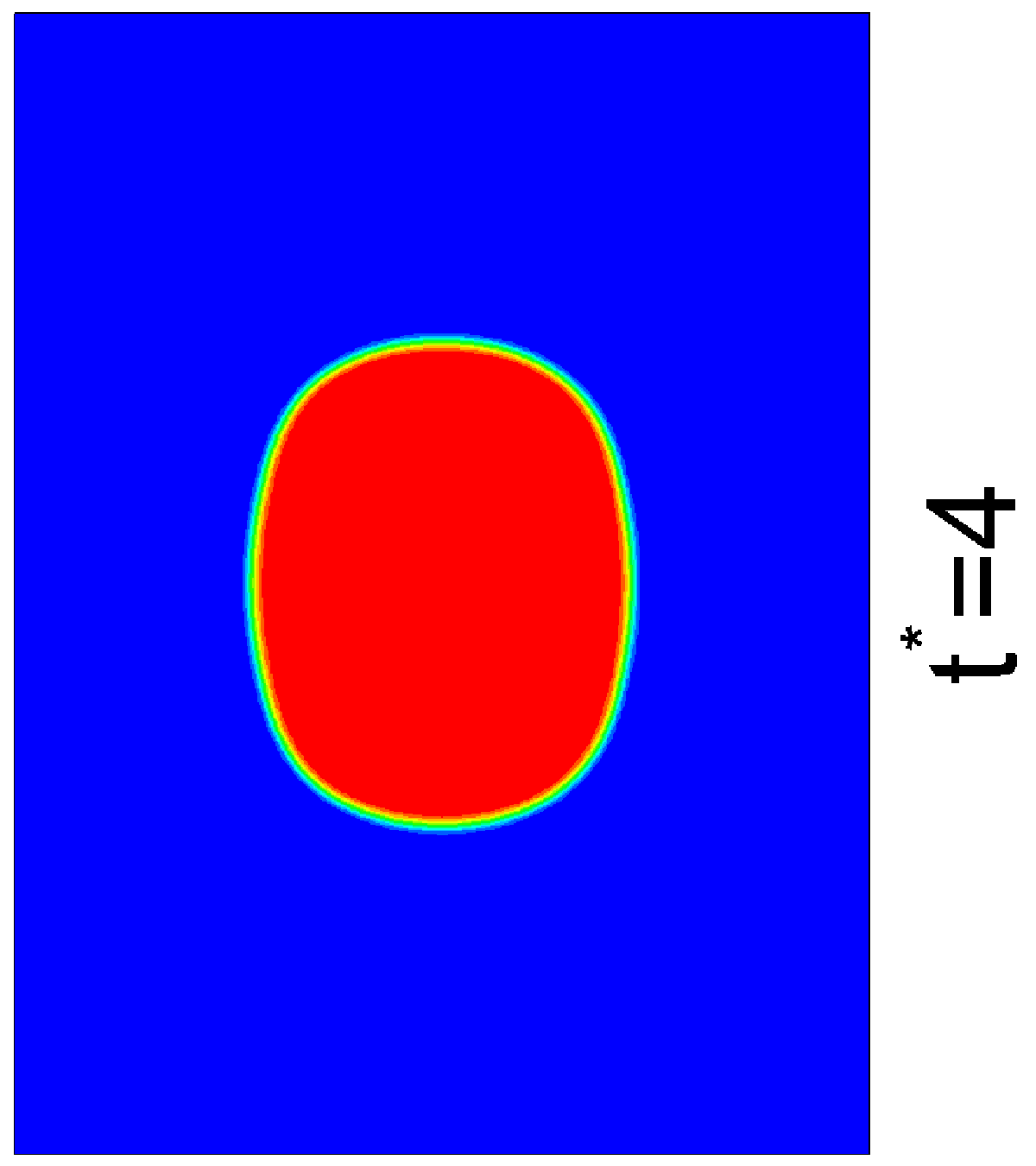}
	   \includegraphics[width=0.14\textwidth,angle=270]{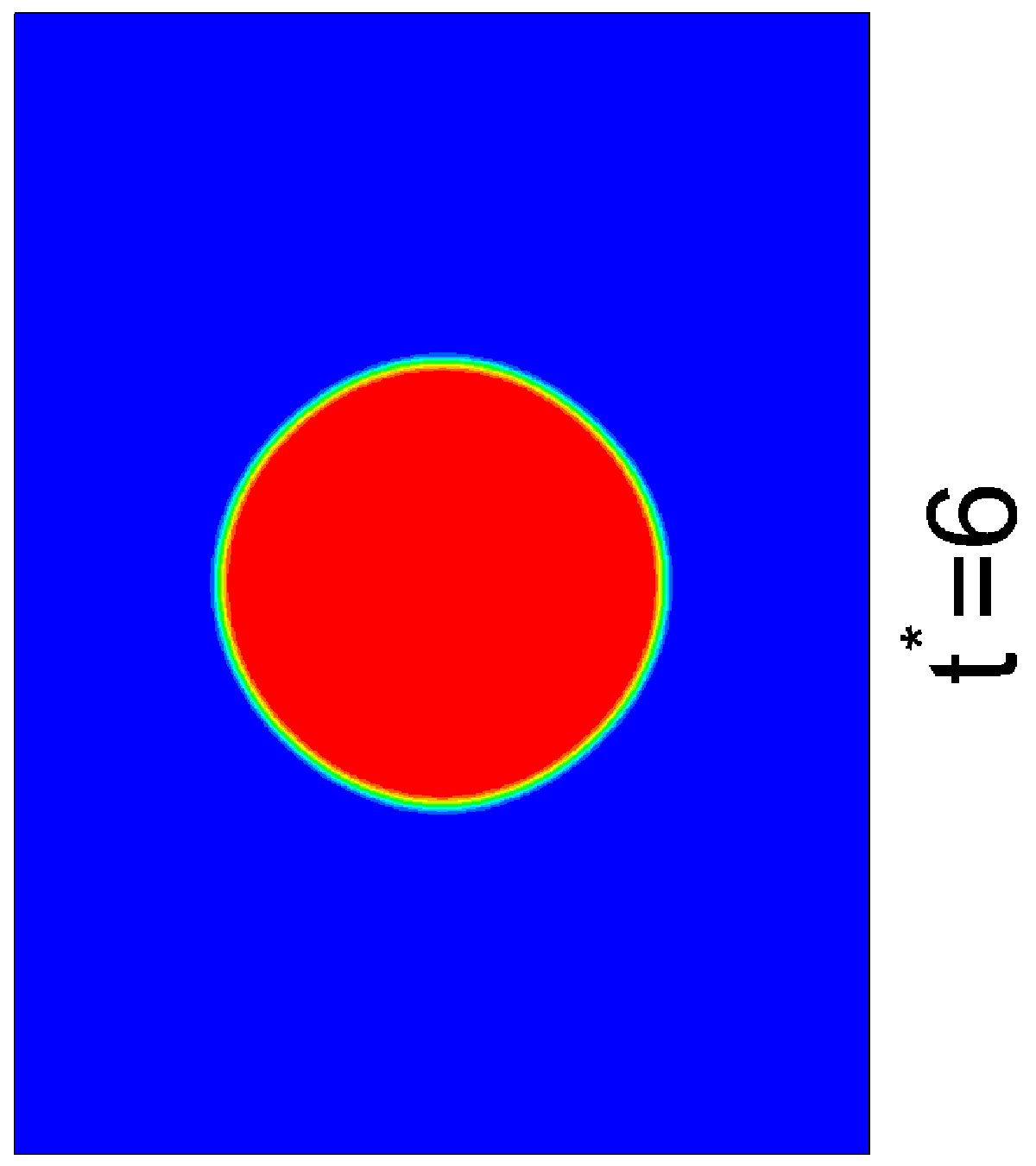}
	   \includegraphics[width=0.14\textwidth,angle=270]{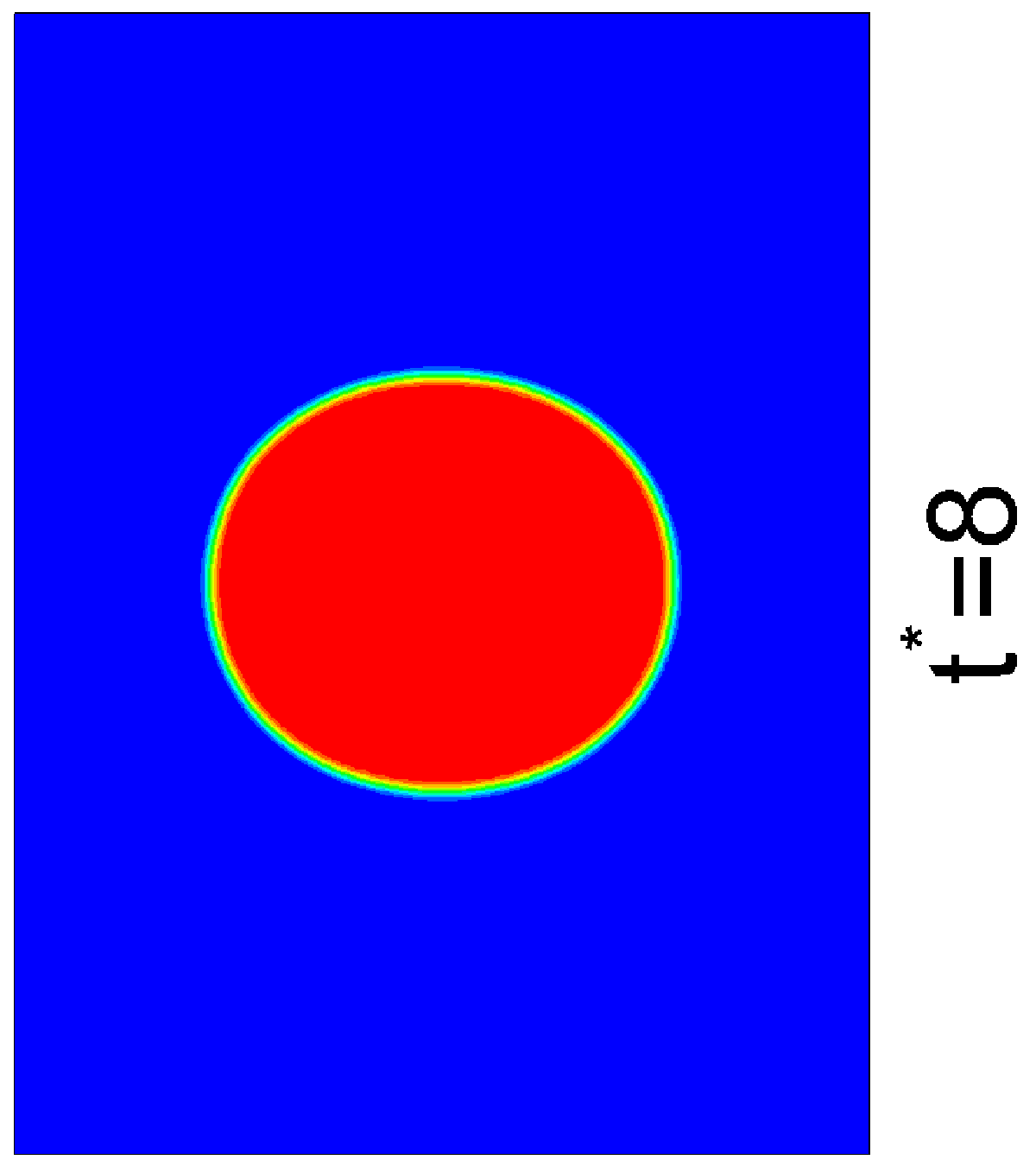}
	   \\
	    {(a)}
	    \vspace{0.3cm}\\
	    \includegraphics[width=0.14\textwidth,angle=270]{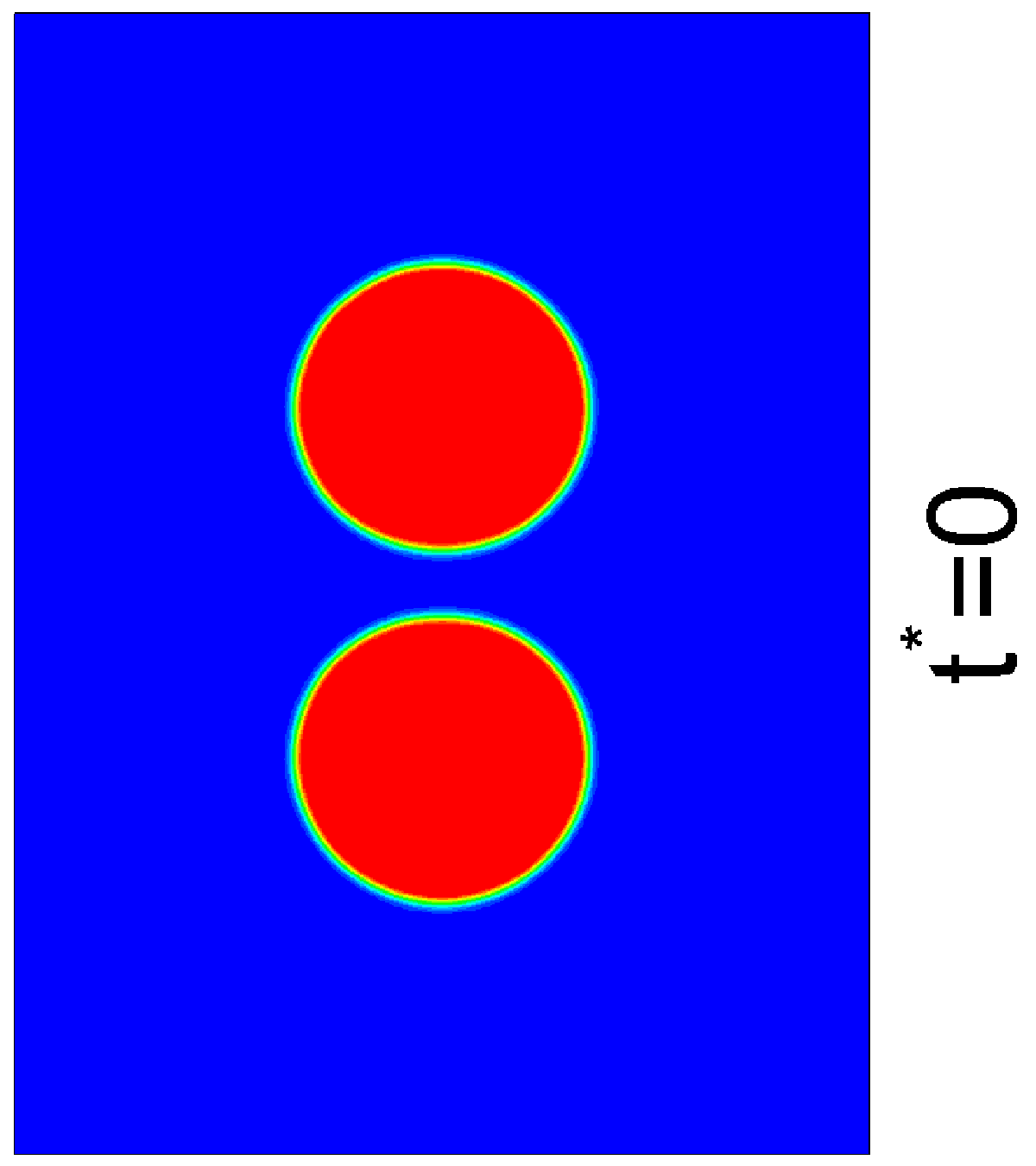}
	    \includegraphics[width=0.14\textwidth,angle=270]{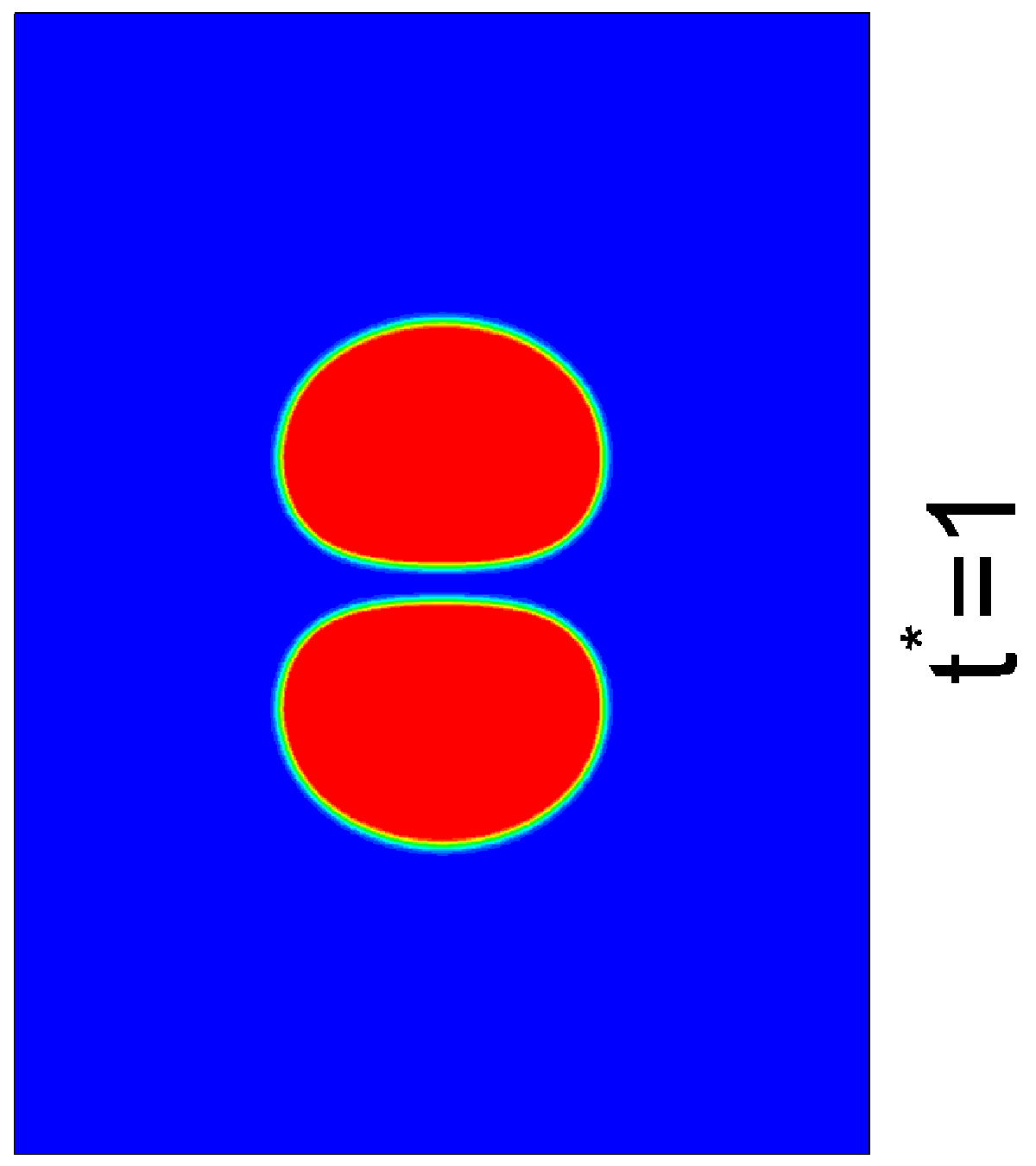}
	    \includegraphics[width=0.14\textwidth,angle=270]{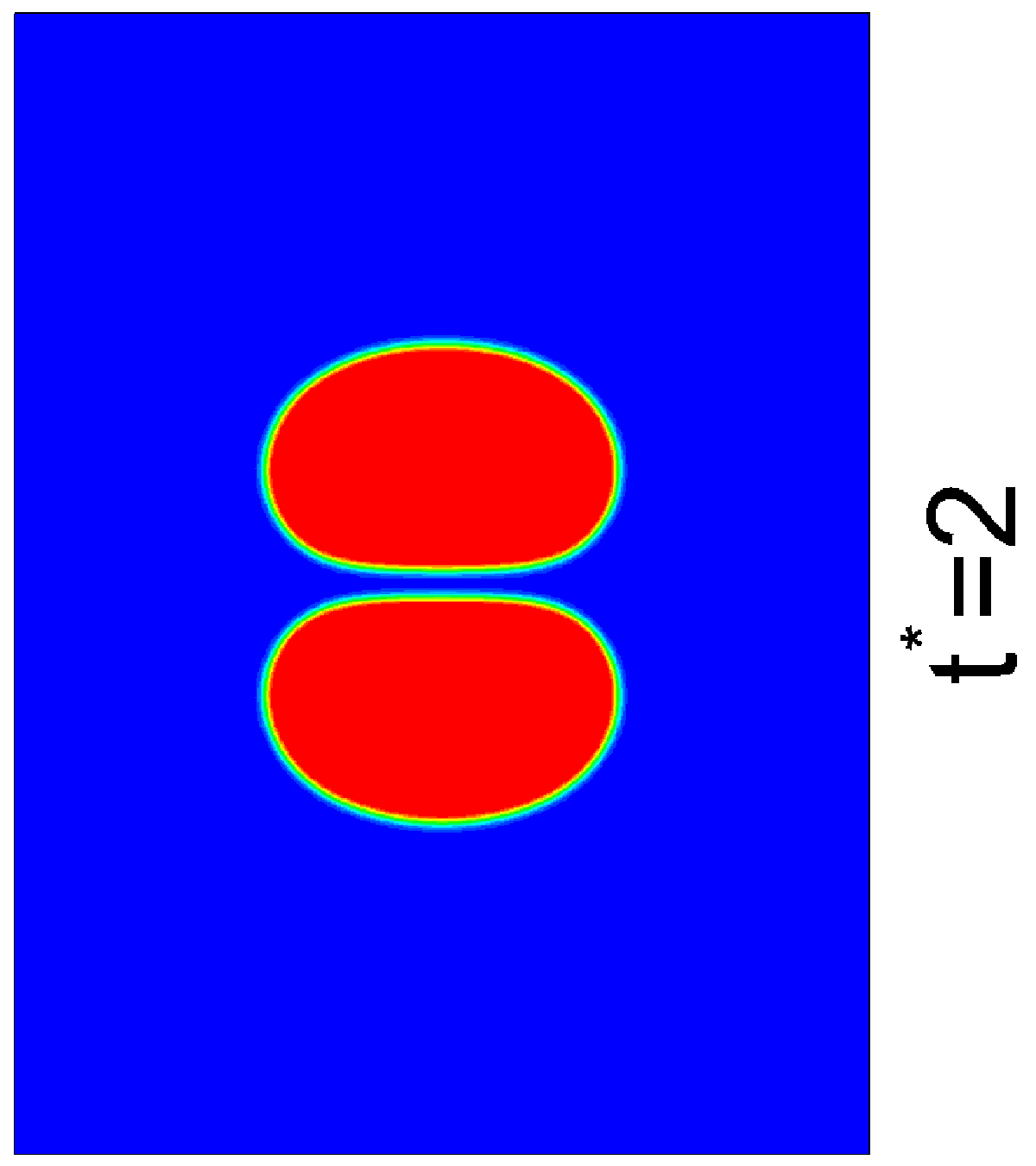}
	    \includegraphics[width=0.14\textwidth,angle=270]{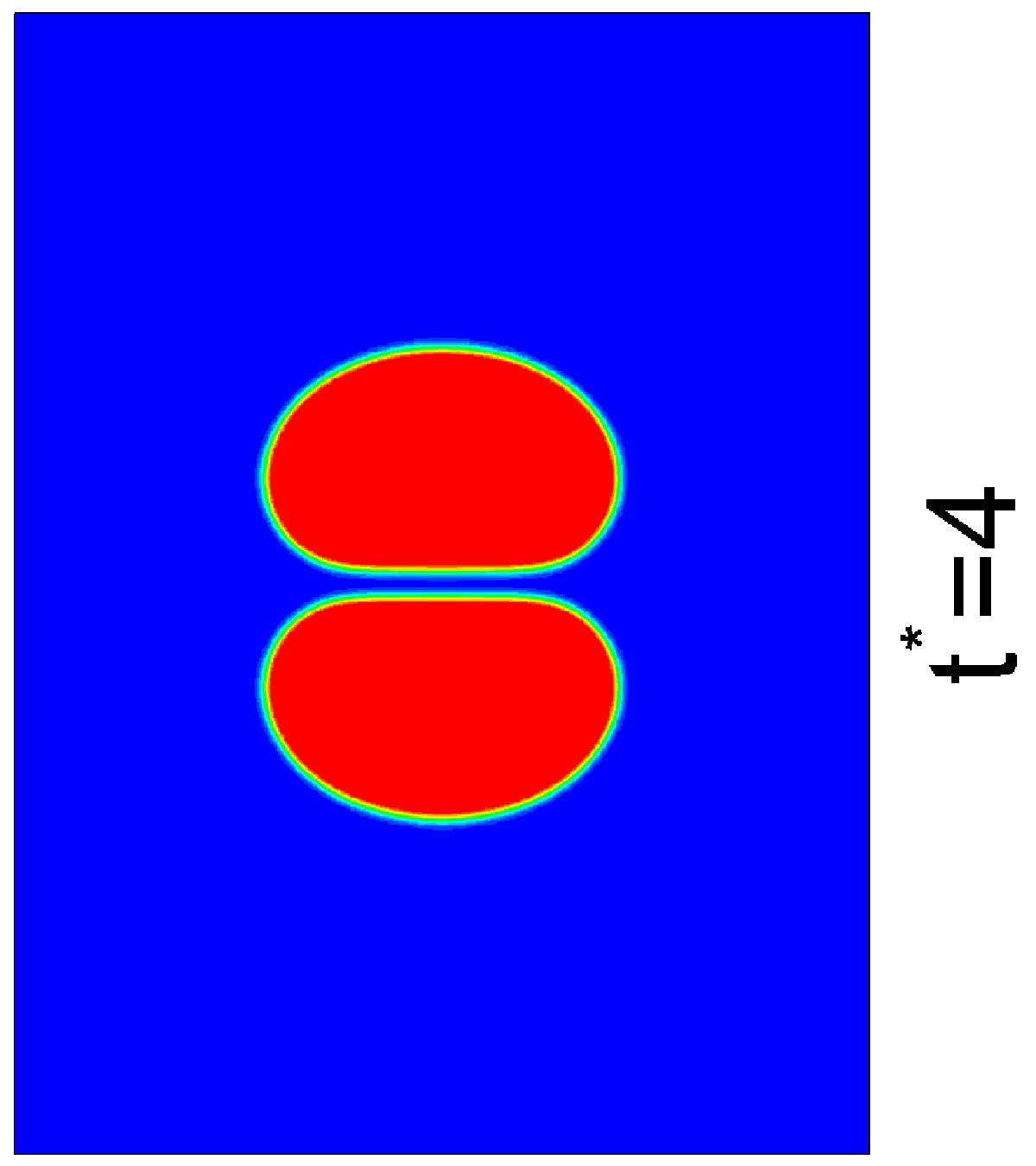}
	    \includegraphics[width=0.14\textwidth,angle=270]{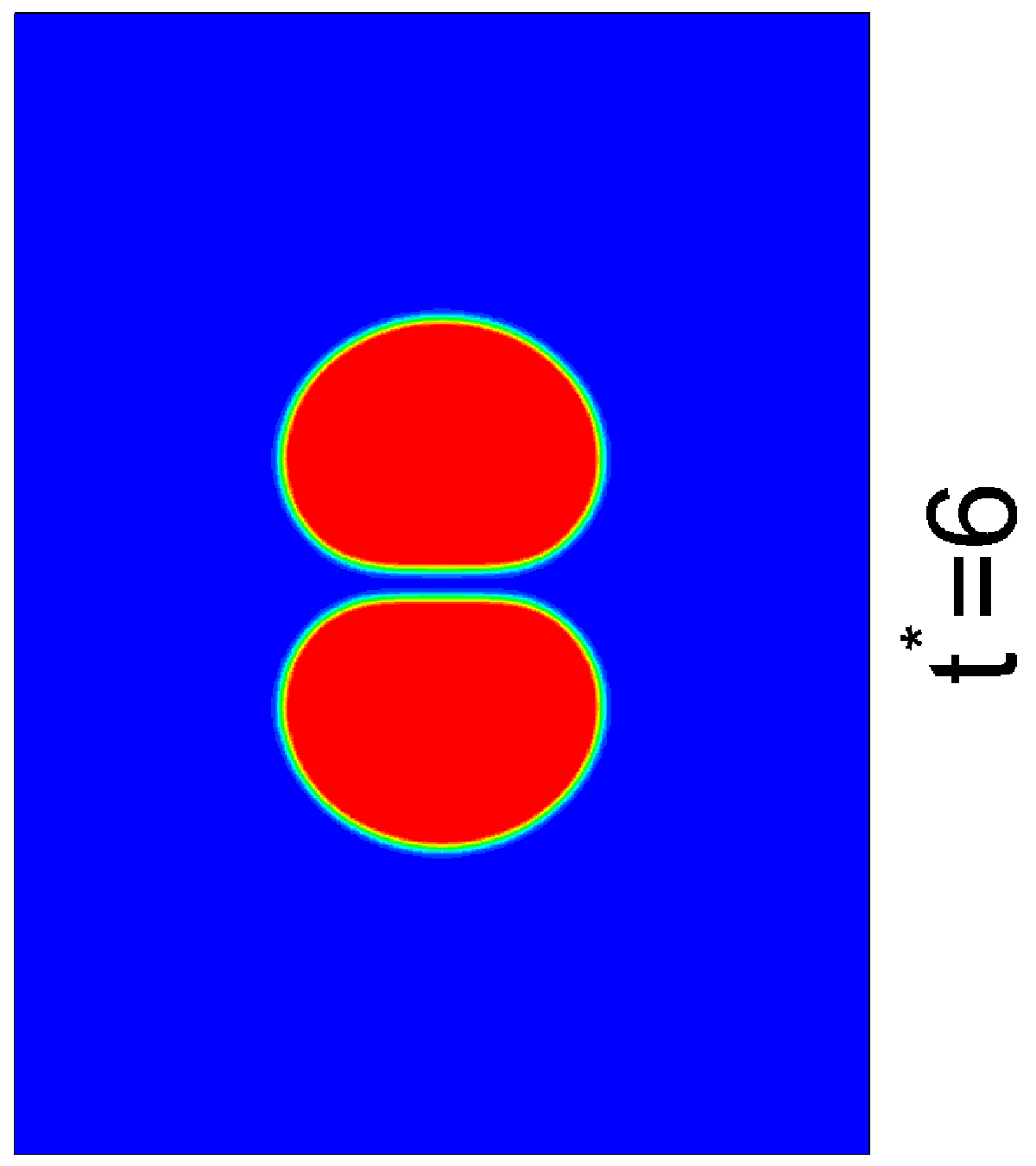}
	    \includegraphics[width=0.14\textwidth,angle=270]{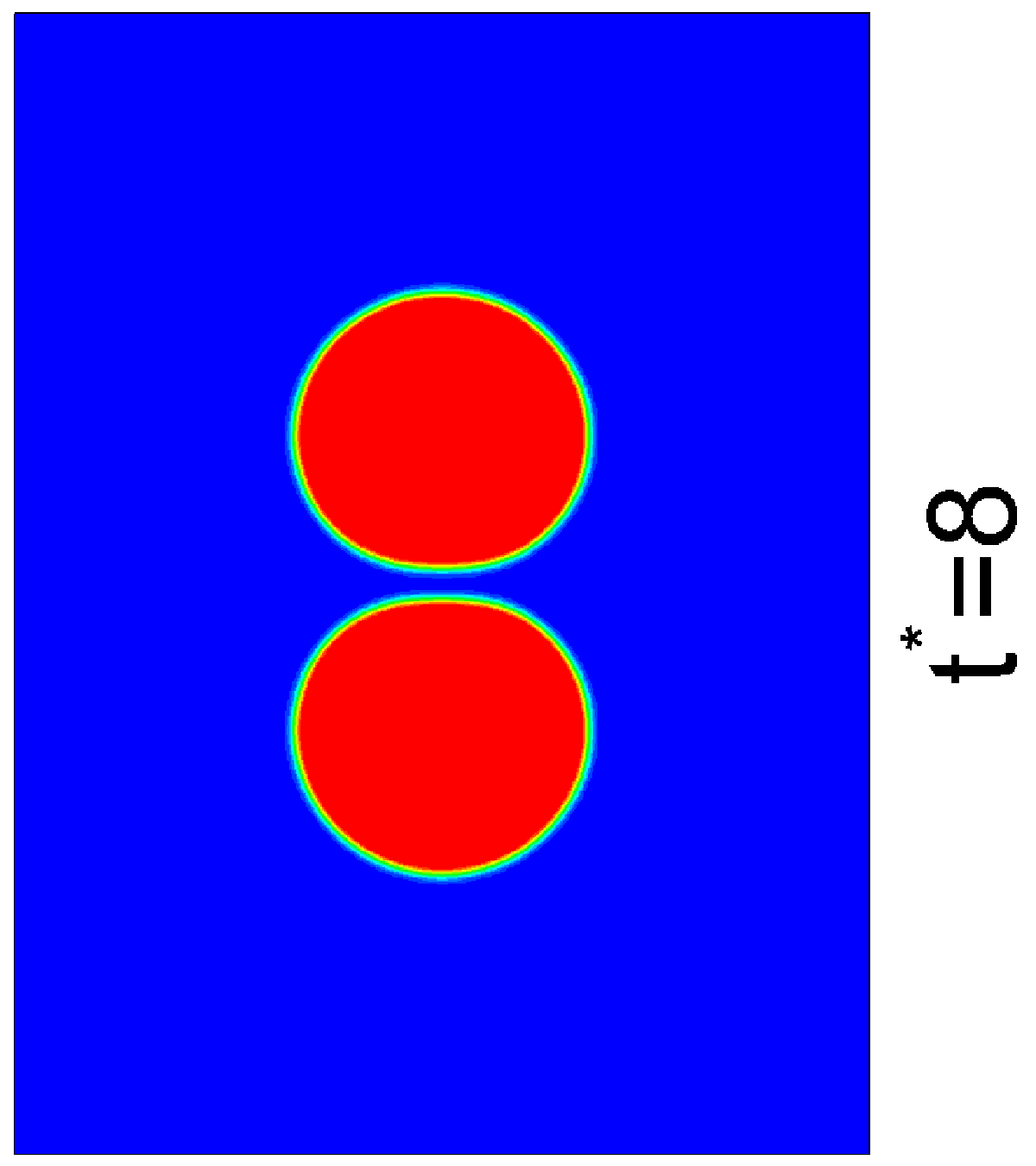}
		\\
		{(b)}
		\\
	}
	\caption{
		Snapshots $ ({t^*} = tU/D) $ of two equal-sized droplets collision at viscosity ratio $ M=3 $: (a) the original two-range model and (b) present two-range model.
	}
	\label{FIG6}
\end{figure*}

Then we compare the original two-range \cite{benzi2009mesoscopic,benzi2009mesoscopic2} and the present model in Fig.~\ref{FIG3b}. For the original model, some $\tau$-dependent terms are reproduced in $ P_{ab}^{kin} $ (see Eq. (\ref{ekab})), thus the surface tension is dependent on the viscosity for each component. In this simulation, the measured surface tensions at viscosity ratio 
$M = \left\{ {1/2,1,3} \right\}$
 are 0.056, 0.043 and 0.039, respectively. Remarkably, the $ \tau$-dependent terms in $ P_{ab}^{kin} $ are removed by using the present forcing scheme. Under different viscosity ratio 
$M = \left\{ {1/10,1,10} \right\}$
, the surface tension is approximately constant (with a maximum relative error of $ 1.2\% $ ). As a result, the disjoining pressure  depends on $ M $ significantly for the original model in Refs. \cite{benzi2009mesoscopic,benzi2009mesoscopic2}, while it is independent of $ M $ for the present model. 
To highlight this point, we consider collisions of droplets of component $ A $ in a liquid matrix of component $ B $. We wish to point out that the configuration here is consistent with Fig.~\ref{interface}, because a thin film (component $ B $) is produced between the two droplets (component $ A $) when the two droplets are approaching each other.
Two equal-sized droplets with a diameter $ D$ are initialized with a relative velocity 
$ U $, which leads to a Capillary number $ Ca = \frac{{{\mu _A}U}}{\gamma } \approx 0.3 $ and 
a Weber number $ We = \frac{{{\rho _A}D{U^2}}}{\gamma } \approx 15 $. 
 First, we choose a unity viscosity ratio $ {\mu _A} = {\mu _B} = 0.1 $. Due to the positive disjoining pressure, a stable thin film between the droplets is supported, and the two droplets bounce back in the end, as shown in Fig.~\ref{FIG5}. Then we change 
$ {\mu _B} $ to 0.3. For the present model, the disjoining pressure is viscosity-independent,
thus the two droplets still bounce back in Fig.~\ref{FIG6}. However, the disjoining pressure changes to negative for the original model when $ M = \frac{{{\mu _B}}}{{{\mu _A}}}=3 $. As a result, the thin film cannot be supported any longer, and the two droplets eventually coalesce.

The novelty of the present model, which displays a disjoining pressure independent of the viscosity ratio, represents a significant improvement
with respect to the original model and makes it of major significance for many applications where non unit viscosity ratios are needed \cite{cubaud2008capillary,farhat2011suppressing,bai2016droplet,foglino2017flow,montessori2018regularized,stolovicki2018throughput}.
\subsection{Pressure-driven emulsion in a planar channel}\label{sec.3d}

\begin{figure}
	 \includegraphics[width=0.42\textwidth,angle=270]{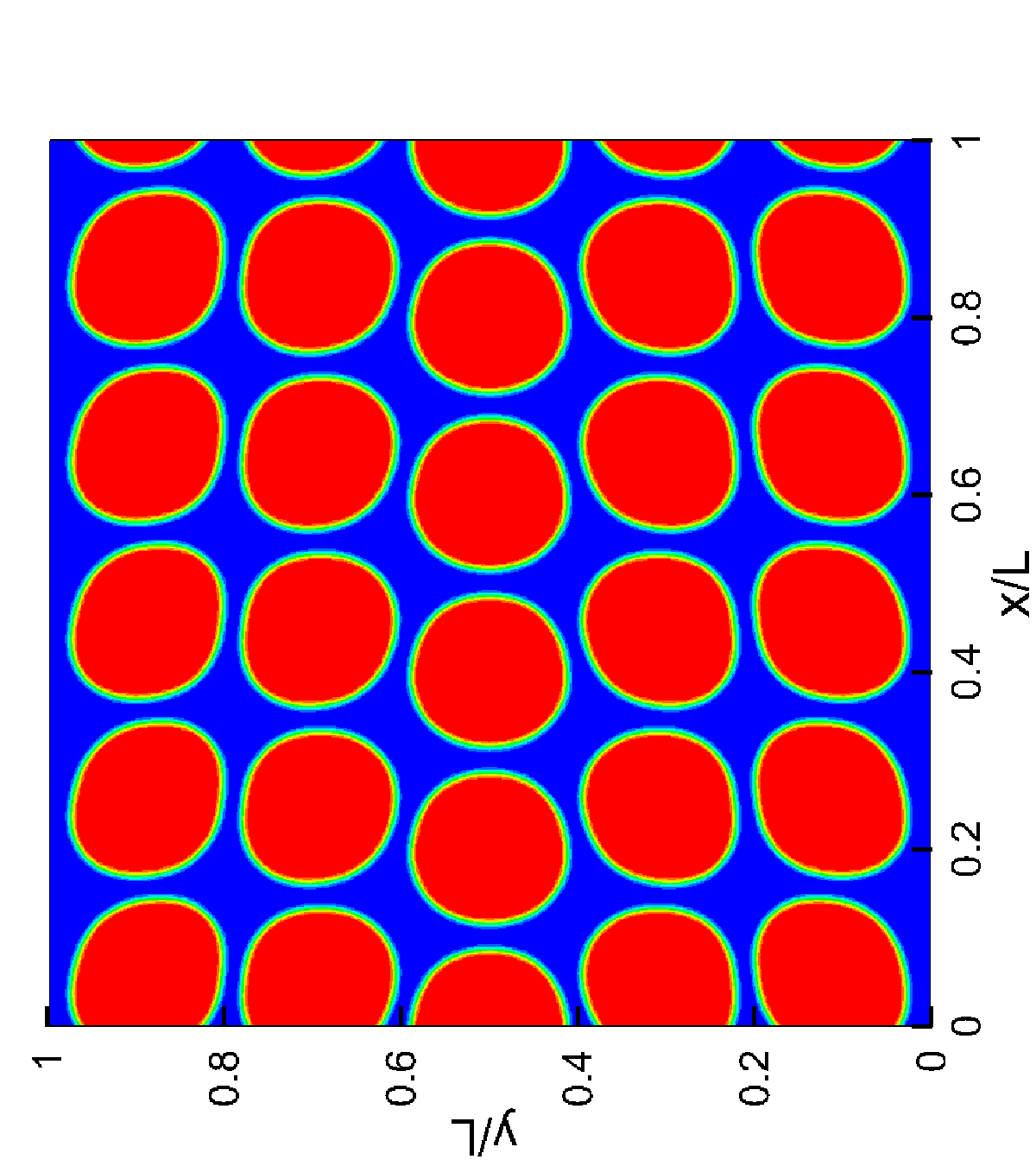}
	 \includegraphics[width=0.42\textwidth,angle=270]{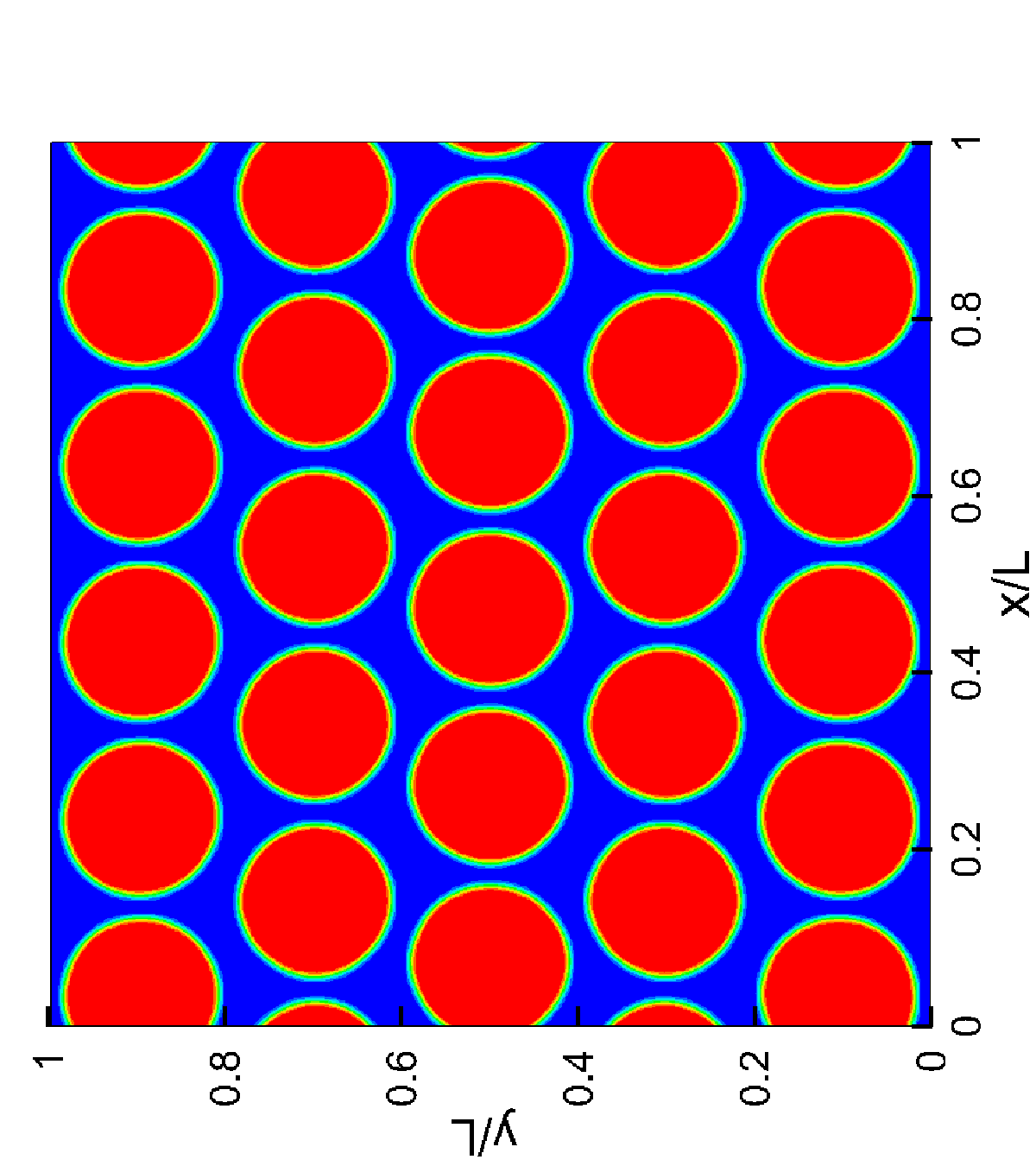}
	\caption{\label{FIG4a} Snapshots of pressure-driven emulsion in a planar channel at volume fraction $\Phi  = 0.64$ with $M = {\mu _B}/{\mu _A} = 1$
		(left panel), and $M = 1/10$ (right panel).		
	}
\end{figure}

The method we have presented, thanks to its built-in properties, results particularly suitable to 
simulate the hydrodynamics of emulsions, especially in dense situations. We move therefore here to simulate a multi-droplet situation.
We consider a pressure-driven flow in a planar channel of a monodisperse emulsion, made of a regular
  arrangement of equal size droplets (component $ A $) dispersed in a continuous matrix (component $ B $). 
The simulations are performed on a $L \times L$ domain, with
  $L=220\Delta x$; no-slip boundary conditions for the velocity are imposed on the top ($y=L$) and bottom ($y=0$) walls; 
non-wetting boundary conditions for droplets apply, i.e. a contact angle of $\theta = 180^o$ is set for component $ A $ on both walls. 
A body force $F_b$ in the $x$-direction is imposed to mimic a constant pressure gradient,
which can be expressed in non-dimensional form as 
$\overline{F}_b = F_b L^3\rho_B/(8 \mu_B^2)$ (let us notice that such expression coincides with the Reynolds number 
that would be achieved in the corresponding Poiseuille flow in the pure continuous phase, for that given forcing $F_b$).
The coupling parameters are ${G_{AB}} = 3.0$, ${G_{A,1}}{\rm{ = -}}7.4$, ${G_{A,2}}{\rm{ = }}6.4$, giving $ \gamma  = 0.04 $, and the
droplet radius is set within $10\Delta x < R < 20\Delta x$ to obtain different volume fractions $\Phi$ of the dispersed phase, $\Phi \in [0.18,0.64]$
  (typical snapshots of $\Phi  = 0.64$  are shown in Fig.~\ref{FIG4a}). 

\begin{figure*}[!ht]	        
 	\center {			
 		{\epsfig{file=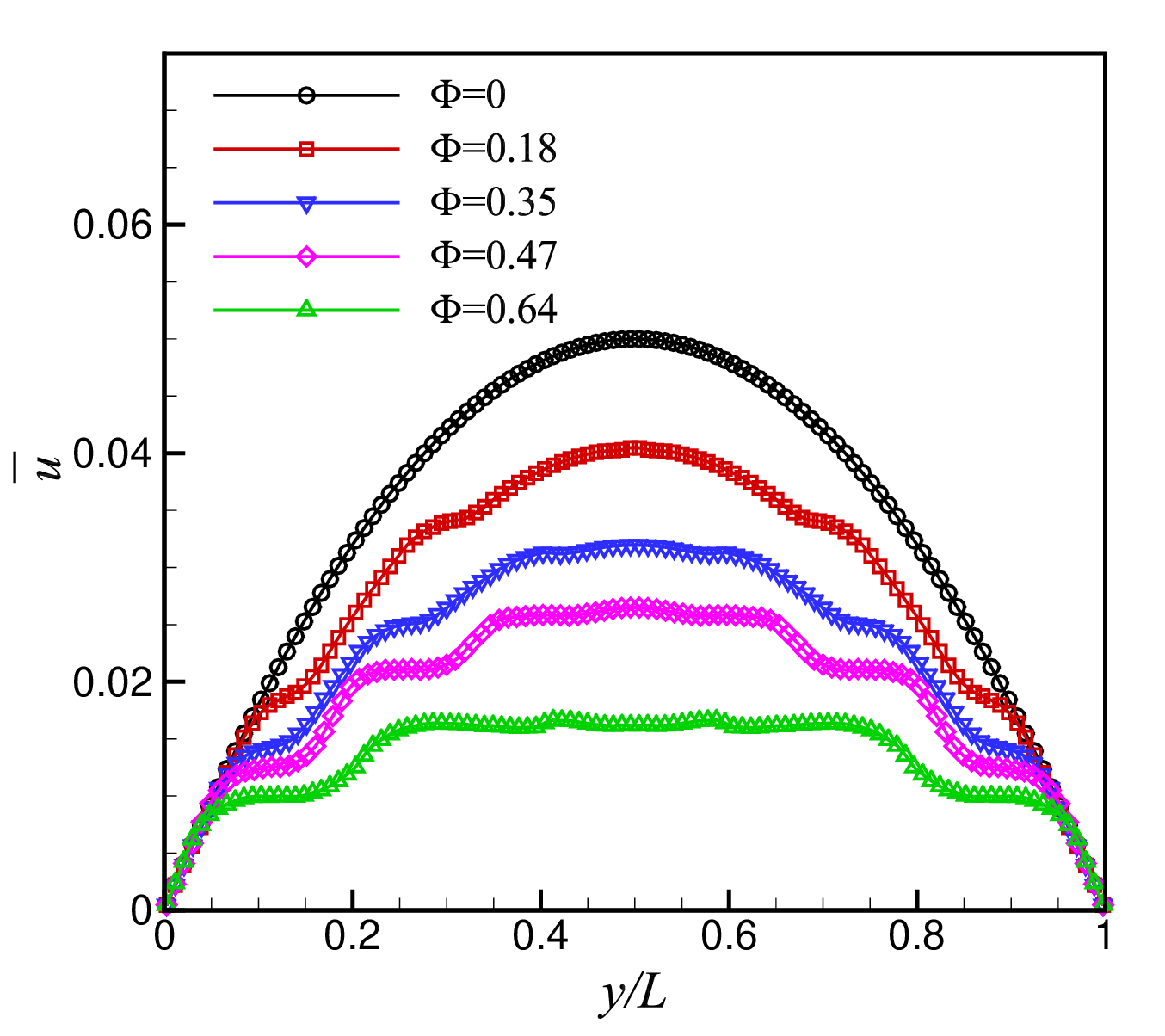,width=0.45\textwidth,clip=}}\hspace{0.0cm}  
 		{\epsfig{file=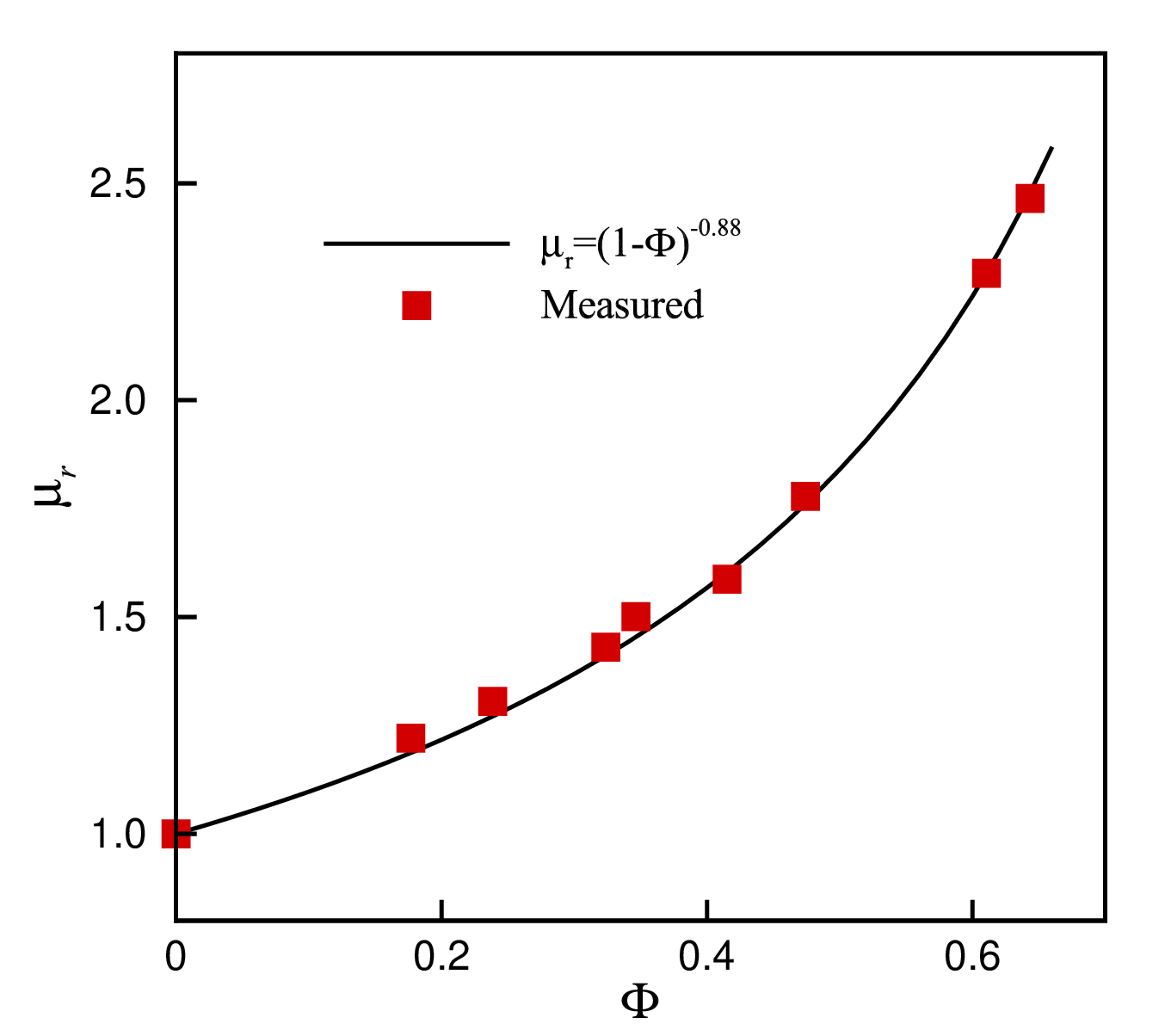,width=0.45\textwidth,clip=}}\hspace{0.0cm}
 	}
 	\caption{\label{FIG4b} 
 		Rheological feature of pressure-driven emulsion in a planar channel at
 		${\bar F_b} = 37$. Left panel: $ x-$direction average velocity profile as a function of $ y $ at different droplets volume fractions $\Phi$. Right panel: measured $\mu_r$ (symbols) as a function of droplets volume fraction $\Phi$, where 
 		the solid line is a fit by a model based on the differential effective medium theory \cite{bullard2009comparison}.
 	}
 \end{figure*}

When no droplets are present in the system ($\Phi  = 0$), 
the usual Poiseuille flow profile for a pure fluid is obtained, as shown in the left panel of Fig.~\ref{FIG4b}. 
In Fig.~\ref{FIG4b}, left panel, we show the velocity profile for different volume fractions at ${\bar F_b} = 37$ and $M = 1$,
where $ \bar u $ is the average velocity along $ x $ direction. As expected, when $\Phi$ is increased, the velocity profile flattens gradually in the central region of the channel, which is consistent with previous results \cite{sbragaglia2012emergence,foglino2017flow}.
To quantify the effect of increasing the volume fraction, we estimate the relative effective viscosity $\mu_r \equiv \mu_{\tiny{\mbox{eff}}}/\mu_B$
as the ratio of flow rates $\mu_r(\Phi) = Q(\Phi=0)/Q(\Phi)$, where $Q=\int_0^L {\bar u(y)dy}$. A plot of  $\mu_r$ as a function of $\Phi$ at
 ${\bar F_b} = 37$
is shown in the right panel of Fig.~\ref{FIG4b}. It can be seen that the relative effective viscosity increases nonlinearly with $ \Phi $,
as observed also for a similar system with neutral wetting boundaries for the droplets \cite{foglino2017flow}.
A number of expressions for the effective viscosity of an emulsion as a function of the volume fraction are available in the literature
\cite{derkach2009rheology,bullard2009comparison}.
Based on the differential effective medium theory, Bullard \textit{et al.} \cite{bullard2009comparison} proposed the following model,
\begin{equation}\label{demt}
{\mu _r} = {(1 - K\Phi )^{ - [\eta ]/K}},
\end{equation}
where the factor $ K $ is set to be $1.0$, and $ {[\eta ]} $ represent the \textit{intrinsic viscosity}. For the present system with slightly deformable droplets (see Fig. \ref{FIG4a}), the \textit{intrinsic viscosity} $ {[\eta ]} $ is restricted between the undeformable limit of Taylor \cite{taylor1932viscosity} ${[\eta ]_T}{\rm{ = }}({\rm{1/}}M{\rm{ + }}{[\eta ]_\infty }){\rm{/}}({\rm{1/}}M{\rm{ + 1}})$
and the freely deformable limit of Douglas \textit{et al.} \cite{douglas1995intrinsic} ${[\eta ]_D}{\rm{ = (1}} - 1/M{\rm{)/[}}(1 - {\rm{1/}}M)/{[\eta ]_\infty }{\rm{ + 1/}}M]$, where $ {[\eta ]_\infty } = 2.5 $. In Fig. \ref{FIG4b}, we show a very good fit by Eq. (\ref{demt}) with $[\eta ] = 0.88.$

\begin{figure}
 	\includegraphics[width=0.45\textwidth]{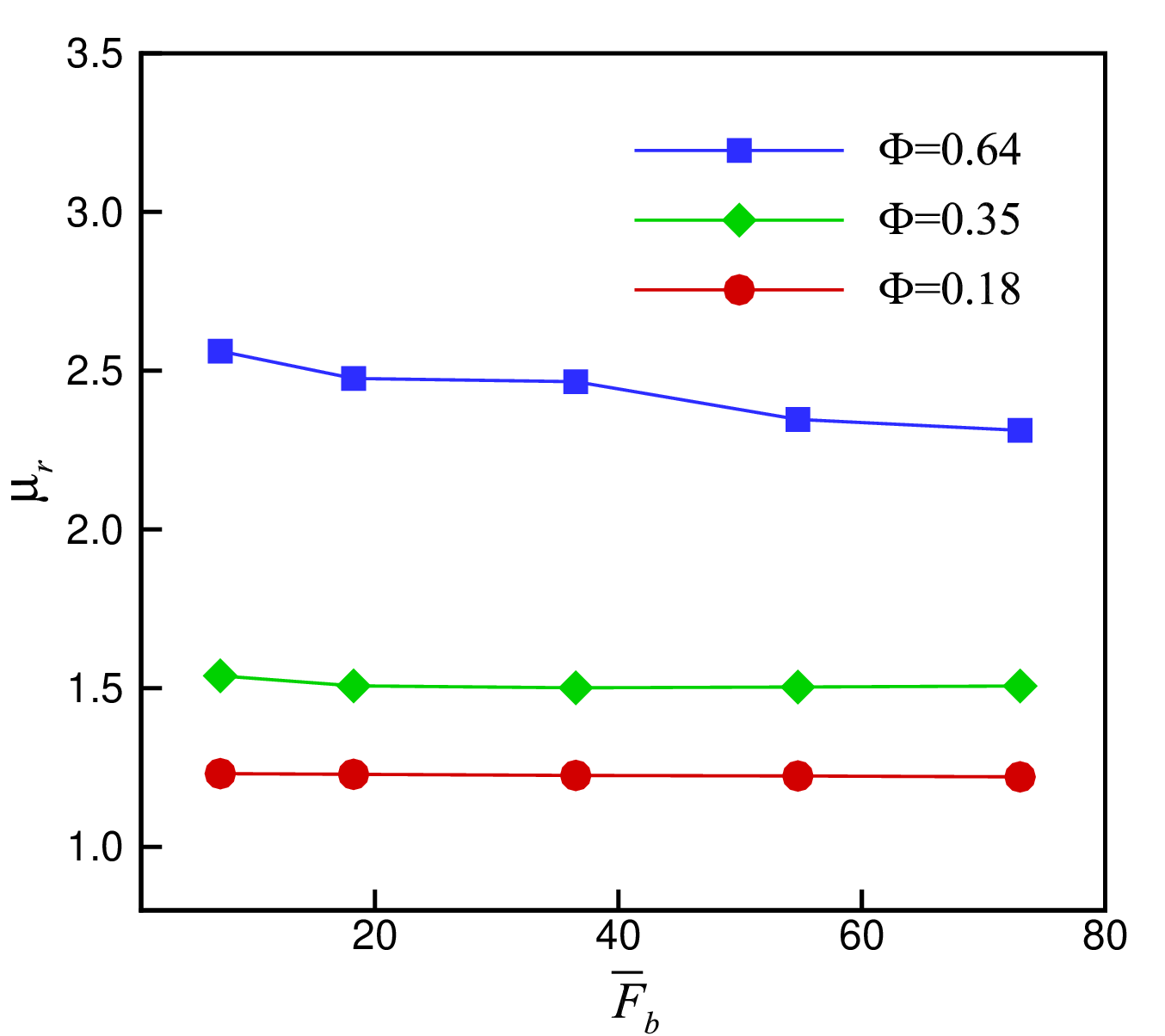}
 	\caption{\label{FIG4c} Relative effective viscosity $\mu_r$ as a function of $\overline{F}_b$ at different volume fractions  
 $\Phi$. }
 \end{figure}

 We now proceed to investigate the effect of body force by exploring a range of non-dimensional body forces $7 \le \overline{F}_b  \le 73 $.
 Figure.~\ref{FIG4c} plots {$\mu_r$} as a function of 
$\overline{F}_b$ at different volume fractions. 
From Fig.~\ref{FIG4c}, we can see that {$\mu_r$} is almost independent of 
 the forcing in the parameter range we considered, as expected for a Newtonian fluid. 
For the highest volume fraction tested, $\Phi  = 0.64$, instead, a slight decrease of $\mu_r$ with the increasing forcing
can be appreciated, hinting at a moderate shear thinning behaviour, as expectable for this kind of systems.  

\begin{figure}
	\includegraphics[width=0.48\textwidth]{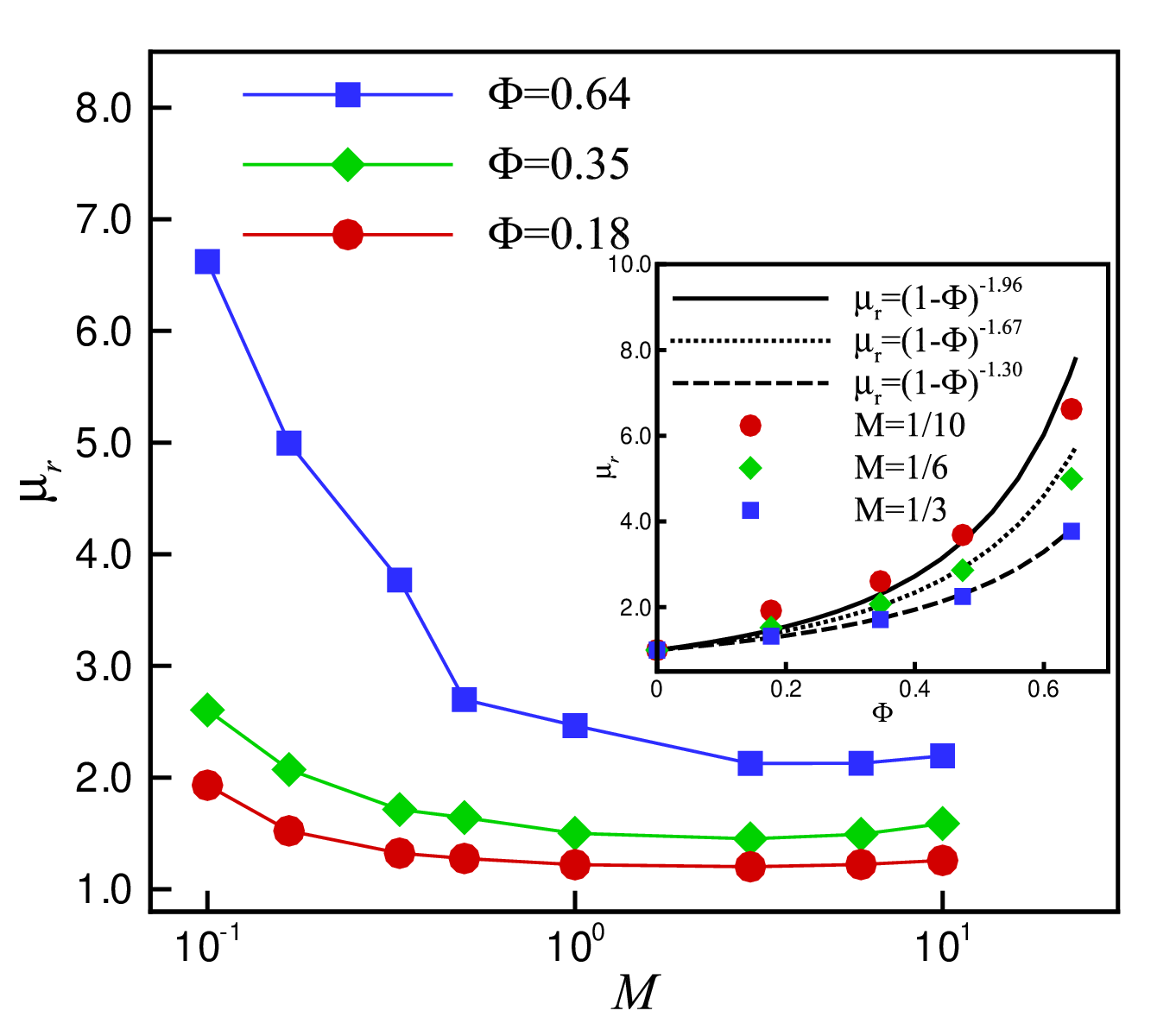}
	\caption{\label{FIG4d} Relative effective viscosity $\mu_r$ as a function of viscosity ratio $ M={\mu _B}/{\mu _A} $ at different volume fractions  $\Phi$ and ${\bar F_b} = 37$. The left branch of the figure, $M<1$, corresponds to more viscous, hence less deformable, droplets 
		($M \rightarrow 0$ is the solid sphere limit), while in the right branch $M>1$ they are less viscous than the matrix, 
		and correspondingly more deformable ($M \gg 1$ is the bubble case). This figure can be also seen, then, as the effective 
		viscosity of a suspension of soft deformable particles, as a function of their stiffness.
		Inset: Plot of $\mu_r$ versus $\Phi $ for $M = 1/10$, 1/6 and 1/3.
	}
\end{figure}

We address now the effect of the viscosity ratio $M = {\mu _B}/{\mu _A} $ on the rheology of the emulsion.
We stress, here, incidentally that the capability of simulating viscosity ratios other than one extends the applicability of 
our method to study the physics of more general soft flowing systems. 
In this sense, the limit $M \rightarrow 0$ is that of suspensions, while $M \gg 1$ would correspond to foams.
In Fig.~\ref{FIG4d} we plot the relative effective viscosity $\mu_r$ as a function of the viscosity ratio $M$.
We can see an asymmetric effect of $ M $: for $ M > 1 $, $\mu_r$ is approximately independent of $ M $, 
while for $M < 1$, $\mu_r$ increases with the decrease of $M$ and this effect is more significant at larger volume fractions. 
As shown in the inset, we also find good fits by Eq. (\ref{demt}) for $M = 1/10, 1/6, 1/3$, where
the \textit{intrinsic viscosity} $ {[\eta ]} $ in each fitting line is restricted between ${[\eta ]_T}$ and ${[\eta ]_D}$ 
for the specified condition.

\section{Conclusions}\label{sec.4}

A mesoscopic numerical method for the simulation of soft flowing systems based on a two-range pseudopotential lattice Boltzmann
has been proposed. 
Our method is unique as it features adjustable surface tension, 
positive disjoining pressure, tunable viscosity ratio and fully resolved hydrodynamics, 
unlike any other existing alternative (such as boundary integral methods \cite{loewenberg1996numerical} or ``bubble models'' \cite{durian1995foam}). In contrast to the previous literature, the present model removes the viscosity-dependence in the original two-range pseudo-potential LBM, 
thus opening the way to  the simulation of multicomponent fluids with non-unit viscosity ratios and a viscosity-independent disjoining pressure.
Such capability is demonstrated by computing the  relative effective viscosity of a pressure-driven emulsion in a planar flow as a function of the
viscosity ratio between the disperse and continuum phases. 

It is hoped that the present upgrade will permit to extend the range of applications of the LBM for complex soft-flowing systems.

\begin{acknowledgments}
	
The authors thank Prof. Mauro Sbragaglia and Prof. Anthony Ladd for fruitful discussions. The research leading to these results has received
funding from the MOST National Key Research and Development Programme (Project No. 2016YFB0600805),
the European Research Council under
the European Union’s Horizon 2020 Framework Programme(No. FP/2014-2020)/ERC Grant Agreement
No. 739964 (“COPMAT”), and the China Scholarship Council (CSC, No. 201706210262). Supercomputing time on ARCHER is provided by the “UK Consortium on Mesoscale Engineering Sciences (UKCOMES)” under the UK Engineering and Physical Sciences Research Council Grant No. EP/R029598/1.
\end{acknowledgments}
\appendix
\section{Spurious currents}
\begin{figure*}[!ht]	        
		\center {			
			{\epsfig{file=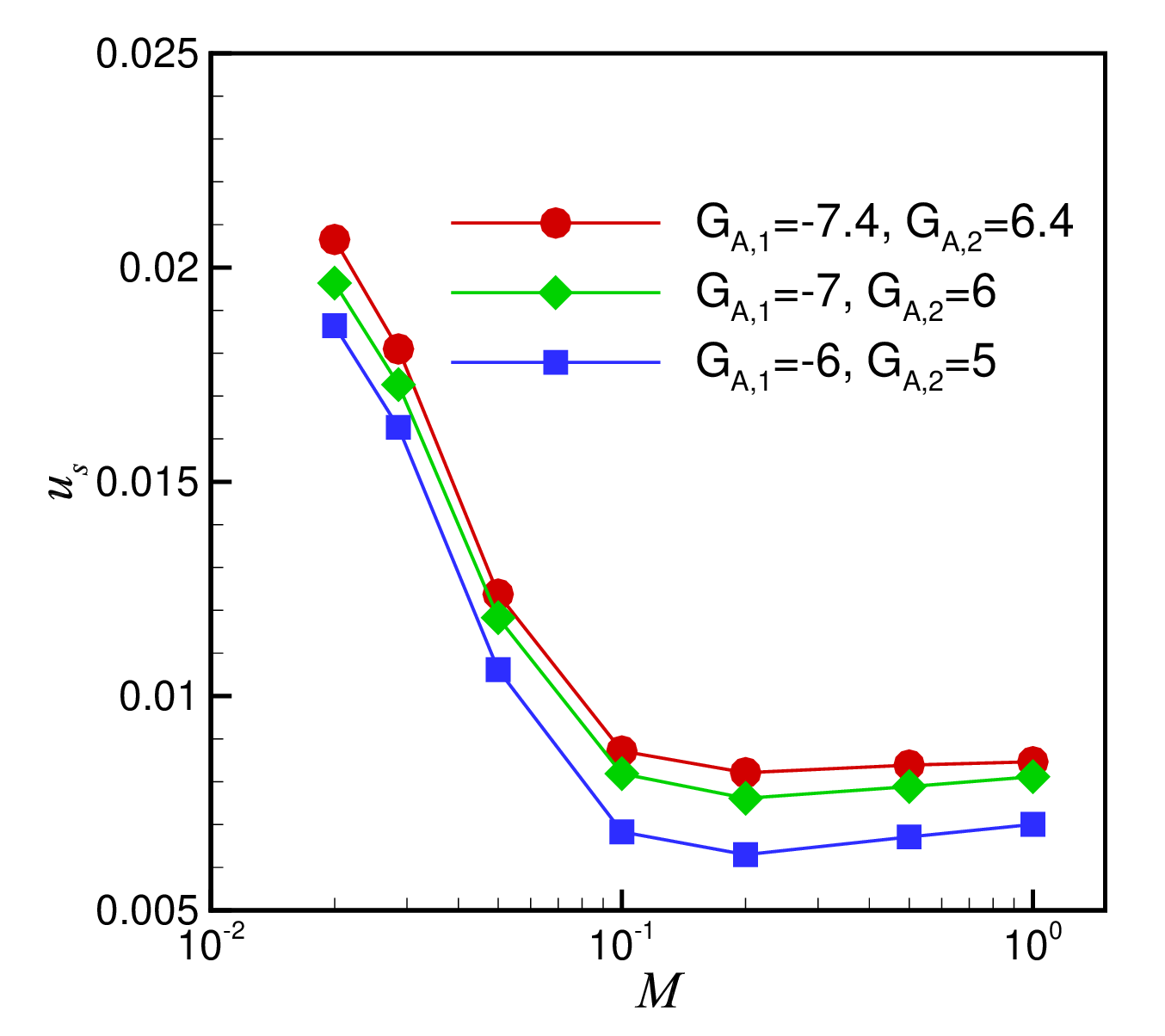,width=0.45\textwidth,clip=}}\hspace{0.0cm}  
		}
		\caption{\label{FIG5a} Maximum spurious currents $ {u_s} $ measured from a steady droplet in a continuous matrix at different viscosity ratios $M$ and strength coefficients $ {G_{A,1}} $. 
		}
	\end{figure*}

It is known that  the diffusive interface models lead to spurious velocity currents around interfaces due to the numerical imbalances of various discretized forces. The maximum spurious currents, $ {u_s} $, produced for a steady droplet  ($R = 20\Delta x $ ) under different conditions are shown in Fig.~\ref{FIG5a}. The parameters used in the simulations are the same as those in Sec. \ref{sec.3d}. As can be seen, the maximum spurious current is relatively small for viscosity ratio $M > 1/10$ (the range considered in Sec. \ref{sec.3d}), while
it increases gradually when $M$ is further decreased. For a given viscosity ratio,  $ {u_s} $ is reduced when the strength coefficient
	$ {G_{A,1}} $ is increased, which corresponds to smaller disjoining pressure (see Fig.~\ref{FIG3a}). For the time being, we mainly focus on the removal of the viscosity-dependence in the original two-range model, while strategies or parameter optimization methods \cite{li2016lattice,lycett2014multiphase,lycett2014binary,montessori2018regularized} for reducing the spurious currents can be adopted for future applications with quite small (or quite large) viscosity ratios.	

\bibliography{Improved_two_range}

\begin{thebibliography}{64}%
\makeatletter
\providecommand \@ifxundefined [1]{%
 \@ifx{#1\undefined}
}%
\providecommand \@ifnum [1]{%
 \ifnum #1\expandafter \@firstoftwo
 \else \expandafter \@secondoftwo
 \fi
}%
\providecommand \@ifx [1]{%
 \ifx #1\expandafter \@firstoftwo
 \else \expandafter \@secondoftwo
 \fi
}%
\providecommand \natexlab [1]{#1}%
\providecommand \enquote  [1]{``#1''}%
\providecommand \bibnamefont  [1]{#1}%
\providecommand \bibfnamefont [1]{#1}%
\providecommand \citenamefont [1]{#1}%
\providecommand \href@noop [0]{\@secondoftwo}%
\providecommand \href [0]{\begingroup \@sanitize@url \@href}%
\providecommand \@href[1]{\@@startlink{#1}\@@href}%
\providecommand \@@href[1]{\endgroup#1\@@endlink}%
\providecommand \@sanitize@url [0]{\catcode `\\12\catcode `\$12\catcode
  `\&12\catcode `\#12\catcode `\^12\catcode `\_12\catcode `\%12\relax}%
\providecommand \@@startlink[1]{}%
\providecommand \@@endlink[0]{}%
\providecommand \url  [0]{\begingroup\@sanitize@url \@url }%
\providecommand \@url [1]{\endgroup\@href {#1}{\urlprefix }}%
\providecommand \urlprefix  [0]{URL }%
\providecommand \Eprint [0]{\href }%
\providecommand \doibase [0]{http://dx.doi.org/}%
\providecommand \selectlanguage [0]{\@gobble}%
\providecommand \bibinfo  [0]{\@secondoftwo}%
\providecommand \bibfield  [0]{\@secondoftwo}%
\providecommand \translation [1]{[#1]}%
\providecommand \BibitemOpen [0]{}%
\providecommand \bibitemStop [0]{}%
\providecommand \bibitemNoStop [0]{.\EOS\space}%
\providecommand \EOS [0]{\spacefactor3000\relax}%
\providecommand \BibitemShut  [1]{\csname bibitem#1\endcsname}%
\let\auto@bib@innerbib\@empty
\bibitem [{\citenamefont {Larson}(1999)}]{larson1999structure}%
  \BibitemOpen
  \bibfield  {author} {\bibinfo {author} {\bibfnamefont {R.G.}\ \bibnamefont
  {Larson}},\ }\href {https://books.google.it/books?id=Vt9fw\_pf1LUC} {\emph
  {\bibinfo {title} {The Structure and Rheology of Complex Fluids}}},\ Topics
  in Chemical Engineering\ (\bibinfo  {publisher} {OUP USA},\ \bibinfo {year}
  {1999})\BibitemShut {NoStop}%
\bibitem [{\citenamefont {Lyklema}(2005)}]{lyklema2005fundamentals}%
  \BibitemOpen
  \bibfield  {author} {\bibinfo {author} {\bibfnamefont {Johannes}\
  \bibnamefont {Lyklema}},\ }\href@noop {} {\emph {\bibinfo {title}
  {Fundamentals of interface and colloid science: soft colloids}}},\
  Vol.~\bibinfo {volume} {5}\ (\bibinfo  {publisher} {Elsevier},\ \bibinfo
  {year} {2005})\BibitemShut {NoStop}%
\bibitem [{\citenamefont {Coussot}(2005)}]{coussot2005rheometry}%
  \BibitemOpen
  \bibfield  {author} {\bibinfo {author} {\bibfnamefont {Philippe}\
  \bibnamefont {Coussot}},\ }\href@noop {} {\emph {\bibinfo {title} {Rheometry
  of pastes, suspensions, and granular materials: applications in industry and
  environment}}}\ (\bibinfo  {publisher} {John Wiley \& Sons},\ \bibinfo {year}
  {2005})\BibitemShut {NoStop}%
\bibitem [{\citenamefont {Weaire}\ and\ \citenamefont
  {Hutzler}(2001)}]{weaire2001physics}%
  \BibitemOpen
  \bibfield  {author} {\bibinfo {author} {\bibfnamefont {Denis~L}\ \bibnamefont
  {Weaire}}\ and\ \bibinfo {author} {\bibfnamefont {Stefan}\ \bibnamefont
  {Hutzler}},\ }\href@noop {} {\emph {\bibinfo {title} {The physics of
  foams}}}\ (\bibinfo  {publisher} {Oxford University Press},\ \bibinfo {year}
  {2001})\BibitemShut {NoStop}%
\bibitem [{\citenamefont {Anna}\ and\ \citenamefont
  {Mayer}(2006)}]{anna2006microscale}%
  \BibitemOpen
  \bibfield  {author} {\bibinfo {author} {\bibfnamefont {Shelley~L}\
  \bibnamefont {Anna}}\ and\ \bibinfo {author} {\bibfnamefont {Hans~C}\
  \bibnamefont {Mayer}},\ }\bibfield  {title} {\enquote {\bibinfo {title}
  {Microscale tipstreaming in a microfluidic flow focusing device},}\
  }\href@noop {} {\bibfield  {journal} {\bibinfo  {journal} {Physics of
  Fluids}\ }\textbf {\bibinfo {volume} {18}},\ \bibinfo {pages} {121512}
  (\bibinfo {year} {2006})}\BibitemShut {NoStop}%
\bibitem [{\citenamefont {Liu}\ and\ \citenamefont
  {Zhang}(2011)}]{liu2011droplet}%
  \BibitemOpen
  \bibfield  {author} {\bibinfo {author} {\bibfnamefont {Haihu}\ \bibnamefont
  {Liu}}\ and\ \bibinfo {author} {\bibfnamefont {Yonghao}\ \bibnamefont
  {Zhang}},\ }\bibfield  {title} {\enquote {\bibinfo {title} {Droplet formation
  in microfluidic cross-junctions},}\ }\href@noop {} {\bibfield  {journal}
  {\bibinfo  {journal} {Physics of Fluids}\ }\textbf {\bibinfo {volume} {23}},\
  \bibinfo {pages} {082101} (\bibinfo {year} {2011})}\BibitemShut {NoStop}%
\bibitem [{\citenamefont {Fu}\ \emph {et~al.}(2017)\citenamefont {Fu},
  \citenamefont {Bai}, \citenamefont {Jin},\ and\ \citenamefont
  {Cheng}}]{fu2017theoretical}%
  \BibitemOpen
  \bibfield  {author} {\bibinfo {author} {\bibfnamefont {Yuhang}\ \bibnamefont
  {Fu}}, \bibinfo {author} {\bibfnamefont {Lin}\ \bibnamefont {Bai}}, \bibinfo
  {author} {\bibfnamefont {Yong}\ \bibnamefont {Jin}}, \ and\ \bibinfo {author}
  {\bibfnamefont {Yi}~\bibnamefont {Cheng}},\ }\bibfield  {title} {\enquote
  {\bibinfo {title} {Theoretical analysis and simulation of obstructed breakup
  of micro-droplet in t-junction under an asymmetric pressure difference},}\
  }\href@noop {} {\bibfield  {journal} {\bibinfo  {journal} {Physics of
  Fluids}\ }\textbf {\bibinfo {volume} {29}},\ \bibinfo {pages} {032003}
  (\bibinfo {year} {2017})}\BibitemShut {NoStop}%
\bibitem [{\citenamefont {Bhateja}\ and\ \citenamefont
  {Khakhar}(2018)}]{bhateja2018rheology}%
  \BibitemOpen
  \bibfield  {author} {\bibinfo {author} {\bibfnamefont {Ashish}\ \bibnamefont
  {Bhateja}}\ and\ \bibinfo {author} {\bibfnamefont {Devang~V}\ \bibnamefont
  {Khakhar}},\ }\bibfield  {title} {\enquote {\bibinfo {title} {Rheology of
  dense granular flows in two dimensions: Comparison of fully two-dimensional
  flows to unidirectional shear flow},}\ }\href@noop {} {\bibfield  {journal}
  {\bibinfo  {journal} {Physical Review Fluids}\ }\textbf {\bibinfo {volume}
  {3}},\ \bibinfo {pages} {062301} (\bibinfo {year} {2018})}\BibitemShut
  {NoStop}%
\bibitem [{\citenamefont {Sollich}\ \emph {et~al.}(1997)\citenamefont
  {Sollich}, \citenamefont {Lequeux}, \citenamefont {H{\'e}braud},\ and\
  \citenamefont {Cates}}]{sollich1997rheology}%
  \BibitemOpen
  \bibfield  {author} {\bibinfo {author} {\bibfnamefont {Peter}\ \bibnamefont
  {Sollich}}, \bibinfo {author} {\bibfnamefont {Fran{\c{c}}ois}\ \bibnamefont
  {Lequeux}}, \bibinfo {author} {\bibfnamefont {Pascal}\ \bibnamefont
  {H{\'e}braud}}, \ and\ \bibinfo {author} {\bibfnamefont {Michael~E}\
  \bibnamefont {Cates}},\ }\bibfield  {title} {\enquote {\bibinfo {title}
  {Rheology of soft glassy materials},}\ }\href@noop {} {\bibfield  {journal}
  {\bibinfo  {journal} {Physical review letters}\ }\textbf {\bibinfo {volume}
  {78}},\ \bibinfo {pages} {2020} (\bibinfo {year} {1997})}\BibitemShut
  {NoStop}%
\bibitem [{\citenamefont {Li}\ \emph {et~al.}(2016{\natexlab{a}})\citenamefont
  {Li}, \citenamefont {Luo}, \citenamefont {Kang}, \citenamefont {He},
  \citenamefont {Chen},\ and\ \citenamefont {Liu}}]{li2016lattice}%
  \BibitemOpen
  \bibfield  {author} {\bibinfo {author} {\bibfnamefont {Qing}\ \bibnamefont
  {Li}}, \bibinfo {author} {\bibfnamefont {Kai~Hong}\ \bibnamefont {Luo}},
  \bibinfo {author} {\bibfnamefont {QJ}~\bibnamefont {Kang}}, \bibinfo {author}
  {\bibfnamefont {YL}~\bibnamefont {He}}, \bibinfo {author} {\bibfnamefont
  {Q}~\bibnamefont {Chen}}, \ and\ \bibinfo {author} {\bibfnamefont
  {Q}~\bibnamefont {Liu}},\ }\bibfield  {title} {\enquote {\bibinfo {title}
  {Lattice boltzmann methods for multiphase flow and phase-change heat
  transfer},}\ }\href@noop {} {\bibfield  {journal} {\bibinfo  {journal}
  {Progress in Energy and Combustion Science}\ }\textbf {\bibinfo {volume}
  {52}},\ \bibinfo {pages} {62--105} (\bibinfo {year}
  {2016}{\natexlab{a}})}\BibitemShut {NoStop}%
\bibitem [{\citenamefont {McNamara}\ and\ \citenamefont
  {Zanetti}(1988)}]{mcnamara1988use}%
  \BibitemOpen
  \bibfield  {author} {\bibinfo {author} {\bibfnamefont {Guy~R}\ \bibnamefont
  {McNamara}}\ and\ \bibinfo {author} {\bibfnamefont {Gianluigi}\ \bibnamefont
  {Zanetti}},\ }\bibfield  {title} {\enquote {\bibinfo {title} {Use of the
  boltzmann equation to simulate lattice-gas automata},}\ }\href@noop {}
  {\bibfield  {journal} {\bibinfo  {journal} {Physical review letters}\
  }\textbf {\bibinfo {volume} {61}},\ \bibinfo {pages} {2332} (\bibinfo {year}
  {1988})}\BibitemShut {NoStop}%
\bibitem [{\citenamefont {Besold}\ \emph {et~al.}(2000)\citenamefont {Besold},
  \citenamefont {Vattulainen}, \citenamefont {Karttunen},\ and\ \citenamefont
  {Polson}}]{besold2000towards}%
  \BibitemOpen
  \bibfield  {author} {\bibinfo {author} {\bibfnamefont {Gerhard}\ \bibnamefont
  {Besold}}, \bibinfo {author} {\bibfnamefont {Ilpo}\ \bibnamefont
  {Vattulainen}}, \bibinfo {author} {\bibfnamefont {Mikko}\ \bibnamefont
  {Karttunen}}, \ and\ \bibinfo {author} {\bibfnamefont {James~M}\ \bibnamefont
  {Polson}},\ }\bibfield  {title} {\enquote {\bibinfo {title} {Towards better
  integrators for dissipative particle dynamics simulations},}\ }\href@noop {}
  {\bibfield  {journal} {\bibinfo  {journal} {Physical Review E}\ }\textbf
  {\bibinfo {volume} {62}},\ \bibinfo {pages} {R7611} (\bibinfo {year}
  {2000})}\BibitemShut {NoStop}%
\bibitem [{\citenamefont {Qian}\ \emph {et~al.}(1992)\citenamefont {Qian},
  \citenamefont {d'Humi{\`e}res},\ and\ \citenamefont
  {Lallemand}}]{qian1992lattice}%
  \BibitemOpen
  \bibfield  {author} {\bibinfo {author} {\bibfnamefont {YH}~\bibnamefont
  {Qian}}, \bibinfo {author} {\bibfnamefont {Dominique}\ \bibnamefont
  {d'Humi{\`e}res}}, \ and\ \bibinfo {author} {\bibfnamefont {Pierre}\
  \bibnamefont {Lallemand}},\ }\bibfield  {title} {\enquote {\bibinfo {title}
  {Lattice bgk models for navier-stokes equation},}\ }\href@noop {} {\bibfield
  {journal} {\bibinfo  {journal} {EPL (Europhysics Letters)}\ }\textbf
  {\bibinfo {volume} {17}},\ \bibinfo {pages} {479} (\bibinfo {year}
  {1992})}\BibitemShut {NoStop}%
\bibitem [{\citenamefont {Benzi}\ \emph {et~al.}(1992)\citenamefont {Benzi},
  \citenamefont {Succi},\ and\ \citenamefont {Vergassola}}]{benzi1992lattice}%
  \BibitemOpen
  \bibfield  {author} {\bibinfo {author} {\bibfnamefont {Roberto}\ \bibnamefont
  {Benzi}}, \bibinfo {author} {\bibfnamefont {Sauro}\ \bibnamefont {Succi}}, \
  and\ \bibinfo {author} {\bibfnamefont {Massimo}\ \bibnamefont {Vergassola}},\
  }\bibfield  {title} {\enquote {\bibinfo {title} {The lattice boltzmann
  equation: theory and applications},}\ }\href@noop {} {\bibfield  {journal}
  {\bibinfo  {journal} {Physics Reports}\ }\textbf {\bibinfo {volume} {222}},\
  \bibinfo {pages} {145--197} (\bibinfo {year} {1992})}\BibitemShut {NoStop}%
\bibitem [{\citenamefont {Shan}\ and\ \citenamefont
  {Chen}(1993)}]{shan1993lattice}%
  \BibitemOpen
  \bibfield  {author} {\bibinfo {author} {\bibfnamefont {Xiaowen}\ \bibnamefont
  {Shan}}\ and\ \bibinfo {author} {\bibfnamefont {Hudong}\ \bibnamefont
  {Chen}},\ }\bibfield  {title} {\enquote {\bibinfo {title} {Lattice boltzmann
  model for simulating flows with multiple phases and components},}\
  }\href@noop {} {\bibfield  {journal} {\bibinfo  {journal} {Physical Review
  E}\ }\textbf {\bibinfo {volume} {47}},\ \bibinfo {pages} {1815} (\bibinfo
  {year} {1993})}\BibitemShut {NoStop}%
\bibitem [{\citenamefont {Bernaschi}\ \emph {et~al.}(2010)\citenamefont
  {Bernaschi}, \citenamefont {Fatica}, \citenamefont {Melchionna},
  \citenamefont {Succi},\ and\ \citenamefont
  {Kaxiras}}]{bernaschi2010flexible}%
  \BibitemOpen
  \bibfield  {author} {\bibinfo {author} {\bibfnamefont {Massimo}\ \bibnamefont
  {Bernaschi}}, \bibinfo {author} {\bibfnamefont {Massimiliano}\ \bibnamefont
  {Fatica}}, \bibinfo {author} {\bibfnamefont {Simone}\ \bibnamefont
  {Melchionna}}, \bibinfo {author} {\bibfnamefont {Sauro}\ \bibnamefont
  {Succi}}, \ and\ \bibinfo {author} {\bibfnamefont {Efthimios}\ \bibnamefont
  {Kaxiras}},\ }\bibfield  {title} {\enquote {\bibinfo {title} {A flexible
  high-performance lattice boltzmann gpu code for the simulations of fluid
  flows in complex geometries},}\ }\href@noop {} {\bibfield  {journal}
  {\bibinfo  {journal} {Concurrency and Computation: Practice and Experience}\
  }\textbf {\bibinfo {volume} {22}},\ \bibinfo {pages} {1--14} (\bibinfo {year}
  {2010})}\BibitemShut {NoStop}%
\bibitem [{\citenamefont {Guo}\ and\ \citenamefont
  {Shu}(2013)}]{guo2013lattice}%
  \BibitemOpen
  \bibfield  {author} {\bibinfo {author} {\bibfnamefont {Zhaoli}\ \bibnamefont
  {Guo}}\ and\ \bibinfo {author} {\bibfnamefont {Chang}\ \bibnamefont {Shu}},\
  }\href@noop {} {\emph {\bibinfo {title} {Lattice Boltzmann method and its
  applications in engineering}}},\ Vol.~\bibinfo {volume} {3}\ (\bibinfo
  {publisher} {World Scientific},\ \bibinfo {year} {2013})\BibitemShut
  {NoStop}%
\bibitem [{\citenamefont {Succi}(2015)}]{succi2015lattice}%
  \BibitemOpen
  \bibfield  {author} {\bibinfo {author} {\bibfnamefont {Sauro}\ \bibnamefont
  {Succi}},\ }\bibfield  {title} {\enquote {\bibinfo {title} {Lattice boltzmann
  2038},}\ }\href@noop {} {\bibfield  {journal} {\bibinfo  {journal} {EPL
  (Europhysics Letters)}\ }\textbf {\bibinfo {volume} {109}},\ \bibinfo {pages}
  {50001} (\bibinfo {year} {2015})}\BibitemShut {NoStop}%
\bibitem [{\citenamefont {Succi}(2018)}]{succi2018lattice}%
  \BibitemOpen
  \bibfield  {author} {\bibinfo {author} {\bibfnamefont {Sauro}\ \bibnamefont
  {Succi}},\ }\href@noop {} {\emph {\bibinfo {title} {The Lattice Boltzmann
  Equation: For Complex States of Flowing Matter}}}\ (\bibinfo  {publisher}
  {Oxford University Press},\ \bibinfo {year} {2018})\BibitemShut {NoStop}%
\bibitem [{\citenamefont {Fei}\ \emph {et~al.}(2018{\natexlab{a}})\citenamefont
  {Fei}, \citenamefont {Luo}, \citenamefont {Lin},\ and\ \citenamefont
  {Li}}]{fei2018modeling}%
  \BibitemOpen
  \bibfield  {author} {\bibinfo {author} {\bibfnamefont {Linlin}\ \bibnamefont
  {Fei}}, \bibinfo {author} {\bibfnamefont {Kai~Hong}\ \bibnamefont {Luo}},
  \bibinfo {author} {\bibfnamefont {Chuandong}\ \bibnamefont {Lin}}, \ and\
  \bibinfo {author} {\bibfnamefont {Qing}\ \bibnamefont {Li}},\ }\bibfield
  {title} {\enquote {\bibinfo {title} {Modeling incompressible thermal flows
  using a central-moments-based lattice boltzmann method},}\ }\href@noop {}
  {\bibfield  {journal} {\bibinfo  {journal} {International Journal of Heat and
  Mass Transfer}\ }\textbf {\bibinfo {volume} {120}},\ \bibinfo {pages}
  {624--634} (\bibinfo {year} {2018}{\natexlab{a}})}\BibitemShut {NoStop}%
\bibitem [{\citenamefont {Lauricella}\ \emph {et~al.}(2018)\citenamefont
  {Lauricella}, \citenamefont {Melchionna}, \citenamefont {Montessori},
  \citenamefont {Pisignano}, \citenamefont {Pontrelli},\ and\ \citenamefont
  {Succi}}]{lauricella2018entropic}%
  \BibitemOpen
  \bibfield  {author} {\bibinfo {author} {\bibfnamefont {Marco}\ \bibnamefont
  {Lauricella}}, \bibinfo {author} {\bibfnamefont {Simone}\ \bibnamefont
  {Melchionna}}, \bibinfo {author} {\bibfnamefont {Andrea}\ \bibnamefont
  {Montessori}}, \bibinfo {author} {\bibfnamefont {Dario}\ \bibnamefont
  {Pisignano}}, \bibinfo {author} {\bibfnamefont {Giuseppe}\ \bibnamefont
  {Pontrelli}}, \ and\ \bibinfo {author} {\bibfnamefont {Sauro}\ \bibnamefont
  {Succi}},\ }\bibfield  {title} {\enquote {\bibinfo {title} {Entropic lattice
  boltzmann model for charged leaky dielectric multiphase fluids in electrified
  jets},}\ }\href@noop {} {\bibfield  {journal} {\bibinfo  {journal} {Physical
  Review E}\ }\textbf {\bibinfo {volume} {97}},\ \bibinfo {pages} {033308}
  (\bibinfo {year} {2018})}\BibitemShut {NoStop}%
\bibitem [{\citenamefont {Falcucci}\ \emph {et~al.}(2007)\citenamefont
  {Falcucci}, \citenamefont {Bella}, \citenamefont {Chiatti}, \citenamefont
  {Chibbaro}, \citenamefont {Sbragaglia},\ and\ \citenamefont
  {Succi}}]{falcucci2007lattice}%
  \BibitemOpen
  \bibfield  {author} {\bibinfo {author} {\bibfnamefont {Giacomo}\ \bibnamefont
  {Falcucci}}, \bibinfo {author} {\bibfnamefont {Gino}\ \bibnamefont {Bella}},
  \bibinfo {author} {\bibfnamefont {Giancarlo}\ \bibnamefont {Chiatti}},
  \bibinfo {author} {\bibfnamefont {Sergio}\ \bibnamefont {Chibbaro}}, \bibinfo
  {author} {\bibfnamefont {Mauro}\ \bibnamefont {Sbragaglia}}, \ and\ \bibinfo
  {author} {\bibfnamefont {Sauro}\ \bibnamefont {Succi}},\ }\bibfield  {title}
  {\enquote {\bibinfo {title} {Lattice boltzmann models with mid-range
  interactions},}\ }\href@noop {} {\bibfield  {journal} {\bibinfo  {journal}
  {Commun. Comput. Phys}\ }\textbf {\bibinfo {volume} {2}},\ \bibinfo {pages}
  {1071--1084} (\bibinfo {year} {2007})}\BibitemShut {NoStop}%
\bibitem [{\citenamefont {Benzi}\ \emph
  {et~al.}(2009{\natexlab{a}})\citenamefont {Benzi}, \citenamefont {Chibbaro},\
  and\ \citenamefont {Succi}}]{benzi2009mesoscopic}%
  \BibitemOpen
  \bibfield  {author} {\bibinfo {author} {\bibfnamefont {Roberto}\ \bibnamefont
  {Benzi}}, \bibinfo {author} {\bibfnamefont {Sergio}\ \bibnamefont
  {Chibbaro}}, \ and\ \bibinfo {author} {\bibfnamefont {Sauro}\ \bibnamefont
  {Succi}},\ }\bibfield  {title} {\enquote {\bibinfo {title} {Mesoscopic
  lattice boltzmann modeling of flowing soft systems},}\ }\href@noop {}
  {\bibfield  {journal} {\bibinfo  {journal} {Physical review letters}\
  }\textbf {\bibinfo {volume} {102}},\ \bibinfo {pages} {026002} (\bibinfo
  {year} {2009}{\natexlab{a}})}\BibitemShut {NoStop}%
\bibitem [{\citenamefont {Benzi}\ \emph
  {et~al.}(2009{\natexlab{b}})\citenamefont {Benzi}, \citenamefont
  {Sbragaglia}, \citenamefont {Succi}, \citenamefont {Bernaschi},\ and\
  \citenamefont {Chibbaro}}]{benzi2009mesoscopic2}%
  \BibitemOpen
  \bibfield  {author} {\bibinfo {author} {\bibfnamefont {Roberto}\ \bibnamefont
  {Benzi}}, \bibinfo {author} {\bibfnamefont {Mauro}\ \bibnamefont
  {Sbragaglia}}, \bibinfo {author} {\bibfnamefont {Sauro}\ \bibnamefont
  {Succi}}, \bibinfo {author} {\bibfnamefont {Massimo}\ \bibnamefont
  {Bernaschi}}, \ and\ \bibinfo {author} {\bibfnamefont {Sergio}\ \bibnamefont
  {Chibbaro}},\ }\bibfield  {title} {\enquote {\bibinfo {title} {Mesoscopic
  lattice boltzmann modeling of soft-glassy systems: theory and simulations},}\
  }\href@noop {} {\bibfield  {journal} {\bibinfo  {journal} {The Journal of
  Chemical Physics}\ }\textbf {\bibinfo {volume} {131}},\ \bibinfo {pages}
  {104903} (\bibinfo {year} {2009}{\natexlab{b}})}\BibitemShut {NoStop}%
\bibitem [{\citenamefont {Sbragaglia}\ \emph {et~al.}(2012)\citenamefont
  {Sbragaglia}, \citenamefont {Benzi}, \citenamefont {Bernaschi},\ and\
  \citenamefont {Succi}}]{sbragaglia2012emergence}%
  \BibitemOpen
  \bibfield  {author} {\bibinfo {author} {\bibfnamefont {M}~\bibnamefont
  {Sbragaglia}}, \bibinfo {author} {\bibfnamefont {R}~\bibnamefont {Benzi}},
  \bibinfo {author} {\bibfnamefont {M}~\bibnamefont {Bernaschi}}, \ and\
  \bibinfo {author} {\bibfnamefont {S}~\bibnamefont {Succi}},\ }\bibfield
  {title} {\enquote {\bibinfo {title} {The emergence of supramolecular forces
  from lattice kinetic models of non-ideal fluids: applications to the rheology
  of soft glassy materials},}\ }\href@noop {} {\bibfield  {journal} {\bibinfo
  {journal} {Soft Matter}\ }\textbf {\bibinfo {volume} {8}},\ \bibinfo {pages}
  {10773--10782} (\bibinfo {year} {2012})}\BibitemShut {NoStop}%
\bibitem [{\citenamefont {Benzi}\ \emph {et~al.}(2015)\citenamefont {Benzi},
  \citenamefont {Sbragaglia}, \citenamefont {Scagliarini}, \citenamefont
  {Perlekar}, \citenamefont {Bernaschi}, \citenamefont {Succi},\ and\
  \citenamefont {Toschi}}]{benzi2015internal}%
  \BibitemOpen
  \bibfield  {author} {\bibinfo {author} {\bibfnamefont {R}~\bibnamefont
  {Benzi}}, \bibinfo {author} {\bibfnamefont {M}~\bibnamefont {Sbragaglia}},
  \bibinfo {author} {\bibfnamefont {A}~\bibnamefont {Scagliarini}}, \bibinfo
  {author} {\bibfnamefont {P}~\bibnamefont {Perlekar}}, \bibinfo {author}
  {\bibfnamefont {M}~\bibnamefont {Bernaschi}}, \bibinfo {author}
  {\bibfnamefont {S}~\bibnamefont {Succi}}, \ and\ \bibinfo {author}
  {\bibfnamefont {F}~\bibnamefont {Toschi}},\ }\bibfield  {title} {\enquote
  {\bibinfo {title} {Internal dynamics and activated processes in soft-glassy
  materials},}\ }\href@noop {} {\bibfield  {journal} {\bibinfo  {journal} {Soft
  Matter}\ }\textbf {\bibinfo {volume} {11}},\ \bibinfo {pages} {1271--1280}
  (\bibinfo {year} {2015})}\BibitemShut {NoStop}%
\bibitem [{\citenamefont {Dollet}\ \emph {et~al.}(2015)\citenamefont {Dollet},
  \citenamefont {Scagliarini},\ and\ \citenamefont
  {Sbragaglia}}]{dollet2015two}%
  \BibitemOpen
  \bibfield  {author} {\bibinfo {author} {\bibfnamefont {Benjamin}\
  \bibnamefont {Dollet}}, \bibinfo {author} {\bibfnamefont {A}~\bibnamefont
  {Scagliarini}}, \ and\ \bibinfo {author} {\bibfnamefont {M}~\bibnamefont
  {Sbragaglia}},\ }\bibfield  {title} {\enquote {\bibinfo {title}
  {Two-dimensional plastic flow of foams and emulsions in a channel:
  experiments and lattice boltzmann simulations},}\ }\href@noop {} {\bibfield
  {journal} {\bibinfo  {journal} {Journal of Fluid Mechanics}\ }\textbf
  {\bibinfo {volume} {766}},\ \bibinfo {pages} {556--589} (\bibinfo {year}
  {2015})}\BibitemShut {NoStop}%
\bibitem [{\citenamefont {Scagliarini}\ \emph {et~al.}(2015)\citenamefont
  {Scagliarini}, \citenamefont {Dollet},\ and\ \citenamefont
  {Sbragaglia}}]{scagliarini2015non}%
  \BibitemOpen
  \bibfield  {author} {\bibinfo {author} {\bibfnamefont {Andrea}\ \bibnamefont
  {Scagliarini}}, \bibinfo {author} {\bibfnamefont {Benjamin}\ \bibnamefont
  {Dollet}}, \ and\ \bibinfo {author} {\bibfnamefont {Mauro}\ \bibnamefont
  {Sbragaglia}},\ }\bibfield  {title} {\enquote {\bibinfo {title} {Non-locality
  and viscous drag effects on the shear localisation in soft-glassy
  materials},}\ }\href@noop {} {\bibfield  {journal} {\bibinfo  {journal}
  {Colloids and Surfaces A: Physicochemical and Engineering Aspects}\ }\textbf
  {\bibinfo {volume} {473}},\ \bibinfo {pages} {133--140} (\bibinfo {year}
  {2015})}\BibitemShut {NoStop}%
\bibitem [{\citenamefont {Guo}\ \emph {et~al.}(2002)\citenamefont {Guo},
  \citenamefont {Zheng},\ and\ \citenamefont {Shi}}]{guo2002discrete}%
  \BibitemOpen
  \bibfield  {author} {\bibinfo {author} {\bibfnamefont {Zhaoli}\ \bibnamefont
  {Guo}}, \bibinfo {author} {\bibfnamefont {Chuguang}\ \bibnamefont {Zheng}}, \
  and\ \bibinfo {author} {\bibfnamefont {Baochang}\ \bibnamefont {Shi}},\
  }\bibfield  {title} {\enquote {\bibinfo {title} {Discrete lattice effects on
  the forcing term in the lattice boltzmann method},}\ }\href@noop {}
  {\bibfield  {journal} {\bibinfo  {journal} {Physical Review E}\ }\textbf
  {\bibinfo {volume} {65}},\ \bibinfo {pages} {046308} (\bibinfo {year}
  {2002})}\BibitemShut {NoStop}%
\bibitem [{\citenamefont {Shan}\ and\ \citenamefont
  {Doolen}(1995)}]{shan1995multicomponent}%
  \BibitemOpen
  \bibfield  {author} {\bibinfo {author} {\bibfnamefont {Xiaowen}\ \bibnamefont
  {Shan}}\ and\ \bibinfo {author} {\bibfnamefont {Gary}\ \bibnamefont
  {Doolen}},\ }\bibfield  {title} {\enquote {\bibinfo {title} {Multicomponent
  lattice-boltzmann model with interparticle interaction},}\ }\href@noop {}
  {\bibfield  {journal} {\bibinfo  {journal} {Journal of Statistical Physics}\
  }\textbf {\bibinfo {volume} {81}},\ \bibinfo {pages} {379--393} (\bibinfo
  {year} {1995})}\BibitemShut {NoStop}%
\bibitem [{\citenamefont {Shan}\ and\ \citenamefont
  {Chen}(1994)}]{shan1994simulation}%
  \BibitemOpen
  \bibfield  {author} {\bibinfo {author} {\bibfnamefont {Xiaowen}\ \bibnamefont
  {Shan}}\ and\ \bibinfo {author} {\bibfnamefont {Hudong}\ \bibnamefont
  {Chen}},\ }\bibfield  {title} {\enquote {\bibinfo {title} {Simulation of
  nonideal gases and liquid-gas phase transitions by the lattice boltzmann
  equation},}\ }\href@noop {} {\bibfield  {journal} {\bibinfo  {journal}
  {Physical Review E}\ }\textbf {\bibinfo {volume} {49}},\ \bibinfo {pages}
  {2941} (\bibinfo {year} {1994})}\BibitemShut {NoStop}%
\bibitem [{\citenamefont {Huang}\ \emph {et~al.}(2011)\citenamefont {Huang},
  \citenamefont {Krafczyk},\ and\ \citenamefont {Lu}}]{huang2011forcing}%
  \BibitemOpen
  \bibfield  {author} {\bibinfo {author} {\bibfnamefont {Haibo}\ \bibnamefont
  {Huang}}, \bibinfo {author} {\bibfnamefont {Manfred}\ \bibnamefont
  {Krafczyk}}, \ and\ \bibinfo {author} {\bibfnamefont {Xiyun}\ \bibnamefont
  {Lu}},\ }\bibfield  {title} {\enquote {\bibinfo {title} {Forcing term in
  single-phase and shan-chen-type multiphase lattice boltzmann models},}\
  }\href@noop {} {\bibfield  {journal} {\bibinfo  {journal} {Physical Review
  E}\ }\textbf {\bibinfo {volume} {84}},\ \bibinfo {pages} {046710} (\bibinfo
  {year} {2011})}\BibitemShut {NoStop}%
\bibitem [{\citenamefont {Li}\ \emph {et~al.}(2012)\citenamefont {Li},
  \citenamefont {Luo},\ and\ \citenamefont {Li}}]{li2012forcing}%
  \BibitemOpen
  \bibfield  {author} {\bibinfo {author} {\bibfnamefont {Q}~\bibnamefont {Li}},
  \bibinfo {author} {\bibfnamefont {Kai~Hong}\ \bibnamefont {Luo}}, \ and\
  \bibinfo {author} {\bibfnamefont {X~J}\ \bibnamefont {Li}},\ }\bibfield
  {title} {\enquote {\bibinfo {title} {Forcing scheme in pseudopotential
  lattice boltzmann model for multiphase flows},}\ }\href@noop {} {\bibfield
  {journal} {\bibinfo  {journal} {Physical Review E}\ }\textbf {\bibinfo
  {volume} {86}},\ \bibinfo {pages} {016709} (\bibinfo {year}
  {2012})}\BibitemShut {NoStop}%
\bibitem [{\citenamefont {Lycett-Brown}\ and\ \citenamefont
  {Luo}(2015)}]{lycett2015improved}%
  \BibitemOpen
  \bibfield  {author} {\bibinfo {author} {\bibfnamefont {Daniel}\ \bibnamefont
  {Lycett-Brown}}\ and\ \bibinfo {author} {\bibfnamefont {Kai~H}\ \bibnamefont
  {Luo}},\ }\bibfield  {title} {\enquote {\bibinfo {title} {Improved forcing
  scheme in pseudopotential lattice boltzmann methods for multiphase flow at
  arbitrarily high density ratios},}\ }\href@noop {} {\bibfield  {journal}
  {\bibinfo  {journal} {Physical Review E}\ }\textbf {\bibinfo {volume} {91}},\
  \bibinfo {pages} {023305} (\bibinfo {year} {2015})}\BibitemShut {NoStop}%
\bibitem [{\citenamefont {He}\ \emph {et~al.}(1998{\natexlab{a}})\citenamefont
  {He}, \citenamefont {Chen},\ and\ \citenamefont {Doolen}}]{he1998novel}%
  \BibitemOpen
  \bibfield  {author} {\bibinfo {author} {\bibfnamefont {Xiaoyi}\ \bibnamefont
  {He}}, \bibinfo {author} {\bibfnamefont {Shiyi}\ \bibnamefont {Chen}}, \ and\
  \bibinfo {author} {\bibfnamefont {Gary~D}\ \bibnamefont {Doolen}},\
  }\bibfield  {title} {\enquote {\bibinfo {title} {A novel thermal model for
  the lattice boltzmann method in incompressible limit},}\ }\href@noop {}
  {\bibfield  {journal} {\bibinfo  {journal} {Journal of Computational
  Physics}\ }\textbf {\bibinfo {volume} {146}},\ \bibinfo {pages} {282--300}
  (\bibinfo {year} {1998}{\natexlab{a}})}\BibitemShut {NoStop}%
\bibitem [{\citenamefont {He}\ \emph {et~al.}(1998{\natexlab{b}})\citenamefont
  {He}, \citenamefont {Shan},\ and\ \citenamefont {Doolen}}]{he1998discrete}%
  \BibitemOpen
  \bibfield  {author} {\bibinfo {author} {\bibfnamefont {Xiaoyi}\ \bibnamefont
  {He}}, \bibinfo {author} {\bibfnamefont {Xiaowen}\ \bibnamefont {Shan}}, \
  and\ \bibinfo {author} {\bibfnamefont {Gary~D}\ \bibnamefont {Doolen}},\
  }\bibfield  {title} {\enquote {\bibinfo {title} {Discrete boltzmann equation
  model for nonideal gases},}\ }\href@noop {} {\bibfield  {journal} {\bibinfo
  {journal} {Physical Review E}\ }\textbf {\bibinfo {volume} {57}},\ \bibinfo
  {pages} {R13} (\bibinfo {year} {1998}{\natexlab{b}})}\BibitemShut {NoStop}%
\bibitem [{\citenamefont {Dellar}(2013)}]{dellar2013interpretation}%
  \BibitemOpen
  \bibfield  {author} {\bibinfo {author} {\bibfnamefont {Paul~J}\ \bibnamefont
  {Dellar}},\ }\bibfield  {title} {\enquote {\bibinfo {title} {An
  interpretation and derivation of the lattice boltzmann method using strang
  splitting},}\ }\href@noop {} {\bibfield  {journal} {\bibinfo  {journal}
  {Computers \& Mathematics with Applications}\ }\textbf {\bibinfo {volume}
  {65}},\ \bibinfo {pages} {129--141} (\bibinfo {year} {2013})}\BibitemShut
  {NoStop}%
\bibitem [{\citenamefont {Hajabdollahi}\ and\ \citenamefont
  {Premnath}(2018)}]{hajabdollahi2018symmetrized}%
  \BibitemOpen
  \bibfield  {author} {\bibinfo {author} {\bibfnamefont {Farzaneh}\
  \bibnamefont {Hajabdollahi}}\ and\ \bibinfo {author} {\bibfnamefont
  {Kannan~N}\ \bibnamefont {Premnath}},\ }\bibfield  {title} {\enquote
  {\bibinfo {title} {Symmetrized operator split schemes for force and source
  modeling in cascaded lattice boltzmann methods for flow and scalar
  transport},}\ }\href@noop {} {\bibfield  {journal} {\bibinfo  {journal}
  {Physical Review E}\ }\textbf {\bibinfo {volume} {97}},\ \bibinfo {pages}
  {063303} (\bibinfo {year} {2018})}\BibitemShut {NoStop}%
\bibitem [{\citenamefont {Li}\ \emph {et~al.}(2016{\natexlab{b}})\citenamefont
  {Li}, \citenamefont {Zhou},\ and\ \citenamefont {Yan}}]{li2016revised}%
  \BibitemOpen
  \bibfield  {author} {\bibinfo {author} {\bibfnamefont {Q}~\bibnamefont {Li}},
  \bibinfo {author} {\bibfnamefont {P}~\bibnamefont {Zhou}}, \ and\ \bibinfo
  {author} {\bibfnamefont {HJ}~\bibnamefont {Yan}},\ }\bibfield  {title}
  {\enquote {\bibinfo {title} {Revised chapman-enskog analysis for a class of
  forcing schemes in the lattice boltzmann method},}\ }\href@noop {} {\bibfield
   {journal} {\bibinfo  {journal} {Physical Review E}\ }\textbf {\bibinfo
  {volume} {94}},\ \bibinfo {pages} {043313} (\bibinfo {year}
  {2016}{\natexlab{b}})}\BibitemShut {NoStop}%
\bibitem [{\citenamefont {Fei}\ and\ \citenamefont
  {Luo}(2017)}]{fei2017consistent}%
  \BibitemOpen
  \bibfield  {author} {\bibinfo {author} {\bibfnamefont {Linlin}\ \bibnamefont
  {Fei}}\ and\ \bibinfo {author} {\bibfnamefont {Kai~Hong}\ \bibnamefont
  {Luo}},\ }\bibfield  {title} {\enquote {\bibinfo {title} {Consistent forcing
  scheme in the cascaded lattice boltzmann method},}\ }\href@noop {} {\bibfield
   {journal} {\bibinfo  {journal} {Physical Review E}\ }\textbf {\bibinfo
  {volume} {96}},\ \bibinfo {pages} {053307} (\bibinfo {year}
  {2017})}\BibitemShut {NoStop}%
\bibitem [{\citenamefont {Huang}\ \emph {et~al.}(2018)\citenamefont {Huang},
  \citenamefont {Wu},\ and\ \citenamefont {Adams}}]{huang2018eliminating}%
  \BibitemOpen
  \bibfield  {author} {\bibinfo {author} {\bibfnamefont {Rongzong}\
  \bibnamefont {Huang}}, \bibinfo {author} {\bibfnamefont {Huiying}\
  \bibnamefont {Wu}}, \ and\ \bibinfo {author} {\bibfnamefont {Nikolaus~A}\
  \bibnamefont {Adams}},\ }\bibfield  {title} {\enquote {\bibinfo {title}
  {Eliminating cubic terms in the pseudopotential lattice boltzmann model for
  multiphase flow},}\ }\href@noop {} {\bibfield  {journal} {\bibinfo  {journal}
  {Physical Review E}\ }\textbf {\bibinfo {volume} {97}},\ \bibinfo {pages}
  {053308} (\bibinfo {year} {2018})}\BibitemShut {NoStop}%
\bibitem [{\citenamefont {Fei}\ \emph {et~al.}(2018{\natexlab{b}})\citenamefont
  {Fei}, \citenamefont {Luo},\ and\ \citenamefont {Li}}]{fei2018three}%
  \BibitemOpen
  \bibfield  {author} {\bibinfo {author} {\bibfnamefont {Linlin}\ \bibnamefont
  {Fei}}, \bibinfo {author} {\bibfnamefont {Kai~H}\ \bibnamefont {Luo}}, \ and\
  \bibinfo {author} {\bibfnamefont {Qing}\ \bibnamefont {Li}},\ }\bibfield
  {title} {\enquote {\bibinfo {title} {Three-dimensional cascaded lattice
  boltzmann method: Improved implementation and consistent forcing scheme},}\
  }\href@noop {} {\bibfield  {journal} {\bibinfo  {journal} {Physical Review
  E}\ }\textbf {\bibinfo {volume} {97}},\ \bibinfo {pages} {053309} (\bibinfo
  {year} {2018}{\natexlab{b}})}\BibitemShut {NoStop}%
\bibitem [{\citenamefont {Falcucci}\ \emph {et~al.}(2010)\citenamefont
  {Falcucci}, \citenamefont {Ubertini},\ and\ \citenamefont
  {Succi}}]{falcucci2010lattice}%
  \BibitemOpen
  \bibfield  {author} {\bibinfo {author} {\bibfnamefont {Giacomo}\ \bibnamefont
  {Falcucci}}, \bibinfo {author} {\bibfnamefont {Stefano}\ \bibnamefont
  {Ubertini}}, \ and\ \bibinfo {author} {\bibfnamefont {Sauro}\ \bibnamefont
  {Succi}},\ }\bibfield  {title} {\enquote {\bibinfo {title} {Lattice boltzmann
  simulations of phase-separating flows at large density ratios: the case of
  doubly-attractive pseudo-potentials},}\ }\href@noop {} {\bibfield  {journal}
  {\bibinfo  {journal} {Soft Matter}\ }\textbf {\bibinfo {volume} {6}},\
  \bibinfo {pages} {4357--4365} (\bibinfo {year} {2010})}\BibitemShut {NoStop}%
\bibitem [{\citenamefont {Shan}(2008)}]{shan2008pressure}%
  \BibitemOpen
  \bibfield  {author} {\bibinfo {author} {\bibfnamefont {Xiaowen}\ \bibnamefont
  {Shan}},\ }\bibfield  {title} {\enquote {\bibinfo {title} {Pressure tensor
  calculation in a class of nonideal gas lattice boltzmann models},}\
  }\href@noop {} {\bibfield  {journal} {\bibinfo  {journal} {Physical Review
  E}\ }\textbf {\bibinfo {volume} {77}},\ \bibinfo {pages} {066702} (\bibinfo
  {year} {2008})}\BibitemShut {NoStop}%
\bibitem [{\citenamefont {Sbragaglia}\ and\ \citenamefont
  {Belardinelli}(2013)}]{sbragaglia2013interaction}%
  \BibitemOpen
  \bibfield  {author} {\bibinfo {author} {\bibfnamefont {M}~\bibnamefont
  {Sbragaglia}}\ and\ \bibinfo {author} {\bibfnamefont {D}~\bibnamefont
  {Belardinelli}},\ }\bibfield  {title} {\enquote {\bibinfo {title}
  {Interaction pressure tensor for a class of multicomponent lattice boltzmann
  models},}\ }\href@noop {} {\bibfield  {journal} {\bibinfo  {journal}
  {Physical Review E}\ }\textbf {\bibinfo {volume} {88}},\ \bibinfo {pages}
  {013306} (\bibinfo {year} {2013})}\BibitemShut {NoStop}%
\bibitem [{\citenamefont {Derjaguin}\ and\ \citenamefont
  {Churaev}(1978)}]{derjaguin1978question}%
  \BibitemOpen
  \bibfield  {author} {\bibinfo {author} {\bibfnamefont {BV}~\bibnamefont
  {Derjaguin}}\ and\ \bibinfo {author} {\bibfnamefont {NV}~\bibnamefont
  {Churaev}},\ }\bibfield  {title} {\enquote {\bibinfo {title} {On the question
  of determining the concept of disjoining pressure and its role in the
  equilibrium and flow of thin films},}\ }\href@noop {} {\bibfield  {journal}
  {\bibinfo  {journal} {Journal of Colloid and Interface Science}\ }\textbf
  {\bibinfo {volume} {66}},\ \bibinfo {pages} {389--398} (\bibinfo {year}
  {1978})}\BibitemShut {NoStop}%
\bibitem [{\citenamefont {Bergeron}(1999)}]{bergeron1999forces}%
  \BibitemOpen
  \bibfield  {author} {\bibinfo {author} {\bibfnamefont {Vance}\ \bibnamefont
  {Bergeron}},\ }\bibfield  {title} {\enquote {\bibinfo {title} {Forces and
  structure in thin liquid soap films},}\ }\href@noop {} {\bibfield  {journal}
  {\bibinfo  {journal} {Journal of Physics: Condensed Matter}\ }\textbf
  {\bibinfo {volume} {11}},\ \bibinfo {pages} {R215} (\bibinfo {year}
  {1999})}\BibitemShut {NoStop}%
\bibitem [{\citenamefont {Porter}\ \emph {et~al.}(2012)\citenamefont {Porter},
  \citenamefont {Coon}, \citenamefont {Kang}, \citenamefont {Moulton},\ and\
  \citenamefont {Carey}}]{porter2012multicomponent}%
  \BibitemOpen
  \bibfield  {author} {\bibinfo {author} {\bibfnamefont {Mark~L}\ \bibnamefont
  {Porter}}, \bibinfo {author} {\bibfnamefont {E~T}\ \bibnamefont {Coon}},
  \bibinfo {author} {\bibfnamefont {Q}~\bibnamefont {Kang}}, \bibinfo {author}
  {\bibfnamefont {J~D}\ \bibnamefont {Moulton}}, \ and\ \bibinfo {author}
  {\bibfnamefont {J~W}\ \bibnamefont {Carey}},\ }\bibfield  {title} {\enquote
  {\bibinfo {title} {Multicomponent interparticle-potential lattice boltzmann
  model for fluids with large viscosity ratios},}\ }\href@noop {} {\bibfield
  {journal} {\bibinfo  {journal} {Physical Review E}\ }\textbf {\bibinfo
  {volume} {86}},\ \bibinfo {pages} {036701} (\bibinfo {year}
  {2012})}\BibitemShut {NoStop}%
\bibitem [{\citenamefont {Huang}\ \emph {et~al.}(2015)\citenamefont {Huang},
  \citenamefont {Sukop},\ and\ \citenamefont {Lu}}]{huang2015multiphase}%
  \BibitemOpen
  \bibfield  {author} {\bibinfo {author} {\bibfnamefont {Haibo}\ \bibnamefont
  {Huang}}, \bibinfo {author} {\bibfnamefont {Michael}\ \bibnamefont {Sukop}},
  \ and\ \bibinfo {author} {\bibfnamefont {Xiyun}\ \bibnamefont {Lu}},\
  }\href@noop {} {\emph {\bibinfo {title} {Multiphase lattice Boltzmann
  methods: Theory and application}}}\ (\bibinfo  {publisher} {John Wiley \&
  Sons},\ \bibinfo {year} {2015})\BibitemShut {NoStop}%
\bibitem [{\citenamefont {Li}\ \emph {et~al.}(2014)\citenamefont {Li},
  \citenamefont {Chai}, \citenamefont {Shi},\ and\ \citenamefont
  {Liang}}]{li2014deformation}%
  \BibitemOpen
  \bibfield  {author} {\bibinfo {author} {\bibfnamefont {Qiuxiang}\
  \bibnamefont {Li}}, \bibinfo {author} {\bibfnamefont {Zhenhua}\ \bibnamefont
  {Chai}}, \bibinfo {author} {\bibfnamefont {Baochang}\ \bibnamefont {Shi}}, \
  and\ \bibinfo {author} {\bibfnamefont {Hong}\ \bibnamefont {Liang}},\
  }\bibfield  {title} {\enquote {\bibinfo {title} {Deformation and breakup of a
  liquid droplet past a solid circular cylinder: A lattice boltzmann study},}\
  }\href@noop {} {\bibfield  {journal} {\bibinfo  {journal} {Physical Review
  E}\ }\textbf {\bibinfo {volume} {90}},\ \bibinfo {pages} {043015} (\bibinfo
  {year} {2014})}\BibitemShut {NoStop}%
\bibitem [{\citenamefont {Cubaud}\ and\ \citenamefont
  {Mason}(2008)}]{cubaud2008capillary}%
  \BibitemOpen
  \bibfield  {author} {\bibinfo {author} {\bibfnamefont {Thomas}\ \bibnamefont
  {Cubaud}}\ and\ \bibinfo {author} {\bibfnamefont {Thomas~G}\ \bibnamefont
  {Mason}},\ }\bibfield  {title} {\enquote {\bibinfo {title} {Capillary threads
  and viscous droplets in square microchannels},}\ }\href@noop {} {\bibfield
  {journal} {\bibinfo  {journal} {Physics of Fluids}\ }\textbf {\bibinfo
  {volume} {20}},\ \bibinfo {pages} {053302} (\bibinfo {year}
  {2008})}\BibitemShut {NoStop}%
\bibitem [{\citenamefont {Farhat}\ and\ \citenamefont
  {Lee}(2011)}]{farhat2011suppressing}%
  \BibitemOpen
  \bibfield  {author} {\bibinfo {author} {\bibfnamefont {H}~\bibnamefont
  {Farhat}}\ and\ \bibinfo {author} {\bibfnamefont {JS}~\bibnamefont {Lee}},\
  }\bibfield  {title} {\enquote {\bibinfo {title} {Suppressing the coalescence
  in the multi-component lattice boltzmann method},}\ }\href@noop {} {\bibfield
   {journal} {\bibinfo  {journal} {Microfluidics and nanofluidics}\ }\textbf
  {\bibinfo {volume} {11}},\ \bibinfo {pages} {137--143} (\bibinfo {year}
  {2011})}\BibitemShut {NoStop}%
\bibitem [{\citenamefont {Bai}\ \emph {et~al.}(2016)\citenamefont {Bai},
  \citenamefont {Fu}, \citenamefont {Zhao},\ and\ \citenamefont
  {Cheng}}]{bai2016droplet}%
  \BibitemOpen
  \bibfield  {author} {\bibinfo {author} {\bibfnamefont {Lin}\ \bibnamefont
  {Bai}}, \bibinfo {author} {\bibfnamefont {Yuhang}\ \bibnamefont {Fu}},
  \bibinfo {author} {\bibfnamefont {Shufang}\ \bibnamefont {Zhao}}, \ and\
  \bibinfo {author} {\bibfnamefont {Yi}~\bibnamefont {Cheng}},\ }\bibfield
  {title} {\enquote {\bibinfo {title} {Droplet formation in a microfluidic
  t-junction involving highly viscous fluid systems},}\ }\href@noop {}
  {\bibfield  {journal} {\bibinfo  {journal} {Chemical Engineering Science}\
  }\textbf {\bibinfo {volume} {145}},\ \bibinfo {pages} {141--148} (\bibinfo
  {year} {2016})}\BibitemShut {NoStop}%
\bibitem [{\citenamefont {Foglino}\ \emph {et~al.}(2017)\citenamefont
  {Foglino}, \citenamefont {Morozov}, \citenamefont {Henrich},\ and\
  \citenamefont {Marenduzzo}}]{foglino2017flow}%
  \BibitemOpen
  \bibfield  {author} {\bibinfo {author} {\bibfnamefont {M.}~\bibnamefont
  {Foglino}}, \bibinfo {author} {\bibfnamefont {A.~N.}\ \bibnamefont
  {Morozov}}, \bibinfo {author} {\bibfnamefont {O.}~\bibnamefont {Henrich}}, \
  and\ \bibinfo {author} {\bibfnamefont {D.}~\bibnamefont {Marenduzzo}},\
  }\bibfield  {title} {\enquote {\bibinfo {title} {Flow of deformable droplets:
  discontinuous shear thinning and velocity oscillations},}\ }\href@noop {}
  {\bibfield  {journal} {\bibinfo  {journal} {Physical review letters}\
  }\textbf {\bibinfo {volume} {119}},\ \bibinfo {pages} {208002} (\bibinfo
  {year} {2017})}\BibitemShut {NoStop}%
\bibitem [{\citenamefont {Montessori}\ \emph {et~al.}(2018)\citenamefont
  {Montessori}, \citenamefont {Lauricella}, \citenamefont {La~Rocca},
  \citenamefont {Succi}, \citenamefont {Stolovicki}, \citenamefont {Ziblat},\
  and\ \citenamefont {Weitz}}]{montessori2018regularized}%
  \BibitemOpen
  \bibfield  {author} {\bibinfo {author} {\bibfnamefont {Andrea}\ \bibnamefont
  {Montessori}}, \bibinfo {author} {\bibfnamefont {Marco}\ \bibnamefont
  {Lauricella}}, \bibinfo {author} {\bibfnamefont {Michele}\ \bibnamefont
  {La~Rocca}}, \bibinfo {author} {\bibfnamefont {Sauro}\ \bibnamefont {Succi}},
  \bibinfo {author} {\bibfnamefont {Elad}\ \bibnamefont {Stolovicki}}, \bibinfo
  {author} {\bibfnamefont {Roy}\ \bibnamefont {Ziblat}}, \ and\ \bibinfo
  {author} {\bibfnamefont {David}\ \bibnamefont {Weitz}},\ }\bibfield  {title}
  {\enquote {\bibinfo {title} {Regularized lattice boltzmann multicomponent
  models for low capillary and reynolds microfluidics flows},}\ }\href@noop {}
  {\bibfield  {journal} {\bibinfo  {journal} {Computers \& Fluids}\ }\textbf
  {\bibinfo {volume} {167}},\ \bibinfo {pages} {33--39} (\bibinfo {year}
  {2018})}\BibitemShut {NoStop}%
\bibitem [{\citenamefont {Stolovicki}\ \emph {et~al.}(2018)\citenamefont
  {Stolovicki}, \citenamefont {Ziblat},\ and\ \citenamefont
  {Weitz}}]{stolovicki2018throughput}%
  \BibitemOpen
  \bibfield  {author} {\bibinfo {author} {\bibfnamefont {Elad}\ \bibnamefont
  {Stolovicki}}, \bibinfo {author} {\bibfnamefont {Roy}\ \bibnamefont
  {Ziblat}}, \ and\ \bibinfo {author} {\bibfnamefont {David~A}\ \bibnamefont
  {Weitz}},\ }\bibfield  {title} {\enquote {\bibinfo {title} {Throughput
  enhancement of parallel step emulsifier devices by shear-free and efficient
  nozzle clearance},}\ }\href@noop {} {\bibfield  {journal} {\bibinfo
  {journal} {Lab on a Chip}\ }\textbf {\bibinfo {volume} {18}},\ \bibinfo
  {pages} {132--138} (\bibinfo {year} {2018})}\BibitemShut {NoStop}%
\bibitem [{\citenamefont {Derkach}(2009)}]{derkach2009rheology}%
  \BibitemOpen
  \bibfield  {author} {\bibinfo {author} {\bibfnamefont {Svetlana~R}\
  \bibnamefont {Derkach}},\ }\bibfield  {title} {\enquote {\bibinfo {title}
  {Rheology of emulsions},}\ }\href@noop {} {\bibfield  {journal} {\bibinfo
  {journal} {Advances in colloid and interface science}\ }\textbf {\bibinfo
  {volume} {151}},\ \bibinfo {pages} {1--23} (\bibinfo {year}
  {2009})}\BibitemShut {NoStop}%
\bibitem [{\citenamefont {Bullard}\ \emph {et~al.}(2009)\citenamefont
  {Bullard}, \citenamefont {Pauli}, \citenamefont {Garboczi},\ and\
  \citenamefont {Martys}}]{bullard2009comparison}%
  \BibitemOpen
  \bibfield  {author} {\bibinfo {author} {\bibfnamefont {Jeffrey~W}\
  \bibnamefont {Bullard}}, \bibinfo {author} {\bibfnamefont {Adam~T}\
  \bibnamefont {Pauli}}, \bibinfo {author} {\bibfnamefont {Edward~J}\
  \bibnamefont {Garboczi}}, \ and\ \bibinfo {author} {\bibfnamefont {Nicos~S}\
  \bibnamefont {Martys}},\ }\bibfield  {title} {\enquote {\bibinfo {title} {A
  comparison of viscosity--concentration relationships for emulsions},}\
  }\href@noop {} {\bibfield  {journal} {\bibinfo  {journal} {Journal of colloid
  and interface science}\ }\textbf {\bibinfo {volume} {330}},\ \bibinfo {pages}
  {186--193} (\bibinfo {year} {2009})}\BibitemShut {NoStop}%
\bibitem [{\citenamefont {Taylor}(1932)}]{taylor1932viscosity}%
  \BibitemOpen
  \bibfield  {author} {\bibinfo {author} {\bibfnamefont {Geoffrey~I}\
  \bibnamefont {Taylor}},\ }\bibfield  {title} {\enquote {\bibinfo {title} {The
  viscosity of a fluid containing small drops of another fluid},}\ }\href@noop
  {} {\bibfield  {journal} {\bibinfo  {journal} {Proceedings of the Royal
  Society of London. Series A}\ }\textbf {\bibinfo {volume} {138}},\ \bibinfo
  {pages} {41--48} (\bibinfo {year} {1932})}\BibitemShut {NoStop}%
\bibitem [{\citenamefont {Douglas}\ and\ \citenamefont
  {Garboczi}(1995)}]{douglas1995intrinsic}%
  \BibitemOpen
  \bibfield  {author} {\bibinfo {author} {\bibfnamefont {Jack~F}\ \bibnamefont
  {Douglas}}\ and\ \bibinfo {author} {\bibfnamefont {Edward~J}\ \bibnamefont
  {Garboczi}},\ }\bibfield  {title} {\enquote {\bibinfo {title} {Intrinsic
  viscosity and the polarizability of particles having a wide range of
  shapes},}\ }\href@noop {} {\bibfield  {journal} {\bibinfo  {journal}
  {Advances in chemical physics}\ }\textbf {\bibinfo {volume} {91}},\ \bibinfo
  {pages} {85--154} (\bibinfo {year} {1995})}\BibitemShut {NoStop}%
\bibitem [{\citenamefont {Loewenberg}\ and\ \citenamefont
  {Hinch}(1996)}]{loewenberg1996numerical}%
  \BibitemOpen
  \bibfield  {author} {\bibinfo {author} {\bibfnamefont {M}~\bibnamefont
  {Loewenberg}}\ and\ \bibinfo {author} {\bibfnamefont {EJ}~\bibnamefont
  {Hinch}},\ }\bibfield  {title} {\enquote {\bibinfo {title} {Numerical
  simulation of a concentrated emulsion in shear flow},}\ }\href@noop {}
  {\bibfield  {journal} {\bibinfo  {journal} {Journal of Fluid Mechanics}\
  }\textbf {\bibinfo {volume} {321}},\ \bibinfo {pages} {395--419} (\bibinfo
  {year} {1996})}\BibitemShut {NoStop}%
\bibitem [{\citenamefont {Durian}(1995)}]{durian1995foam}%
  \BibitemOpen
  \bibfield  {author} {\bibinfo {author} {\bibfnamefont {Douglas~J}\
  \bibnamefont {Durian}},\ }\bibfield  {title} {\enquote {\bibinfo {title}
  {Foam mechanics at the bubble scale},}\ }\href@noop {} {\bibfield  {journal}
  {\bibinfo  {journal} {Physical review letters}\ }\textbf {\bibinfo {volume}
  {75}},\ \bibinfo {pages} {4780} (\bibinfo {year} {1995})}\BibitemShut
  {NoStop}%
\bibitem [{\citenamefont {Lycett-Brown}\ and\ \citenamefont
  {Luo}(2014)}]{lycett2014multiphase}%
  \BibitemOpen
  \bibfield  {author} {\bibinfo {author} {\bibfnamefont {Daniel}\ \bibnamefont
  {Lycett-Brown}}\ and\ \bibinfo {author} {\bibfnamefont {Kai~H}\ \bibnamefont
  {Luo}},\ }\bibfield  {title} {\enquote {\bibinfo {title} {Multiphase cascaded
  lattice boltzmann method},}\ }\href@noop {} {\bibfield  {journal} {\bibinfo
  {journal} {Computers \& Mathematics with Applications}\ }\textbf {\bibinfo
  {volume} {67}},\ \bibinfo {pages} {350--362} (\bibinfo {year}
  {2014})}\BibitemShut {NoStop}%
\bibitem [{\citenamefont {Lycett-Brown}\ \emph {et~al.}(2014)\citenamefont
  {Lycett-Brown}, \citenamefont {Luo}, \citenamefont {Liu},\ and\ \citenamefont
  {Lv}}]{lycett2014binary}%
  \BibitemOpen
  \bibfield  {author} {\bibinfo {author} {\bibfnamefont {Daniel}\ \bibnamefont
  {Lycett-Brown}}, \bibinfo {author} {\bibfnamefont {Kai~H}\ \bibnamefont
  {Luo}}, \bibinfo {author} {\bibfnamefont {Ronghou}\ \bibnamefont {Liu}}, \
  and\ \bibinfo {author} {\bibfnamefont {Pengmei}\ \bibnamefont {Lv}},\
  }\bibfield  {title} {\enquote {\bibinfo {title} {Binary droplet collision
  simulations by a multiphase cascaded lattice boltzmann method},}\ }\href@noop
  {} {\bibfield  {journal} {\bibinfo  {journal} {Physics of Fluids}\ }\textbf
  {\bibinfo {volume} {26}},\ \bibinfo {pages} {023303} (\bibinfo {year}
  {2014})}\BibitemShut {NoStop}%
\end{thebibliography}%
\end{document}